\documentclass[journal=jacsat,manuscript=article,
layout=twocolumn 
]{achemso}
\usepackage{array}
\usepackage{color}
\usepackage{amsmath}
\usepackage{xcolor}

\usepackage[version=3]{mhchem} 

\author{Mirza Wasif Baig}
\affiliation{J. Heyrovsk\'{y} Institute of Physical Chemistry of the Czech Academy of Sciences, Dolej\v{s}kova 3, 18223 Prague 8, Czech Republic}
\alsoaffiliation{Charles University, Faculty of Science, Department of Physical and Macromolecular Chemistry, Hlavova 8, 12840 Prague 2, Czech Republic}

\author{Marek Pederzoli}
\affiliation{J. Heyrovsk\'{y} Institute of Physical Chemistry of the Czech Academy of Sciences, Dolej\v{s}kova 3, 18223 Prague 8, Czech Republic}

\author{Mojm\'{\i}r K\'{y}vala}
\email{mojmir.kyvala@uochb.cas.cz}
\affiliation{Institute of Organic Chemistry and Biochemistry of the Czech Academy of Sciences, Flemingovo n\'{a}m. 2, 16000 Prague 6, Czech Republic}

\author{Ji\v{r}\'{i} Pittner}
\email{jiri.pittner@jh-inst.cas.cz}
\affiliation{J. Heyrovsk\'{y} Institute of Physical Chemistry of the Czech Academy of Sciences, Dolej\v{s}kova 3, 18223 Prague 8, Czech Republic}
\date{\today}

\title{Quantum Chemical and Trajectory Surface Hopping Molecular Dynamics Study of Iodine-based BODIPY Photosensitizer}

\abbreviations{IR,NMR,UV}

\begin{document}


\begin{abstract}
A computational study of I-BODIPY (2-ethyl-4,4-difluoro-6,7-diiodo-1,3-dimethyl-4-bora-3a,4a-diaza-s-indacene)  has  been carried out to investigate its key photophysical properties as a potential triplet photosensitizer capable of generating singlet oxygen.
Multireference CASPT2 and CASSCF methods have been used to calculate vertical excitation energies and spin--orbit couplings (SOCs), respectively, in a model (monoiodinated BODIPY) molecule to assess the applicability of the single-reference second-order algebraic diagrammatic construction, ADC(2), method to this and similar molecules.
Subsequently, time-dependent density functional theory (TD-DFT), possibly within the Tamm--Dancoff approximation (TDA), using several exchange-correlation functionals has been tested on I-BODIPY against ADC(2), both employing a basis set with a two-component pseudopotential on the iodine atoms.
Finally, the magnitudes of SOC between excited electronic states of all types found have thoroughly been discussed using the Slater--Condon rules applied to an arbitrary one-electron one-center effective spin--orbit Hamiltonian.
The geometry dependence of SOCs between the lowest-lying states has also been addressed.
Based on these investigations, the TD-DFT/B3LYP and TD-DFT(TDA)/BHLYP approaches have been selected as the methods of choice for the subsequent nuclear ensemble approach absorption spectra simulations and mixed quantum-classical trajectory surface hopping (TSH) molecular dynamics (MD) simulations, respectively. 
Two bright states in the visible spectrum of I-BODIPY have been found, exhibiting a red shift of the main peak with respect to unsubstituted BODIPY caused by the iodine substituents.
Excited-state MD simulations including both nonadiabatic effects and SOCs have been performed to investigate the relaxation processes in I-BODIPY after its photoexcitation to the S$_1$ state.
The TSH MD simulations revealed that intersystem crossings occur on a time scale comparable to internal conversions and that after an initial phase of triplet population growth a ``saturation'' is reached where the ratio of the net triplet to singlet populations is about 4:1.
The calculated triplet quantum yield of 0.85 is in qualitative agreement with the previously reported experimental singlet oxygen generation yield of 0.99$\pm$0.06.
	
\end{abstract}

\section{Introduction}

Photosensitizers are photoactive molecules that absorb light and transfer the energy to nearby molecules.
They have many applications including their use in photodynamic therapy (PDT) for the treatment of cancer and tumors.
For the last two decades, 4,4-difluoro-4-bora-3a,4a-diaza-s-indacene (BODIPY) has been known for its potential use as a photosensitizer.~\cite{awuah2012boron}
Both qualitative and quantitative assessments of a photosensitizer used in PDT can be made from spin–orbit couplings (SOCs) between states of different spin multiplicities and the energy gaps involved.~\cite{wenger2020bright,mazzone2016can}
Substitution of iodine in organic molecules renders them photoactive by enhancing their phosphorescent activity.~\cite{li2021unexpected}
Iodinated organic compounds are also interesting from a synthetic point of view.~\cite{churakov2021first,jereb2004effective}
Iodinated BODIPY photosensitizers~\cite{zou2017bodipy} have an advantage over arylated~\cite{alberto2014theoretical} or brominated~\cite{sanchez2017towards} BODIPY photosensitizers due to the more pronounced heavy atom effect of iodine compared to that of bromine, which results in much stronger SOC.~\cite{zou2017bodipy}

BODIPY derivatives have been experimentally studied as important candidates for photosensitizers in PDT~\cite{kamkaew2013bodipy} and potential sensitizers in dye-sensitized solar cells,~\cite{singh2014evolution} but they have also been widely studied theoretically.~\cite{menges2015computational,ponte2018bodipy}
In the last decade, there has been a considerable growth of the literature covering quantum chemical studies of BODIPY derivatives.~\cite{alkhatib2022accurate,spiegel2015quantum,manton2014photo,pogonin2020quantum,lin2020toward, nykanen2024toward} 
In addition to quantum chemical studies, BODIPY dyes have also been subjected to extensive molecular dynamics (MD) studies.~\cite{chen2018insight,song2011orientation}
Among BODIPY derivatives, the iodine-substituted ones have gained most attention due to their numerous applications ranging from their use as an efficient photostable metal-free organic photocatalyst,~\cite{li2013iodo} use in preparation of pyrrolo[2,1-a]isoquinoline,~\cite{guo2013porous} assisting the formation of carbon-carbon bonds via oxidative and reductive quenching,~\cite{huang2013iodo} to their use as photosensitizers in PDT.~\cite{piskorz2021bodipy,zhao2013triplet} 

V. Pomogaev \emph{et al}.\cite{pomogaev2020computational} have recently computed the electronic structure, transition probabilities and corresponding quantum yields of fluorescence for dihalogen-tetraphenyl-aza-BODIPY compounds using both time-dependent density functional theory (TD-DFT) and \emph{ab initio} correlation methods. 
They have achieved good agreement between computed and experimental spectral-luminescent properties with the HSE06 functional and the 6-311G* basis set and have also successfully explained the anomalous dependence of fluorescence efficiency on the atomic number of the halogen substituents.

Y. Lee \emph{et al}.~\cite{lee2020halogen} have employed ultrafast transient absorption spectroscopy to study how the presence of pyridine-based halogen bonding solvent molecules facilitates intersystem crossing (ISC) in a diiodinated BODIPY derivative.
Quantum chemical calculations at the B3LYP/6-311++G(d,p) level of theory have shown how halogen bonding alters both the relative energies of the singlet and triplet states and the SOCs between them.

J. T. Ly \emph{et al}.~\cite{ly2021impact} have synthesized four core and six distyryl-extended methylated-\emph{meso}-phenyl-BODIPY dyes with varying iodine content and have explored the crucial substitution positions for iodine atoms that can induce maximum intersystem crossing rates.
Their experimental studies have been complemented with DFT and TD-DFT calculations.

E. Bassan \emph{et al}.~\cite{bassan2022effect} have prepared a new BODIPY derivative containing an iodine atom in the \emph{ortho} position of the \emph{meso}-linked phenyl group. Their experimental investigation has revealed that this molecule has an efficient population transfer to the triplet state and, unlike core-iodinated derivatives, also maintains the electrochemical properties of unsubstituted BODIPYs. A theoretical investigation has also been carried out using DFT and TD-DFT with the B3LYP and M06-2X functionals and a pseudopotential for iodine atoms.
Solvent (MeCN) effects were included by means of the polarizable continuum model (PCM).

Recently, two new BODIPY derivatives with a general formula MnBr(CO)$_3$(bpy-X-BODIPY), X = H or I, have been synthesized of which the iodine containing one, in particular, can act as both a photoCORM (photo-activated CO-releasing molecule) and a photosensitizer.
Spectral properties of the newly-reported Mn complexes have been investigated using several experimental methods complemented by TD-DFT calculations with the B3LYP functional, the SDD pseudopotentials for Mn, Br and I, and the 6-311G(d) basis set for the other elements.~\cite{pordel2021release}

In another study, \emph{meso}-mesityl-2,6-iodine substituted BODIPY dye has been evaluated as a photosensitizer; TD-DFT excitation energies have been benchmarked with multi-state restricted active space second-order perturbation theory (MS-RASPT2). TD-DFT with the PBE0 functional slightly overestimates the bright HOMO-LUMO transition by 0.17 eV in comparison with MS-RASPT2 but it still allows a qualitative evaluation of the
low-lying excited states of the \emph{meso}-mesityl-2,6-iodine substituted BODIPY.~\cite{ziems2018photo}

Z. Wang \emph{et al}.~\cite{wang2021insight} studied a series of iodine-substituted BODIPY derivatives computationally and experimentally by the time-resolved electron paramagnetic resonance (TREPR) spectroscopy with the aim to explain the drastically different triplet lifetimes observed in different species.
They found that the short triplet lifetimes in some species were caused by strong spin--orbit coupling between one sublevel of the zero-field splitted triplet state with the singlet ground state, which was in accordance with a strong anisotropy of the decay rates.
This anisotropy manifests itself by the electron spin polarization inversion in the TREPR spectra. On the other hand, in the species with long triplet lifetimes no such anisotropy was present.

Y. Chen \emph{et al}.~\cite{chen2018insight} have investigated molecular interactions  between iodine-substituted BODIPY photosensitizers and human serum albumin (HSA) in a combined computational/experimental study. Formation of stable BODIPY-HSA complexes
has been confirmed spectroscopically and it turned out that these complexes exhibit better water solubility and singlet oxygen generation efficiency than the BODIPY molecule itself, which makes them promising biocompatible photosensitizers.

Iodine-substituted BODIPY was also considered as a long wavelength light sensitizer for the near-infrared emission of the ytterbium (III) ion.
It formed a ligand chelating the ytterbium atom, where an energy transfer from the triplet state of the iodinated BODIPY to an excited state of ytterbium (III) occurs, followed by its radiative relaxation to the ground state.~\cite{he2012iodized} Such systems might find an application as novel near-infrared optical probes for sensitive biomedical imaging.

E. Ozcan \emph{et al}.~\cite{ozcan2021halogen} investigated solid-state halogen-bonded frameworks with tunable optical properties based on brominated and iodinated BODIPY derivatives with the --NO$_2$ group as an electron acceptor. 
Molecular crystalline forms of different species exhibited a large variation in their fluorescence properties and optical band gaps depending on the crystal packing, opening possibilities for future design of complex BODIPY-based organic electronic materials.

L. J. Patalag \emph{et al}.~\cite{patalag2022transforming} have presented a systematic study on a series of ethylene-bridged oligo-BODIPYs demonstrating to which extent exciton formation can amplify fluorescence.
They have also explored different BODIPY motifs with regard to the deactivation pathways responsible for a suppressed light emission at the monomeric state.
Diiodo-BODIPY derivatives have been reported to have rapid ISC populating a low-energy triplet state thus effectively quenching fluorescence.

J. Dole\v zel \emph{et al}.~\cite{dolezel2023spin} synthesized and examined, both experimentally and computationally, seven $\pi$-extended BODIPY derivatives with iodine atoms in different positions.
They found that the heavy-atom effect of iodine atoms on the $\text{S}_1\to\text{T}_2$ ISC rate is site-specific, causing high triplet yields in only some positions, and attributed this observation to geometry dependence of SOC.

\begin{figure}
\centering
\includegraphics[width=7cm]{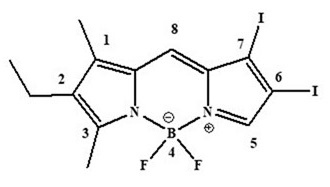}
\caption{Molecular structure of I-BODIPY.}
\label{fig:ibodipy} 
\end{figure}

Recently, we have investigated the excited-state dynamics of bromine- and/or alkyl-substituted BODIPY derivatives employing mixed quantum-classical fewest-switches surface-hopping (FSSH) molecular dynamics (MD) at the TD-DFT level of electronic structure theory, including both nonadiabatic effects and spin--orbit couplings.~\cite{wasif2021theoretical}
In a preceding work, in addition to trajectory surface hopping (TSH) simulations, purely classical MD study was also performed to explore the penetration of Br-BODIPY into biological membranes.~\cite{pederzoli2019photophysics}
To the best of our knowledge, there is presently no such excited-state dynamics study of an iodinated BODIPY derivative available in the literature.
In this work, we employ FSSH MD to investigate the photophysics of I-BODIPY (2-ethyl-4,4-difluoro-6,7-diiodo-1,3-dimethyl-4-bora-3a,4a-diaza-s-indacene) whose molecular structure is shown in Fig.~\ref{fig:ibodipy}.
This specie has been reported as an efficient photosensitizer due to its high quantum yield of singlet oxygen (${}^1$O${}_2$) generation, which is essential for use in PDT.~\cite{sanchez2017towards}

\section{Methodology}

\subsection{Methods and basis sets}

DFT and TD-DFT calculations have been carried out using the program Turbomole v7.0.1~\cite{ahlrichs1989electronic} with the hybrid functionals B3LYP~\cite{becke1993density, stephens1994ab} and BHLYP~\cite{becke1993new} and the hybrid meta functional M06-2X~\cite{zhao2008m06} employing the dhf-TZVP basis set~\cite{weigend1998ri,weigend2010segmented} with the (spin-averaged part of) the multiconfiguration-Dirac--Hartree--Fock-adjusted two-component pseudopotential by K. A. Peterson \emph{et al}.~\cite{peterson2006spectroscopic} on the iodine atoms.
All triplets as well as all excited singlets were represented in linear-response TD-DFT by excitations from the closed-shell S${}_0$ Slater determinant calculated by spin-restricted Kohn--Sham SCF.

Turbomole has been used for geometry optimization of both monoiodinated BODIPY in the position 2 and I-BODIPY in the ground state S$_0$ by DFT with the B3LYP functional employing the dhf-TZVP basis set.
Gaussian 09~\cite{frisch2009gaussian} has been used for geometry optimization of I-BODIPY in the excited states S${}_1$ and T${}_2$ by TD-DFT with the B3LYP functional due to the availability of analytical second derivatives.
The aug-cc-pVDZ basis set~\cite{dunning1989gaussian, kendall1992electron} has been employed for all atoms except iodine, for which we used the aug-cc-pVDZ-PP basis set~\cite{peterson2003systematically, peterson2006spectroscopic}.
For all geometry optimizations, the Grimme’s dispersion correction was taken into account.~\cite{grimme2016dispersion}.

ADC(2) calculations have been carried out by the program Turbomole employing the dhf-TZVP basis set.
DKH2 (second-order Douglas--Kroll--Hess) CASSCF and CASPT2 calculations have been performed using the program Molcas~7~\cite{aquilante2010molcas} employing the TZP contracted ANO-RCC basis set.~\cite{roos2004main}
DMRG calculations have been done using the program MOLMPS\cite{brabec2021massively} employing the dhf-TZVP basis set.
The program BDF~\cite{liu1997beijing, zhang2020bdf} has been used mainly for the calculation of electron densities in the ground and lowest-excited singlet states by the sf-X2C-S-TD-DFT method~\cite{liu2018relativistic} with the B3LYP functional employing the x2c-TZVPPall basis set.~\cite{pollak2017segmented}  

For nuclear ensemble approach (NEA) absorption spectra simulations\cite{crespo2012spectrum} and mixed quantum-classical Tully's FSSH MD simulations\cite{tully1990molecular,wang2016recent,crespo2018recent,nelson2020nonadiabatic} including ISCs~\cite{richter2011sharc,cui2014generalized} the program Newton-X~\cite{barbatti2014newton,barbatti2022newton} with our implementation of
nonadiabatic dynamics with time-derivative couplings~\cite{pittner2009optimization} and
spin--orbit couplings~\cite{pederzoli2017new} applying the 3-step integrator approach~\cite{mai2015general} has been used.
All simulations have been performed using the dhf-TZVP basis set.

In TSH MD, often decoherence corrections are employed~\cite{granucci2007critical,granucci2010including,jain2016efficient}. 
However, they were designed for localized, quickly damped couplings and their use might be problematic when SOCs are included~\cite{granucci2012surface,heller2021exploring}.
We thus did not use any decoherence correction in our simulations.

\subsection{Spin-orbit couplings}

SOCs (as well as the so-called orbital overlaps, see below) have been calculated using a program written by Mojm\'{\i}r K\'{y}vala employing (possibly orthonormalized) configuration interaction singles-like (CIS-like) auxiliary wave functions for the triplets and excited singlets obtained by ADC(2) or linear-response TD-DFT.~\cite{pederzoli2019photophysics,wasif2021theoretical}
Another program by the same author has been used for the calculation of SOCs between CASSCF and CASCIS (MRCIS on top of CASSCF) states as well as for the calculation of the CASCIS states themselves.

The employed effective one-electron spin--orbit Hamiltonian is the sum of
the Breit--Pauli or DKH1 (first-order Douglas--Kroll--Hess) one-electron spin--orbit
Hamiltonian~\cite{hess1995abinitio} (which itself is a sum of
contributions of individual atoms of a molecule) and a basis set independent
approximation to the ``two-electron'' part of the `exact' Breit--Pauli or DKH1 one-electron mean-field spin--orbit Hamiltonian.~\cite{hess1996mean}
The approximation is a gentle modification of the so-called flexible nuclear
screening spin--orbit (FNSSO) approximation~\cite{chalupsky2013flexible}
showing mean relative error below a few percent for virtually all elements
of the periodic table in most bonding situations.
A more descriptive view at the approximation is based
on the explicit form of e.g. the Breit--Pauli one-electron spin--orbit
Hamiltonian~\cite{hess1995abinitio} (in SI units)
\begin{equation}
\nonumber
\hat h(1)=\frac{e^2}{8\pi\varepsilon_0m^2c^2}\sum_CZ_Cr_{C}^{-3}\left(\mathbf{r}_C\times\mathbf{\hat p}\right)\cdot\mathbf{\hat s},
\end{equation}
where
$\varepsilon_0$ is the permittivity of vacuum,
$c$ is the velocity of light in vacuum,
$e$ is the charge and $m$ is the mass of an electron,
$Z_C$ is the number of protons ($Z_Ce$ is the charge) of the nucleus $C$,
$\mathbf{r}_C$ is the position vector of an electron with respect to the
nucleus $C$, $r_C=|\mathbf{r}_C|$, and
$\mathbf{\hat p}$ or $\mathbf{\hat s}$ is the operator of the vector of
linear momentum or spin angular momentum of an electron, respectively.
Accordingly, the employed effective spin--orbit Hamiltonian is essentially
the Breit--Pauli or DKH1 one-electron spin--orbit
Hamiltonian in which, for each nucleus $C$, the true nuclear charge $Z_Ce$
has been replaced with an effective nuclear charge $(1-q_C)Z_Ce$, where
the so-called screening quotient
$q_C\equiv q_C(l,\alpha,\beta)\in\langle 0,1\rangle$
depends not only on the nucleus $C$ but \emph{also} on the two one-electron
primitive Gaussian functions bracketing the operator in a
particular one-center matrix element---through their common azimuthal
quantum number $l$ (describing their angular parts) and individual exponents
$\alpha$ and $\beta$ (describing their radial parts).
Clearly, an electron close enough to the atomic nucleus experiences only the
electric field brought by the positive nuclear charge $Z_Ce$
(as the contributions of all the electrons of the atom to the spherically symmetric electric field at the position of the nucleus mutually cancel)
while an electron in the infinity experiences just the
zero electric field of the neutral atom. Therefore, the quotient $q_C$
should almost vanish for the tightest basis functions and
should approach $1$ for the sufficiently diffuse ones. Moreover, the
quotient $q_C$ should never be lower than $0$ or greater than $1$ as the
fictitious electron can nowhere in space experience
a spherically symmetric electric field of a positive charge greater than
$Z_Ce$ or of a negative charge.

The dependence of $q_C$ on the variables $l$, $\alpha$ and $\beta$ has been determined through nonrelativistic or (scalar relativistic) DKH2 full-valence CASSCF calculations on the ground states of neutral atoms (employing large \emph{uncontracted} one-electron bases) followed by the evaluation of the nonzero matrix elements of the `exact' Breit--Pauli or DKH1 one-electron mean-field spin--orbit Hamiltonian.

The contribution of a heavy atom on which a \emph{two-component} (relativistic, $j$-dependent) pseudopotential~\cite{lee1977ab} is defined (and a reduced number of basis functions are centered to represent just the pseudovalence orbitals instead of the full set of core and valence orbitals) to matrix elements of the effective one-electron spin--orbit Hamiltonian is best evaluated using an operator implicitly included in the two-component pseudopotential itself.~\cite{pitzer1988electronic}

The central idea behind the pseudopotential
approximation~\cite{schwerdtfeger2011pseudopotential,dolg2012relativistic} is to select
from the total number
of $n$ (actually indistinguishable!) electrons of a molecule the $q$ valence
electrons (responsible for all the chemistry) that are effectively moving in
the electric field of the remaining $n-q$ core electrons and the positively
charged nuclei (to which the core electrons are ``stuck'' in exactly the
same way as in the isolated atoms).
Heavy nuclei are thus replaced with spherical cores, while
$Q_Ce$ is the charge of the core whose center is at the nucleus $C$
with the charge $Z_Ce$, $0<Q_C<Z_C$. Consequently,
$$q=n-\sum_C(Z_C-Q_C).$$
Point-like cores can be regarded as the roughest, zero-order, approximation.
Therefore, the valence-only effective Hamiltonian of a molecule (in SI
units)
\begin{eqnarray}
\nonumber
\hat H(1,\dots,q)&=&
\frac{1}{2m}\sum_{i=1}^q\hat p_i^2\\
&&\nonumber
+\frac{e^2}{4\pi\varepsilon_0}\left\{\sum_{i=1}^q\sum_C
\left[-\frac{Q_C}{r_{iC}}+\Delta\mathrm{v}_C(r_{iC})\right]\right.\\
&&\nonumber
\left.{}+\sum_{i=1}^q\sum_{j>i}^q\frac{1}{r_{ij}}+
\sum_C\sum_{D>C}\frac{Q_CQ_D}{ r_{CD}}\right\}
\end{eqnarray}
may be introduced which, in its simplest practicable form, stands for kinetic energy of the
valence electrons and potential energy of the Coulombic (\emph{i})
attraction between the valence electrons and the cores (taking into account
also the correction for the error of the zero-order
approximation by means of pre-parameterized, through all-electron relativistic calculations on atoms or their ions, one-electron operators $\Delta\mathrm{v}_C$ called pseudopotentials),
(\emph{ii}) repulsion between the valence electrons,
and (\emph{iii}) repulsion between the (point-like) cores, where
$\hat{\bf p}_i$ is the linear momentum of the electron $i$,
$\varepsilon_0$ is the permitivity of vacuum,
$e$ is the charge and $m$ is the mass of an electron,
$r_{iC}$ is the distance between the electron $i$ and the nucleus $C$ while
$r_{ij}$ or $r_{CD}$ is the distance between two electrons
or two nuclei (point-like cores), respectively.
The energy of the cores themselves is supposed to be a constant and has been subtracted.

The pseudopotential $\Delta\mathrm{v}_C$ is a function of the distance of an electron from the nucleus $C$ and is generally different for basis functions of different angular symmetries.
It is repulsive in the short range (so as to keep the valence electrons out of the core) and attractive in the long range.
Typically, $Q_C$ differs from $Z_C$ only for the atoms of the heaviest elements, often just one or two atoms in the molecule.
For atoms with $Q_C=Z_C$ the pseudopotential $\Delta\mathrm{v}_C$ clearly vanishes and a full one-electron basis is required.
The contribution of such an atom to matrix elements of the effective one-electron spin--orbit Hamiltonian is then calculated using the (Breit--Pauli variant of the) \emph{one-center} FNSSO approximation described above.
For atoms with $Q_C<Z_C$ the related part of the \emph{one-center} effective spin--orbit Hamiltonian
is extracted from the two-component pseudopotential $\Delta\mathrm{v}_C$ by spin separation.~\cite{pitzer1988electronic}

Adoption of the one-center approximation enabled us, first, to avoid implementing the existing, rather complicated, algorithm for the evaluation of multicenter matrix elements of the effective spin--orbit Hamiltonian implicitly included in a two-component pseudopotential between \emph{redundant} Cartesian Gaussian functions~\cite{pitzer1991spin} and, second, as a nice bonus, to exploit spherical symmetry.
The derivation of the new, truly simple, formulae for the calculation of matrix elements of the one-electron \emph{one-center} effective spin--orbit Hamiltonian implicitly included in a two-component pseudopotential between \emph{nonredundant} Cartesian Gaussian functions is given in SI on pp. 110--115.

It should be mentioned that in the aforesaid typical case, where pseudopotentials are only defined for a subset of the atoms of a molecule,
the one-center approximation to SOC seems to be the natural choice. Clearly, if the multicenter
contributions were evaluated using an all-electron spin--orbit Hamiltonian on lighter atoms, they would be incomplete and thus incorrect
due to the missing tight basis functions describing core orbitals on the heavy atoms with pseudopotentials. Then it is perhaps better not to calculate
them at all. 
However, if the multicenter contributions are not calculated on the lighter atoms, they should probably not be calculated on the heavy
atoms either to avoid unbalanced treatment.

\subsection{Accelerated nonadiabatic MD approach to ISC}

Given that iodinated BODIPY derivatives are known to exhibit ISCs on the picosecond timescale,~\cite{lee2020halogen} we employed an accelerated approach in our simulations. 
Following the works of K. A. Lingerfelt et al.~\cite{lingerfelt2016direct} and A. Nijamudheen \emph{et al.}~\cite{nijamudheen2017excited}, we assume that the rates of slow nonadiabatic transitions are reasonably well captured by the Fermi's golden rule which describes the transition rate $k$ as being proportional to the square of the matrix element $V$ of the related perturbation operator between the initial and final adiabatic states,
\begin{equation}
\nonumber
k \propto V^2. 
\end{equation}
This relation implies that by artificially enhancing the coupling $V$ by a factor $\alpha > 1$, the dynamics of slow nonadiabatic transitions will be sped up by a factor of $\alpha^2$ and the corresponding time constant or lifetime will decrease to $\tau_\alpha = \tau_1 \alpha^{-2}$. 
In practice one performs simulations for several (rather small) values of $\alpha$ and extrapolates the calculated time constants or lifetimes to $\alpha=1$, preferably by a straight line in logarithmic scale in which $\log\tau_\alpha=\log\tau_1-2\log\alpha$.

This approach allows the acceleration of slow relaxation processes to a time window (of about 1~ps) more practical for NAMD simulations.
In Refs.~\cite{lingerfelt2016direct,nijamudheen2017excited} the scaling was applied to nonadiabatic couplings (NACs).
However, in our case the slow process is ISC governed by SOC while the same-spin nonadiabatic transitions are fast enough and need no acceleration.
We have thus applied the scaling to SOCs only, leaving NACs unchanged (as it was done also in our previous works~\cite{pederzoli2019photophysics,wasif2021theoretical}).

\section{Results and discussion}

\subsection{Ground- and excited-state geometries}

The optimized ground-state geometry of I-BODIPY is planar with the terminal methyl group of the ethyl moiety protruding out of the plane of the molecule away from the pyrrole ring by almost 113 degrees,
slightly more than is the valence angle 109.5 degrees of a perfect tetrahedral bond.
The dihedral angle between this methyl group and the plane of the molecule is about 80 degrees
(measured with respect to the carbon atom bound to the closest nitrogen atom).
The bond lengths between the B and N atoms are nearly 1.6~\AA{} while the bond lengths between the N and C atoms as well as between two C atoms forming the rings are about 1.4~\AA.
The bond lengths between the C and I atoms are approximately 2.1~\AA.

The optimized T$_2$ geometry is almost identical to the ground-state geometry, especially if calculated using the same basis set, while the optimized S$_1$ geometry differs mainly in the bond length between the two carbon atoms bearing the iodine substituents: it is by almost 0.1~\AA{} longer in the first excited singlet than in the ground state (or in the lowest triplet).

\subsection{Benchmarking TD-DFT functionals}

M. R. Momeni \emph{et al.}~\cite{momeni2015why} calculated the vertical excitation energies of the lowest singlet S${}_1$ in 17 BODIPY derivatives both by TD-DFT using 9 density functionals and by a plethora of \emph{ab initio} methods including TD-HF, CIS, CIS(D), EOM-CCSD, SAC-CI, CC2, LR-CCSD, CCSDR(T), CASSCF or CASPT2 (the last two within several active spaces up to 12 electrons in 11 orbitals) emploing the cc-pVDZ and cc-pVTZ basis sets and compared them with experimental values.
The observed failure of TD-DFT (mean absolute error greater than 0.3 eV) was 
attributed to the detected different amounts of electron correlation in the ground and excited states as well as to the finding that, while the ground state could in all cases be well treated by single-reference coupled cluster methods, the excited state S${}_1$ had, quite naturally and hardly surprisingly, a multireference character (i.e., was represented by a multiconfiguration wave function) and, in some cases, might not be described accurately enough without including double excitations from the dominant ground-state configuration.

M. de Vetta \emph{et al.}~\cite{devetta2019role} evaluated the performance of ADC(2) on the unsubstituted BODIPY molecule employing the aug-cc-pVDZ basis set against multistate CASPT2 using the active space of 12 electrons in 11 orbitals and the TZ contracted ANO-L basis set and found that the position of the dominant transition in the spectrum obtained by CASPT2 was 0.18 eV blue-shifted from the experimental value while ADC(2) yielded a further 0.13 eV blue shift with respect to CASPT2.
Notice, though, that CASPT2 using a small active space and basis lacking diffuse functions can yield a result inferior to ADC(2).~\cite{devetta2019ctc}.

In a recent paper, V. Postils \emph{et al.}~\cite{postils2021mild} chose basically all the methods and density functionals used in the three papers mentioned above and, employing the cc-pVTZ basis set, calculated the vertical excitation energies of the electronic states S${}_1$ and T${}_1$ of unsubstituted BODIPY to uncover why not only TD-DFT but also some wave-function-based methods cannot accurately predict their energies. The main reason was reported to be a mild open-shell character of (i.e., strong HOMO and LUMO exchange interaction in) the ground state of the molecule.

Based on these observations (as well as on their own TD-DFT and \emph{ab initio} benchmark calculations), J. Dole\v zel \emph{et al}.~\cite{dolezel2023spin} concluded that for a computational treatment of BODIPY dyes a multireference method is necessary and applied DKH2 NEVPT2 to a series of multiply iodinated $\pi$-extended BODIPY derivatives using the active space of 6 electrons in 4 orbitals and an all-electron double-$\zeta$ basis set.

We used a much simpler approach and compared the two leading CASCI weights (squares of expansion coefficients) in the ground states of BODIPY, monoiodinated BODIPY in the position 2, and some (mostly aromatic and presumably ``single-reference'') organic molecules in the basis of state-specific CASSCF \emph{natural} orbitals to find out if the ever lasting effort to study the electronic structure of BODIPY dyes by single-reference methods is really a lost battle.
At first sight, there is no indication that the ground state of BODIPY or monoiodinated BODIPY has a multireference character.
The magnitude of the leading weight naturally decreases with the increasing size of the active space, but if we compare active spaces of similar sizes, the leading weights in BODIPY and monoiodinated BODIPY never fall below those calculated for typical "single-reference" molecules like pyrrole, benzene or naphtalene, see Table S24.

As an additional test of the possible multireference character of the ground states of BODIPY derivatives, we performed a DMRG calculation of the ground state of monoiodinated BODIPY in the position 2 using (frozen) RHF canonical orbitals and an active space consisting of 32 electrons in 30 orbitals with bond dimension 512.
The resulting orbital entropies~\cite{boguslawski2013orbital} were all below 0.3, and only for HOMO and LUMO their values exceeded 0.2, which can be seen in Table S25.
The leading CI expansion coefficient (reconstructed from the MPS wave function) was 0.95 while the next largest one (in absolute value) of $-0.12$ belonged to the double excitation from HOMO to LUMO.
For I-BODIPY we carried out an analogous DMRG calculation (48e in 46o, bond dimension 1024) and the resulting orbital entropies gave a qualitatively identical picture (cf. Table S26).
We thus see no clear reason why single-reference methods like ADC(2) or TD-DFT should be rejected \emph{a priori} for methods applicable to I-BODIPY and similar molecules.

Moreover, multireference methods like CASSCF and CASPT2 (or NEVPT2) should not be overestimated. They usually work well within the so-called full valence approximation, i.e., if \emph{all} valence molecular orbitals (MOs) are taken into account, but may fail when used with some artificially restricted active spaces.
For planar unsaturated organic molecules it is mostly sufficient to include in the active space only the $\pi$-symmetry (out-of-plane) MOs possibly supplemented with some of the $\sigma$-symmetry (in-plane) MOs located on selected heteroatoms, but as soon as the molecule loses its planarity, the application of these methods often starts to be problematic.
It should also be emphasized that while an ADC(2) excitation energy is (for a given molecule and basis set) a unique number (not necessarily correct), a CASPT2 (or NEVPT2) excitation energy is, unfortunately, never a single value but rather a range of possible values depending on the selected active space and MOs---state-specific or state-averaged---and for the latter also on the number (and types) of states included in the averaging (together with their weights as the case might be).
Therefore, if someone claims that some CASPT2 excitation energy has a sharp value, it means that they probably chose from such a range a value most suitable for their purposes.

Concerning NAMD simulations of a molecule as large as I-BODIPY, one is, for efficiency reasons, essentially limited to TD-DFT.
Nevertheless, for the sake of prudence, we first benchmarked the method, using different exchange-correlation functionals, against both ADC(2) and CASPT2.

For a reasonable description of a sufficient number of the lowest-lying electronic states of a halogenated BODIPY derivative by CASSCF and CASPT2, the active space should consist of at least 12 electrons in 11 orbitals (6 $\pi$ bonding and 5 $\pi$ antibonding MOs of the parent BODIPY molecule, i.e., linear combinations of the 11 AOs 2p, all perpendicular to the plane of the rings, of the 9 carbon and 2 nitrogen atoms) plus 6 electrons in 4 orbitals (one C--X $\sigma$ bonding MO, two mutually perpendicular basically valence AOs p of the halogen atom X, both doubly occupied and perpendicular to the C--X bond, one of them perpendicular to the plane of the rings and thus slightly bonding due to its interaction with the less stable $\pi$-symmetry MOs of the parent BODIPY molecule and the other parallel to the plane of the rings and thus essentially nonbonding, and one C--X $\sigma$ antibonding MO) per each halogen substituent.
Such a requirement effectively restricts the aforesaid methods to monohalogenated BODIPY derivatives.
Therefore, (scalar relativistic) DKH2 CASSCF and CASPT2 calculations on monoiodinated BODIPY in the position 2 (without any alkyl substituents), a simplified model of I-BODIPY, in its planar ground-state geometry (belonging to the point group C${}_s$) have been carried out to gain a deeper insight into the electronic structure of I-BODIPY and to assess the accuracy of both ADC(2) and TD-DFT using various exchange-correlation functionals.

The ultimate DKH2 CAS(18,15)SCF calculations of wave functions and energies have been performed in two steps.
First, state-specific CAS(16,14)SCF calculations with the essentially nonbonding doubly occupied AO 5p of iodine parallel to the plain of the rings kept out of the active space have been carried out for the lowest singlets and triplets of symmetries A${}'$ and A${}''$, namely for the states 1SA${}'$, 2SA${}'$, 1SA${}''$, 1TA${}'$, and 1TA${}''$.
Second, one state-specific and four state-averaged frozen-core CAS(18,15)SCF calculations have been done for the ground singlet and several lowest excited states of all four combinations of the two spatial symmetries and two spin multiplicities using the five sets of the inactive (always doubly occupied) MOs obtained in the first step as constant (except for the nonbonding AO 5p of iodine now included in the active space).
Every frozen-core CAS(18,15)SCF calculation has been followed, after reducing the number of frozen, although now in a completely different sense, MOs to just 26 of symmetry A${}'$ and 6 of symmetry A${}''$ (i.e., to those forming the subvalence shells), by the corresponding CAS(18,15)PT2 calculation of electronic energies (or energy).
Finally, singlets of symmetry A${}'$ were orthonormalized through diagonalization of the matrix of the scalar relativistic Hamiltonian in the basis of the mutually nonorthogonal CASSCF states with the CASPT2 energies on the diagonal.
Attempts to carry out any of the CAS(18,15)SCF calculations in a single (unconstrained) step while holding all the important MOs specified above in the active space were not successful.

Four types of electronic excited states have been found among the lowest-lying singlets and triplets, two for each spatial symmetry.
There are many $(\pi,\pi^*)$ states dominated by excitations from a $\pi$ bonding to a $\pi$ antibonding MO and one $(n,\sigma^*)$ state dominated by the excitation from the essentially nonbonding AO 5p of iodine parallel to the plane of the rings to the C--I $\sigma$ antibonding MO.
These types of states belong to the symmetry species A${}'$.
There are also several $(\pi,\sigma^*)$ states dominated by excitations from a $\pi$ bonding MO to the C--I $\sigma$ antibonding MO and a few $(n,\pi^*)$ states dominated by excitations from the nonbonding AO 5p of iodine to a $\pi$ antibonding MO.
These types of states belong to the symmetry species A${}''$.
Electronic states dominated by excitations from the C--I $\sigma$ bonding MO, namely $(\sigma,\sigma^*)$ and $(\sigma,\pi^*)$, have not been found, probably due to the fact that their CASSCF excitation energies are outside the selected windows.

Vertical excitation energies of the low-lying excited states of monoiodinated BODIPY in the position 2 calculated by DKH2 CAS(18,15)PT2 are given in Table S1a.
Symmetries and characters (of the dominant excitations) of the excited electronic states are included as well. For comparison, similar calculations have been done also by the aforementioned less demanding single-reference (i.e., black-box) methods ADC(2) and TD-DFT employing the B3LYP, BHLYP, and M06-2X functionals.
The results of these calculations are given in Tables S1b through S1g.
Besides the excitation energy and oscilator strength the so-called orbital overlap has also been calculated (by numerical integration on a grid) for each excited state as the sum of the overlap integrals of the \emph{absolute values} of a pair of MOs (the one from which an electron was promoted and the one to which the electron was promoted) weighted by the square of the corresponding CIS expansion coefficient of the particular auxiliary wave function.~\cite{peach2008excitation}
The orbital overlaps, whose values lie between 0 and 1, are measures of locality of the (single) excitations dominating the ADC(2) and TD-DFT excited states. Their small values, typically less than 0.3, may indicate either a charge-transfer (CT) or a Rydberg state.

\begin{figure}
\centering
\includegraphics[width=7cm]{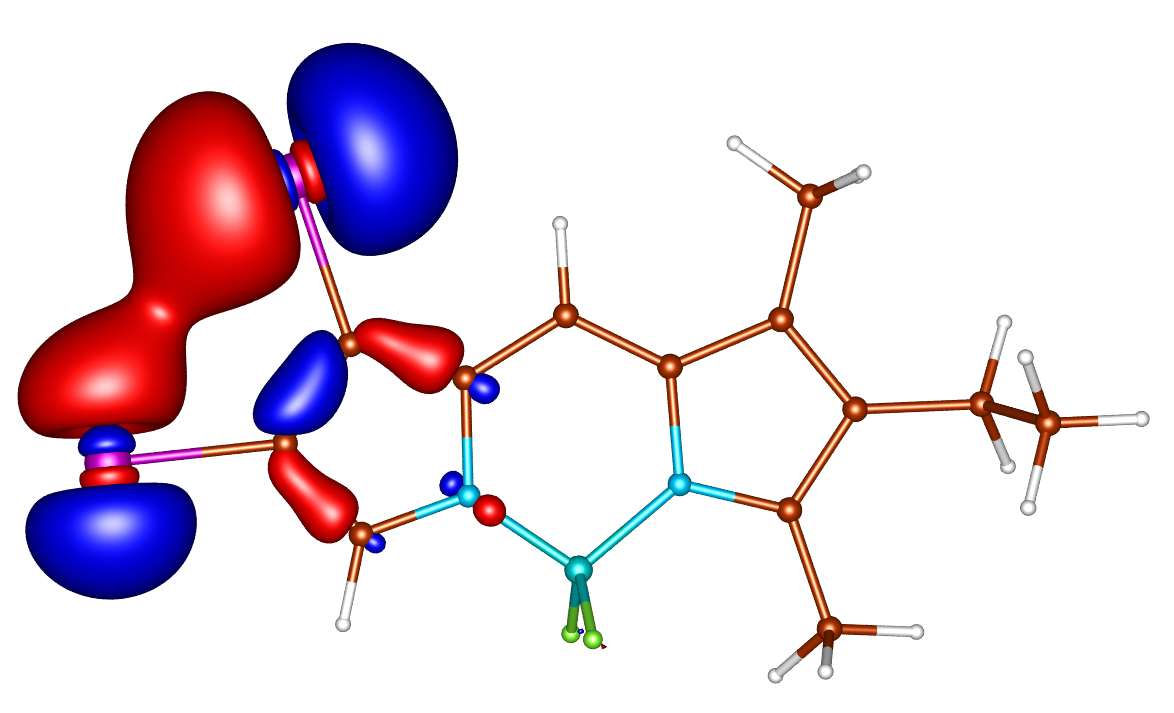}
\caption{In-plane nonbonding ($n$) HOMO-5 of I-BODIPY, orbital energy -7.67 eV.}
\label{fig:homo5_ibodipy} 
\end{figure}

\begin{figure}
\centering
\includegraphics[width=7cm]{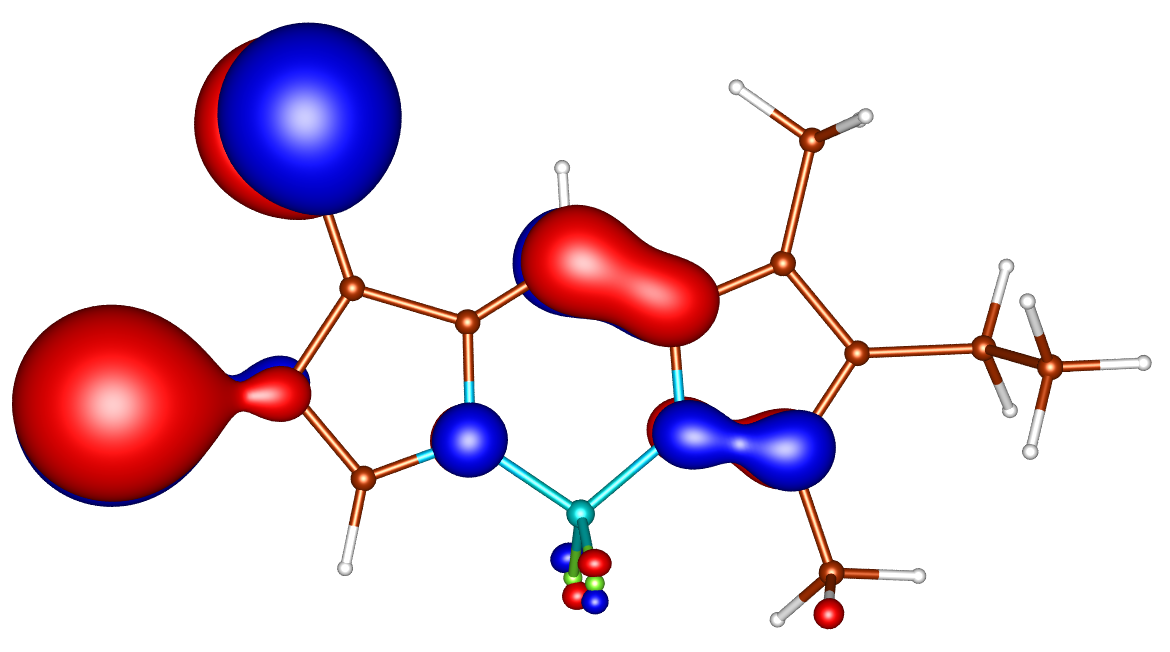}
\caption{Out-of-plane nonbonding/$\pi$ bonding ($\pi$) HOMO-4 of I-BODIPY, orbital energy -7.65 eV.}
\label{fig:homo4_ibodipy} 
\end{figure}

\begin{figure}
\centering
\includegraphics[width=7cm]{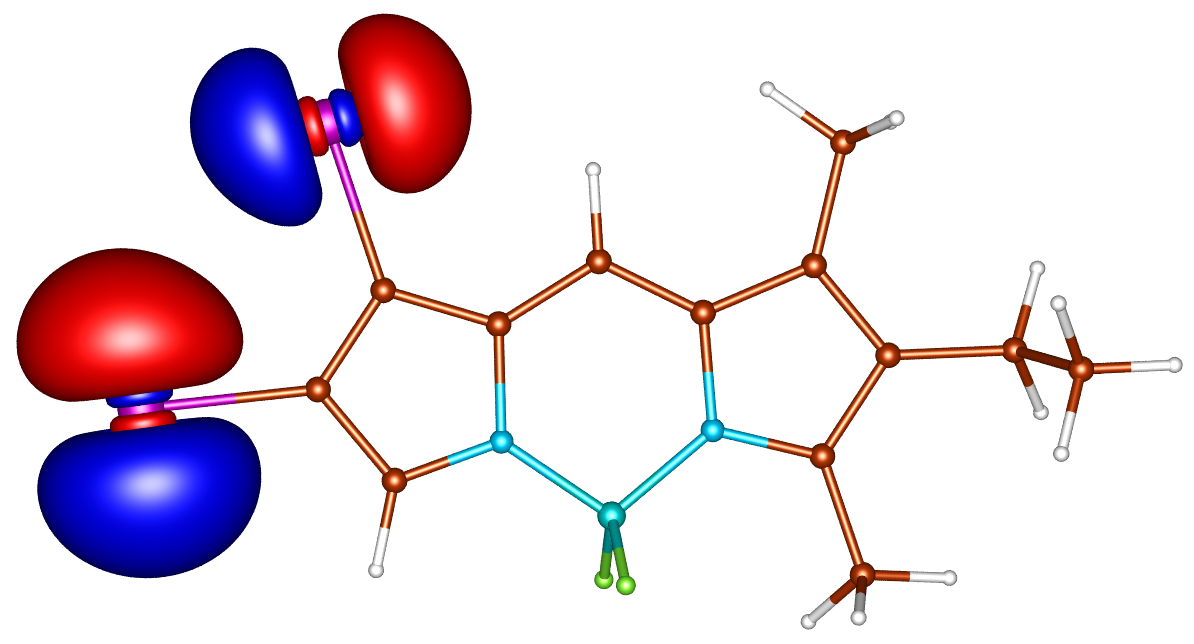}
\caption{In-plane nonbonding ($n$) HOMO-3 of I-BODIPY, orbital energy -7.20 eV.}
\label{fig:homo3_ibodipy} 
\end{figure}

\begin{figure}
\centering
\includegraphics[width=7cm]{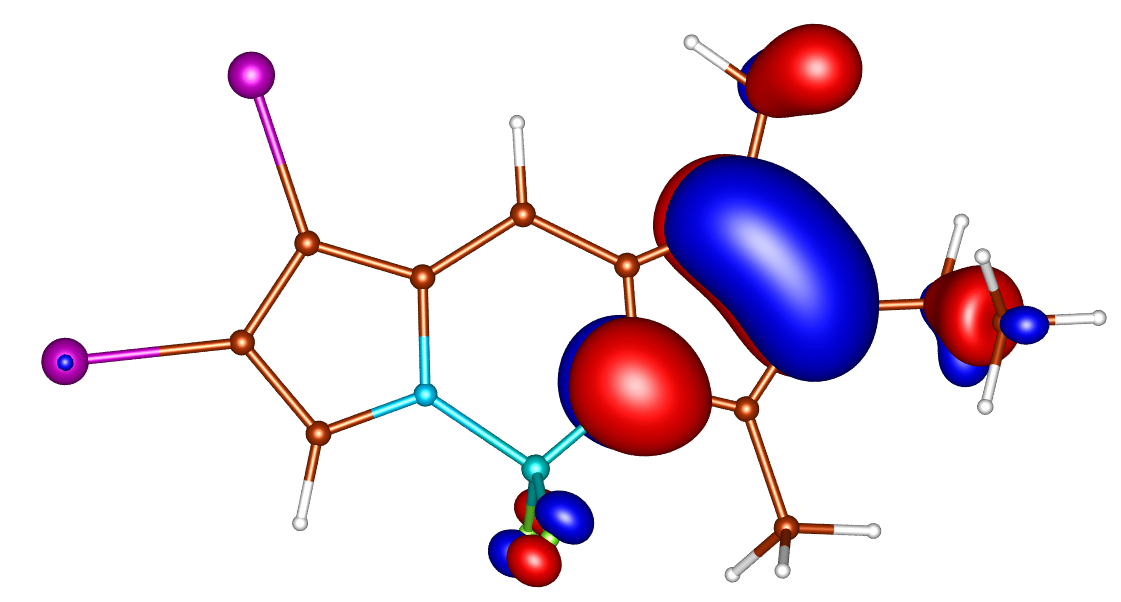}
\caption{$\pi$ bonding ($\pi$) HOMO-2 of I-BODIPY, orbital energy -7.07 eV.}
\label{fig:homo2_ibodipy} 
\end{figure}

\begin{figure}
\centering
\includegraphics[width=7cm]{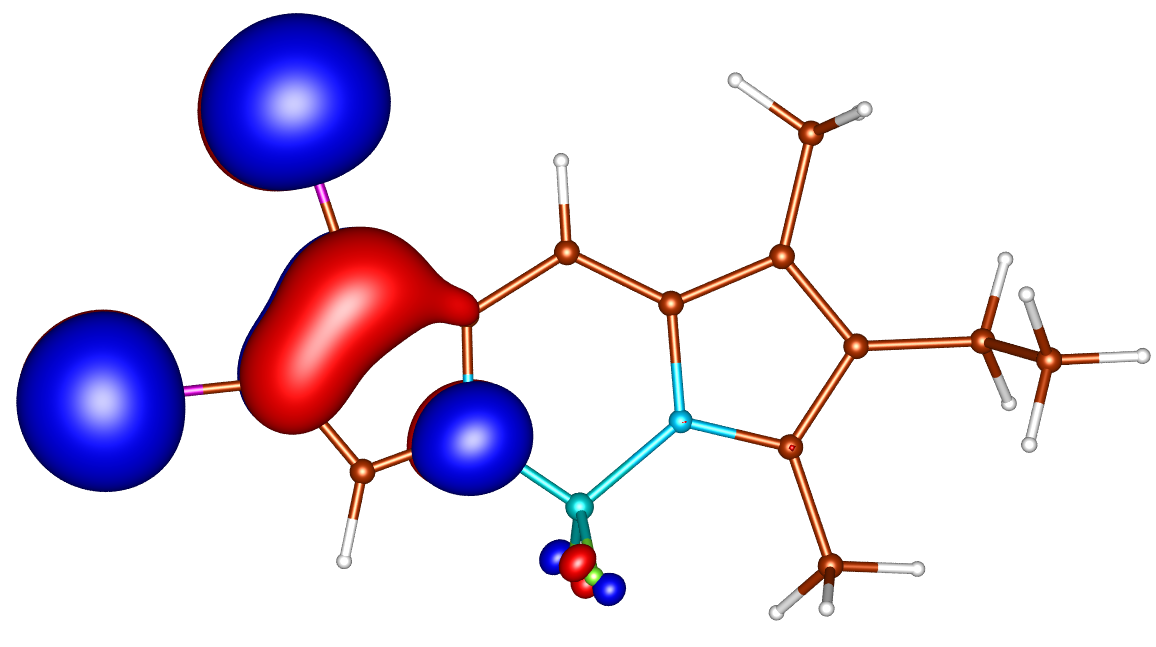}
\caption{Out-of-plane nonbonding/$\pi$ bonding ($\pi$) HOMO-1 of I-BODIPY, orbital energy -6.53 eV.}
\label{fig:homo1_ibodipy} 
\end{figure}

\begin{figure}
\centering
\includegraphics[width=7cm]{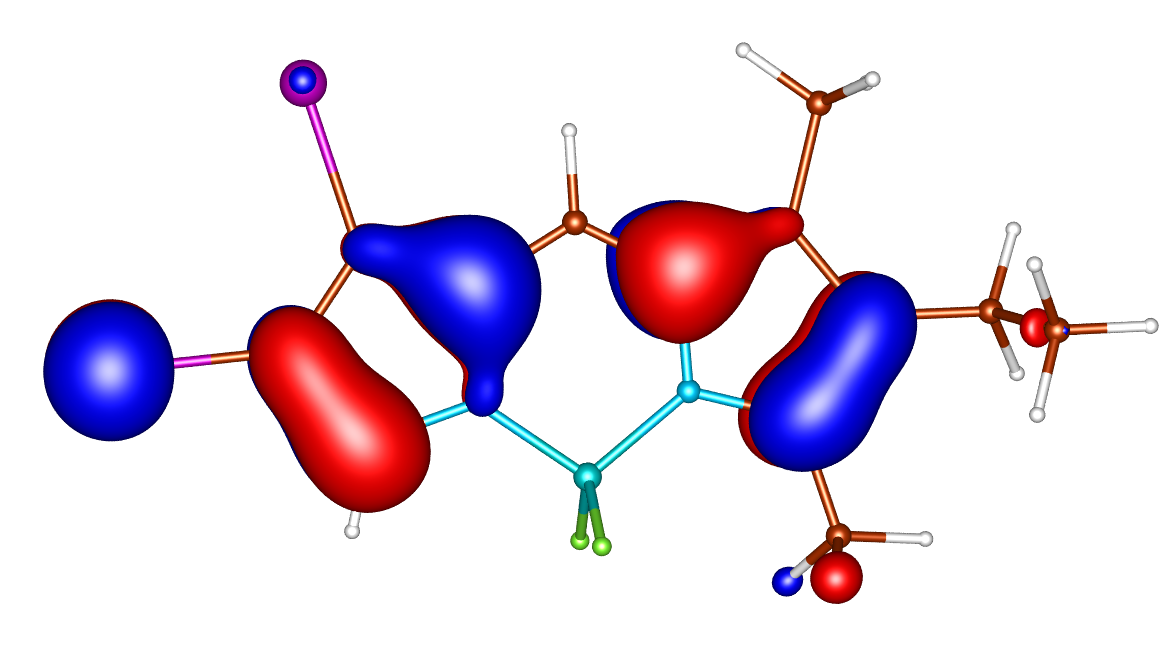}
\caption{$\pi$ bonding ($\pi$) HOMO of I-BODIPY, orbital energy -5.98 eV.}
\label{fig:homo_ibodipy} 
\end{figure}

\begin{figure}
\centering
\includegraphics[width=7cm]{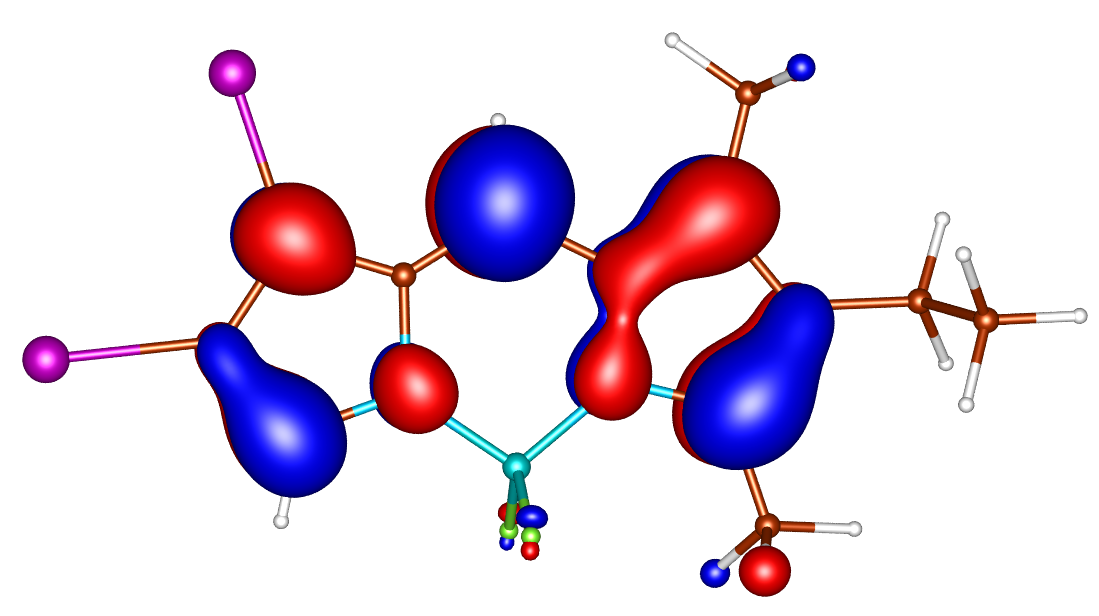}
\caption{$\pi$ antibonding ($\pi^*$) LUMO of I-BODIPY, orbital energy -3.07 eV.}
\label{fig:lumo_ibodipy} 
\end{figure}

\begin{figure}
\centering
\includegraphics[width=7cm]{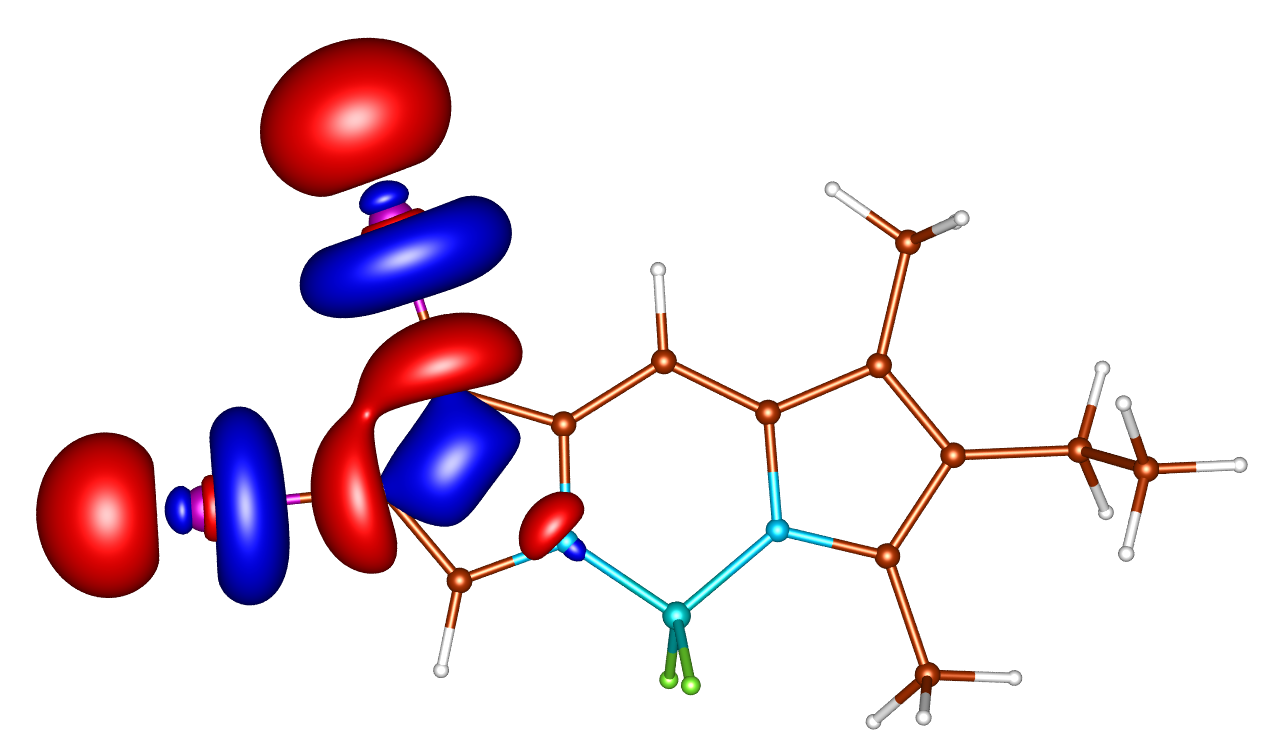}
\caption{C--I $\sigma$ antibonding ($\sigma^*$) LUMO+1 of I-BODIPY, orbital energy -1.38 eV.}
\label{fig:lumo1_ibodipy} 
\end{figure}

\begin{figure}
\centering
\includegraphics[width=7cm]{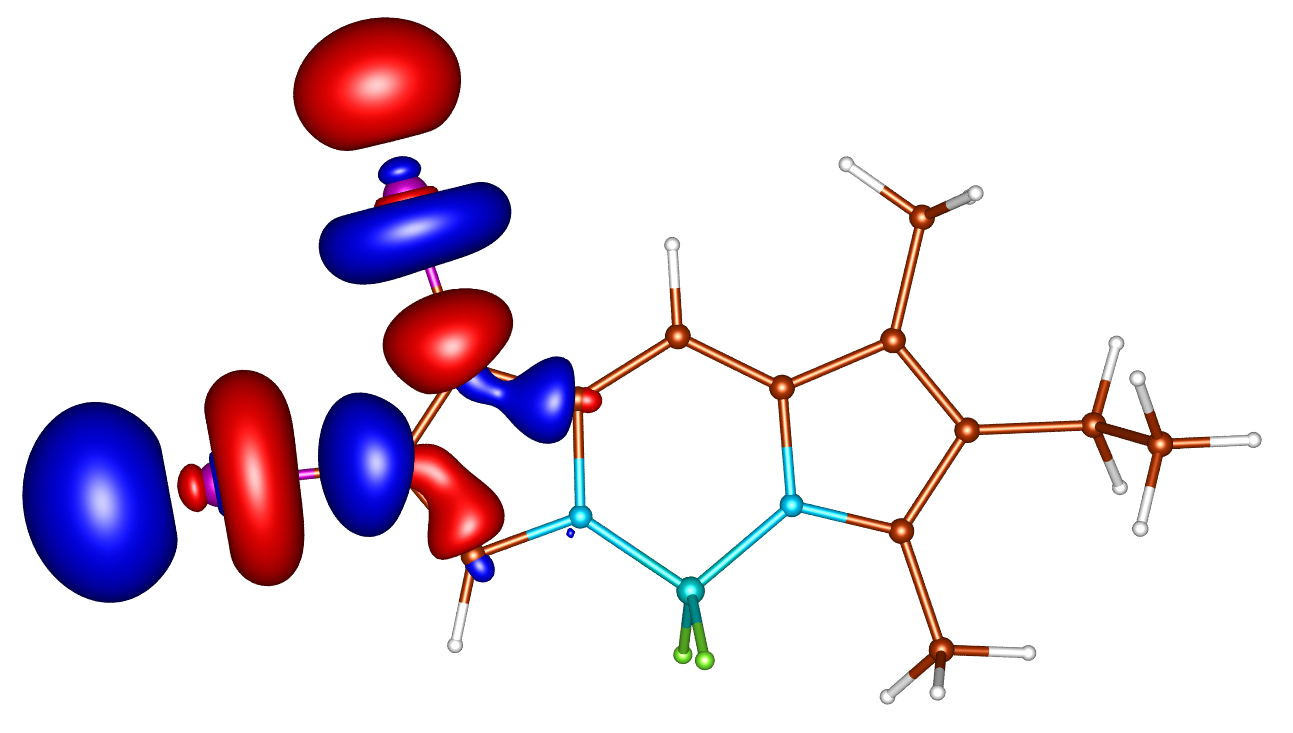}
\caption{C--I $\sigma$ antibonding ($\sigma^*$) LUMO+2 of I-BODIPY, orbital energy -0.55 eV.}
\label{fig:lumo2_ibodipy} 
\end{figure}

Vertical excitation energies, oscilator strengths, and orbital overlaps of the low-lying excited states of I-BODIPY at the optimized S${}_0$, S${}_1$ and T${}_2$ geometries calculated by ADC(2) and by TD-DFT using the B3LYP, BHLYP and M06-2X functionals are summarized in Tables S2a through S8c.
Since there are two iodine substituents in I-BODIPY, there are also twice as many molecular orbitals located on iodine atoms see Figs.~\ref{fig:homo5_ibodipy} through \ref{fig:lumo2_ibodipy} and Table S12 with the shapes and energies of the frontier MOs calculated by KS SCF employing the B3LYP functional), twice as many $(n,\pi^*)$ states, more than twice as many $(\pi,\sigma^*)$ states, and four times as many $(n,\sigma^*)$ states compared to monoiodinated BODIPY.
The number of $(\pi,\pi^*)$ states is also greater due to the additional AO 5p of iodine perpendicular to the plane of the rings.
However, not all of these ``new'' states have always been found in the selected energy windows.
Nevertheless, slightly higher density of excited states is observed in I-BODIPY compared to monoiodinated BODIPY in the position 2.

In all but a few calculations (carried out exclusively by TD-DFT employing the B3LYP functional), for both the BODIPY derivatives, the lowest three excited singlets and the lowest four triplets are $(\pi,\pi^*)$ states.
The order of the higher excited states depends, often dramatically, on the method (and density functional) used.
The dependence is most pronounced for the $(n,\pi^*)$ states whose lowest calculated triplet excitation energy in the monoiodinated BODIPY molecule varies from 3.51 eV obtained by TD-DFT/B3LYP for T${}_5$ to 4.70 eV obtained, coincidentally, by both ADC(2) and TD-DFT/BHLYP (or 4.59 eV obtained by TD-DFT/M06-2X) for T${}_6$ to 4.95 eV obtained by CAS(18,15)PT2 for T${}_7$.
It is nevertheless fair to say that the last value may be slightly overestimated due to the fact that the doubly occupied MOs have been fixed at the values optimal for the lowest $(\pi,\sigma^*)$ triplet (see above).
Such a huge discrepancy, which also occurs for the $(n,\pi^*)$ singlet as well as for the four lowest $(n,\pi^*)$ states of I-BODIPY, can well be explained by the fact that all these states are dominated by excitations in the course of which an electron is promoted from an AO 5p of iodine perpendicular to the C--I bond and parallel to the plane of the rings to a distant $\pi^*$ orbital located above and below the rings (see also Table S12, namely HOMO-5, HOMO-3, and LUMO).
And since it has long been known that linear response TD-DFT fails for CT states when used with ``standard'' exchange-correlation functionals like SVWN, BLYP or B3LYP,~\cite{dreuw2003long,dreuw2004failure} it is no surprise that the variability in the calculated excitation energies is the greatest precisely for the $(n,\pi^*)$ states.
This qualitative observation is in agreement with the calculated orbital overlaps whose values range between 0.45 and 0.75 for the vast majority of $(\pi,\pi^*)$ states, between 0.5 and 0.6 for the $(n,\sigma^*)$ states, and from 0.35 to 0.6 for the $(\pi,\sigma^*)$ states (and a few $(\sigma,\pi^*)$ states), but equal only about 0.15 for the lowest $(n,\pi^*)$ state and are slightly above 0.2 for the second lowest $(n,\pi^*)$ state (in the case of I-BODIPY) of each spin multiplicity.

To avoid the failure of TD-DFT, i.e., a significant underestimation of the excitation energies of CT states, it is usually suggested that either long-range corrected hybrid functionals or hybrid functionals comprising about 50\% of exact exchange (like the BHLYP and M06-2X functionals also included in our comparison) be used.~\cite{jacquemin2010on,dev2012determining,peverati2012performance,li2014comparison,shao2020benchmarking}
Indeed, the order of the lowest eight to ten excited singlets and triplets together with their excitation energies calculated by TD-DFT employing the M06-2X functional and by TD-DFT within the Tamm-Dancoff approximation (TDA)~\cite{hirata1999time} employing the BHLYP functional are in best agreement with those calculated by ADC(2).
Nevertheless, if the focus is on the three lowest excited singlets S${}_1$ through S${}_3$ of I-BODIPY at the ground-state optimized geometry, all dominated by $(\pi,\pi^*)$ excitations, the energies calculated by TD-DFT with the B3LYP functional are the closest to the ADC(2) values.

Actually, ADC(2) provides the same order and similar excitation energies as CASPT2 for the three lowest excited singlets and five triplets of the monoiodinated BODIPY in the position 2, but there is a considerable disagreement between these two \emph{ab initio} methods in the order (and, of course, also the energies) of the fourth and higher excited singlets and the sixth and higher triplets.
The two main differences are the greater number of low-lying $(\pi,\pi^*)$ states found by CASPT2 (at the expense of $(n,\sigma^*)$ and $(n,\pi^*)$ states) and the almost exact degeneracy of each pair of $(n,\pi^*)$ states, singlet and triplet, calculated by ADC(2) (and also by TD-DFT regardless of the employed exchange-correlation functional).
The first discrepancy can partly be explained by the fact that CASSCF and CASPT2, unlike ADC(2) and TD-DFT, are not limited to single excitations.
The second discrepancy is due to the fact that, in the crudest approximation, the energy gap between a singly excited singlet ${}^1(a,b)$ and triplet ${}^3(a,b)$ is twice the exchange integral $(ab|ba)$ which is always positive and, naturally, very small if the MOs $a$ and $b$ are far apart.

Vertical excitation energies and oscillator strengths of the lowest-lying singlets of I-BODIPY at its ground-state optimized geometry computed by ADC(2) and by TD-DFT with the B3LYP and BHLYP functionals have been summarized in Table~\ref{table1}.
The ADC(2) vertical excitation energy 2.56 eV of the brightest state S${}_1$ is quite close to the experimental value of 2.46 eV.~\cite{sanchez2017towards}
Tables~\ref{table1} and S2a through S7c compare the vertical excitation energies obtained by TD-DFT using various functionals with those calculated by ADC(2).
Among the functionals tested, B3LYP gives the best agreement with ADC(2) for the vertical excitation energies of singlets S${}_1$ through S${}_3$, all dominated by $(\pi,\pi^*)$ excitations.
For singlets S${}_4$ through S${}_7$ dominated by either $(\pi,\sigma^*)$ or $(n,\pi^*)$ excitations (rarely in the same order), the energies calculated with the BHLYP and M06-2X functionals are the closest to the ADC(2) values while the energies calculated with the B3LYP functional are fatally underestimated.
For the low-lying triplets, clearly the best agreement with the vertical excitation energies calculated by ADC(2) are observed for TD-DFT employing the M06-2X functional, closely followed by TD-DFT within TDA employing the BHLYP functional.
However, it must be admitted that the agreement between ADC(2) and TD-DFT is never really good.

To benchmark the spin-averaged part of the multiconfiguration-Dirac--Hartree--Fock-adjusted two-component pseudopotential by K. A. Peterson \emph{et al}. (on the iodine atoms) used in our NAMD simulations against the presumably more accurate, but less efficient, X2C spinless Hamiltonian, we evaluated the vertical excitation energies also by the all-electron sf-X2C-S-TD-DFT method employing the B3LYP, BHLYP and M062X functionals.
From the comparison of Tables S3a--c with Tables S8a--c, Tables S7a--c with Tables S9a--c, and Tables S5a--c with Tables S10a--c it follows that the calculated values are almost identical.

\begin{table*}[ht]
\caption{\label{table1}Vertical excitation energies [eV] and oscillator strengths (in parentheses) of the lowest-lying singlets of I-BODIPY computed using ADC(2) and TDDFT methods.}
\centering
\begin{tabular}{|c|c|c|c|}
\hline
Electronic Excitation & ADC(2)/cc-pVTZ & B3LYP/dhf-TZVP & BHLYP/dhf-TZVP \\ \hline
S${}_1$ & 2.56 (0.58) & 2.75 (0.33) & 2.96 (0.74) \\ \hline
S${}_2$ & 3.22 (0.22) & 3.01 (0.37) & 3.66 (0.10) \\ \hline
S${}_3$ & 3.49 (0.11) & 3.36 (0.08) & 3.93 (0.08) \\ \hline
S${}_4$ & 4.39 (0.00) & 3.52 (0.00) & 4.37 (0.00) \\ \hline
S${}_5$ & 4.52 (0.00) & 3.79 (0.00) & 4.54 (0.00) \\ \hline
\end{tabular}
\end{table*}

It is obvious that, for monohalogenated BODIPY derivatives, the active space of 18 electrons in 15 orbitals is only necessary if one needs to include in the (frozen-core) CASSCF calculation the $(n,\sigma^*)$ and/or $(n,\pi^*)$ states.
After excluding the nonbonding valence AO p on the halogen atom X (the one parallel to the plane of the rings) the active space reduces to 16 electrons in 14 orbitals within which all low-lying $(\pi,\pi^*)$, $(\pi,\sigma^*)$ and $(\sigma,\pi^*)$ states can still be calculated.
Subsequent exclusion of the three remaining MOs localized mainly on the atom X (C--X $\sigma$ bonding, largely valence AO p perpendicular to the plane of the rings, and C--X $\sigma$ antibonding) leads to the active space of 12 electrons in 11 orbitals within which not only the $(\pi,\sigma^*)$ and $(\sigma,\pi^*)$ states, but also some of the higher-lying $(\pi,\pi^*)$ states cannot be seen.
Finally, if one includes in the active space just HOMO-2, HOMO-1, HOMO and LUMO (6 electrons in 4 orbitals), only the lowest six excited states S${}_1$ through S${}_3$ and T${}_1$ through T${}_3$ are formally accessible.
The last choice can nevertheless be made for essentially any BODIPY derivative regardless of whether it is planar or not, at least in principle.~\cite{dolezel2023spin}

\begin{figure}
\centering
\includegraphics[width=8.5cm]{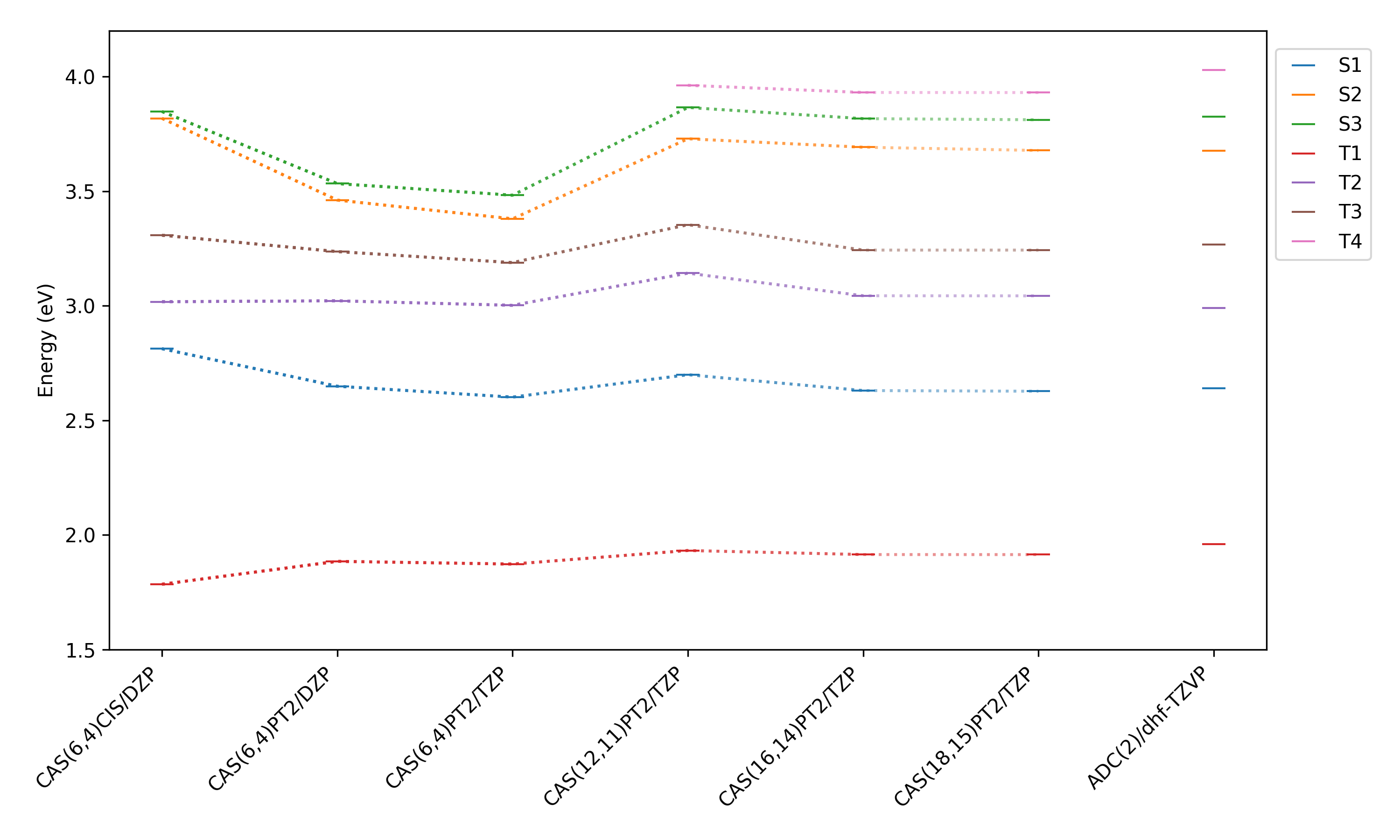}
\caption{Vertical excitation energies of the lowest-lying singlets and triplets of monoiodinated BODIPY in the position 2 calculated by selected multireference methods (employing the DZP or TZP contracted ANO-RCC basis set) and by the single-reference ADC(2) method (employing the dhf-TZVP basis set).}
\label{fig:excitation_energies} 
\end{figure}

To assess the applicability of the various limited active spaces specified above, we carried out a series of multireference calculations of vertical excitation energies of the lowest $(\pi,\pi^*)$ singlets and triplets of monoiodinated BODIPY in the position 2. The results are given in Fig.~\ref{fig:excitation_energies}. This time, the MOs were averaged either over the four lowest singlets S${}_0$ through S${}_3$ or over the four (or three) lowest triplets T${}_1$ through T${}_4$ (or T${}_3$), so the depicted frozen-core CAS(18,15)PT2 excitation energies slightly differ from those given in table S1a. From the figure it follows that even the relatively crude CAS(6,4)PT2 method employing the DZP contracted ANO-RCC basis set may provide quite reasonable excitation energies for the states S${}_1$, T${}_1$, T${}_2$ and T${}_3$, though possibly not for a good reason but rather due to fortuitous cancellation of errors.

Knowledge of the characters (of the dominant excitations) of the low-lying excited states of a molecule is crucial to understanding its nonadiabatic dynamics, especially if SOC plays a significant role.
From the Slater--Condon rules applied to an arbitrary one-electron effective spin--orbit Hamiltonian confined by the one-center approximation it follows that a molecule containing heavy atoms of main group elements can only have large SOC between two states, in our case a singlet and a triplet, if the two states differ in a single pair of MOs that each have a strong admixture of a \emph{different} AO p of the \emph{same} heavy atom (and the two AOs are thus mutually perpendicular).
Using this simple selection rule, one can easily predict, at least qualitatively (but sometimes even semi-quantitatively), the magnitudes of SOCs between low-lying states of planar iodinated BODIPY derivatives.

First of all, it is now clear that SOC is small, actually in units of cm${}^{-1}$, between S${}_0$ and the $(\pi,\pi^*)$ triplets as well as between any two $(\pi,\pi^*)$ states since even if the two states differ in a single pair of MOs that have both a strong admixture of an AO 5p of the same iodine atom, it is always \emph{the same} AO 5p, the one perpendicular to the plane of the rings.
Other implications can be drawn for a monoiodinated BODIPY derivative.
Nevertheless, these implications are valid also for multiply iodinated BODIPY derivatives provided the key singly occupied MOs of the excited states under consideration are located on the same iodine atom.
The greatest SOC, comparable to the value of about 2300~cm${}^{-1}$ observed between individual components of the threefold spatially degenerate ground (scalar relativistic) doublet ${}^2$P of the iodine atom, can be found between S${}_0$ and the $(n,\sigma^*)$ triplet as the two MOs, $n$ and $\sigma^*$, have both essentially atomic character.
Large SOC, about 1000 to 2000 cm${}^{-1}$, can also be expected between S${}_0$ and the $(\pi,\sigma^*)$ triplets as well as between the $(n,\sigma^*)$ singlet or triplet and a $(\pi,\sigma^*)$ triplet or singlet provided the MO $\pi$ from which the electron is promoted has a strong admixture of the AO 5p of iodine (necessarily the one perpendicular to the plane of the rings).
On the contrary, SOC is just moderate (usually in tens of cm${}^{-1}$) between S${}_0$ and the $(n,\pi^*)$ triplets, between the $(n,\sigma^*)$ singlet or triplet and an $(n,\pi^*)$ triplet or singlet, and between the $(\pi,\pi^*)$ states and the $(\pi,\sigma^*)$ states (provided the MO $\pi$ from which the electron is promoted is the same for both states) since the MO $\pi^*$ to which the electron is promoted (typically LUMO) can at best have only weak admixture of the AO 5p of iodine perpendicular to the plane of the rings due to its high one-electron energy.

An intriguing question arises which states, if any, can have large SOC (in hundreds of cm${}^{-1}$ or even greater than 1000 cm${}^{-1}$) with the lowest-lying $(\pi,\pi^*)$ singlets and triplets.
The qualified answer is the low-lying $(n,\pi^*)$ triplets and singlets provided the MO $\pi$ from which the electron is promoted has a strong admixture of the AO 5p of iodine and the MO $\pi^*$ to which the electron is promoted (typically LUMO) is the same for both excited states (otherwise the two states differ in more than a single pair of MOs and SOC between them is much smaller).
This observation, which may also be considered a manifestation of the well-known El-Sayed rules, only underscores the importance of using an exchange-correlation functional for nonadiabatic TD-DFT MD simulations that describes the CT states $(n,\pi^*)$ correctly.
Finally, SOC is quite small between the $(n,\sigma^*)$ singlet or triplet and a $(\pi,\pi^*)$ triplet or singlet and between the $(\pi,\sigma^*)$ states and the $(n,\pi^*)$ states because these states, if they are `pure', always differ in at least two pairs of MOs.
Moreover, SOC is also small between states with nearly the same orbital occupation numbers. Typical example are the calculated pairs of almost exactly degenarete $(n,\pi^*)$ singlet and triplet (see above).
It is obvious that strong SOC between these virtually isoenergetic TD-DFT states could effectively paralyze the TSH simulations.

The commonly quoted El-Sayed rules which were originally formulated for diazenes~\cite{elsayed1962radiationless}, nitrogen heterocyclics~\cite{elsayed1963spin}, carbonyls and halogenated aromatics~\cite{elsayed1968triplet} and later reportedly generalized to (either the same or) some other types of molecules are nothing but a special case of the much more general Slater--Condon rules outlined above and will therefore not be referred to further in this text.

\begin{figure}
\centering
\includegraphics[width=8.5cm]{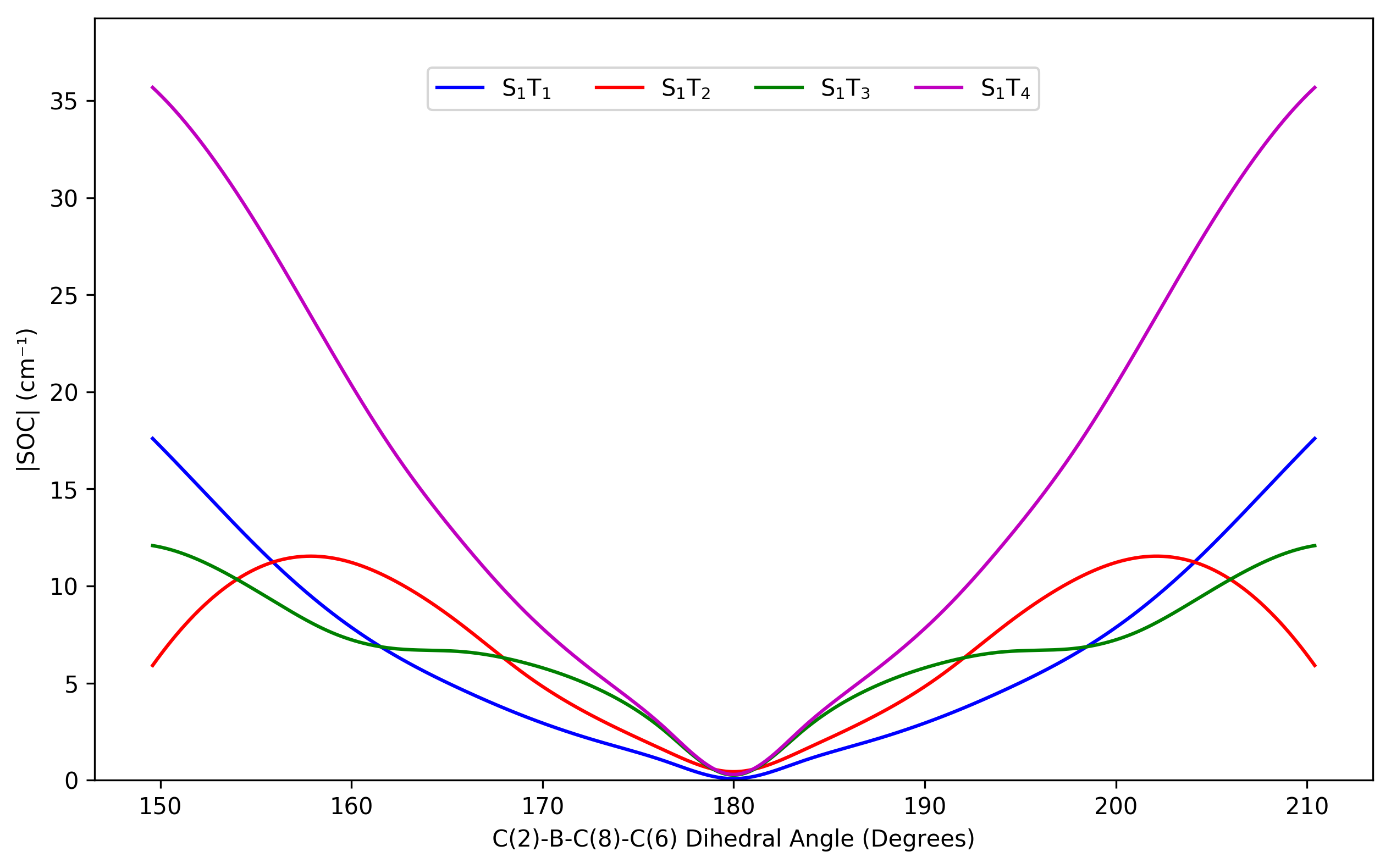}
\caption{Variation of magnitudes of SOC between S${}_1$ and the lowest-lying triplets of monoiodinated BODIPY in the position 2 with the C(2)--B--C(8)--C(6) dihedral angle (calculated at the TD-DFT/M06-2X/dhf-TZVP level of theory).}
\label{fig:soc_variation} 
\end{figure}

It is evident that the magnitudes of SOC between the lowest-lying excited singlets and triplets, all dominated by $(\pi,\pi^*)$ excitations, are very likely to increase as soon as the iodinated BODIPY derivative loses planarity---mainly due to the unavoidable mixing of the originally $(n,\pi^*)$ and $(\pi,\sigma^*)$ excitations into the `zero-order' $(\pi,\pi^*)$ states.
This notorious phenomenon~\cite{perun2008singlet,penfold2010effect} can be lucidly demonstrated e.g. by plotting the SOC magnitudes in the initial state, typically S${}_1$, against the dimensionless (scaled) normal-mode-projected displacement from the planar equilibrium geometry along the normal coordinate of a suitably chosen out-of-plane vibration.
This is almost exactly what we did in Fig.~\ref{fig:soc_variation} for monoiodinated BODIPY in the position 2 (in the ground-state geometry) and its softest normal mode (with harmonic vibrational frequency of about 20--30 cm${}^{-1}$) basically corresponding to bending of the entire molecule along the line connecting the atoms B and C(8), except that we artificially set the same mass to all atoms of the molecule (to effectively restrict the hydrogen atoms in their otherwise rather sweeping harmonic motion) and plotted the C(2)--B--C(8)--C(6) dihedral angle instead of the normal-mode-projected displacement on abscissa.

Accordingly, the existence of a thermally accessible minimum energy crossing point (MEXP) on the crossing seam of the initial singlet and some close-lying triplet located \emph{at a non-planar geometry} would suggest that the corresponding ISC rate may be enhanced.~\cite{dolezel2023spin}
Obviously, a sufficiently long and accurate direct (on-the-fly) spin-forbidden NAMD simulation would provide the definitive answer.
Such a simulation naturally captures not only the geometry dependence of SOC, but also the so-called spin--vibronic coupling (including its own geometry dependence) which is a joint effect of SOC and NAC having its origin in the second order of time-dependent perturbation theory.~\cite{marian2012spin,penfold2018spin} Actually, this classical terminology is probably going to be forgotten as the term spin--vibronic coupling is now frequently being (mis)used for essentially everything beyond single-point SOC.

The magnitudes of SOC between the lowest-lying singlets and triplets of monoiodinated BODIPY in the position 2 and I-BODIPY computed by DKH2 CAS(18,15)SCF (only for the former, using both the full two-electron DKH1 spin--orbit Hamiltonian and some other, less accurate, spin--orbit Hamiltonians to demonstrate, on the one hand, the inadequacy of the Breit--Pauli spin--orbit Hamiltonian in combination with the DKH2 spinless Hamiltonian for the description of iodine-containing molecules resulting in a systematic overestimation of all SOCs by some 20\% and, on the other hand, the negligible errors of the mean-field, one-center, and FNSSO approximations), ADC(2), and TD-DFT with the B3LYP, BHLYP and M06-2X functionals (using the Breit--Pauli variant of the one-center FNSSO spin--orbit Hamiltonian combined with a two-component pseudopotential on the iodine atoms) at their ground-state optimized geometries are given in tables S13 through S21.
Although the calculated values vary with the method and/or density functional used, the differences are not extremely large if states of the same characters are aligned.

In the rest of our study, only B3LYP and BHLYP functionals (the latter exclusively within TDA) are used since analytical TD-DFT gradients for the M06-2X functional are not available in Turbomole v7.0.1.
Actually, the use of TDA has certain advantages.
Unlike full TD-DFT, TD-DFT(TDA) does not suffer from numerical instabilities near conical intersections, making the method comparatively more stable in these regions.~\cite{yang2016conical, matsika2021electronic}
It can thus at least approximately reproduce the topology of the potential energy surface (PES) as obtained by multireference methods. 
Moreover, TDA offers a remedy for triplet instabilities that impact on TD-DFT.~\cite{peach2011influence}

\subsection{Potential energy cuts}

A potential energy cut (PEC) was generated by linear interpolation in the internal coordinates  between the S$_1$ and T$_2$ optimized geometries of I-BODIPY to benchmark TD-DFT with the B3LYP and BHLYP functionals (possibly within TDA) against ADC(2) for a wider range of geometries. 
We first performed static quantum chemical calculations of electronic excitation energies along the PEC by ADC(2) and TD-DFT/TDA with the B3LYP and BHLYP functionals to pick out a set of suitable low-lying excited states. 
The three lowest $(\pi,\pi^*)$ singlets and the lowest $(n,\pi^*)$ singlet
together with the four lowest $(\pi,\pi^*)$ triplets, the lowest $(n,\pi^*)$ triplet, and the lowest $(\pi,\sigma^*)$ triplet have been included in this selection, see Figs. S1--S4.

To facilitate the comparison, we have also calculated the so-called non-parallelities (NPs) between the potential energy curves.~\cite{dutta2003full}. 
For two curves, $f(R)$ and $g(R)$, NP is defined as
\begin{equation}
\nonumber
\mathrm{NP}(f,g)=\underset{R}{\max} \left|f(R)-g(R)\right|-\underset{R}{\min}\left|f(R)-g(R)\right|.
\end{equation} 
The NP values between ADC(2) and TD-DFT with the B3LYP and BHLYP functionals (the latter within TDA) are summarized in Table S11, while the underlying PECs can be seen in Figs.~S1 through S4.
For the first excited singlet, NP for the B3LYP functional is 0.19 eV, significantly lower than the value of 0.43 eV for the BHLYP functional.
For the rest of the lowest-lying singlets dominated by $(\pi,\pi^*)$ excitations, the NP values for the B3LYP and BHLYP functionals are quite similar, ranging from 0.2 to 0.4 eV, slightly lower (i.e., better) for B3LYP.
For the lowest singlet dominated by $(n,\pi^*)$ excitations, the NP value reaches about 0.3 eV for both the functionals. The same value is typical, with a few exceptions, also for the lowest-lying triplets regardless of the density functional used.
The moderate NP values between ADC(2) and TD-DFT with the selected functionals only confirmed our decision to use the computationally much less expensive TD-DFT(TDA) method with the BHLYP functional for the excited-state MD study of I-BODIPY.

Using the same methods, i.e., ADC(2) and TD-DFT/TDA with the B3LYP and BHLYP functionals, we have also computed SOCs between selected low-lying singlets and triplets along the PEC.
As can be seen from Figs.~S5 through S24, both the magnitudes and the trends vary only moderately between the methods and density functionals used.

Despite the fact that we have eventually selected exchange-correlation functionals more or less suitable for the subsequent NEA absorption spectra and mixed quantum-classical TSH MD simulations on I-BODIPY, the search for a TD-DFT functional \emph{ideal} for these purposes has evidently not come to its end.

\subsection{Absorption spectra}

\begin{figure}
\centering
\includegraphics[width=7cm]{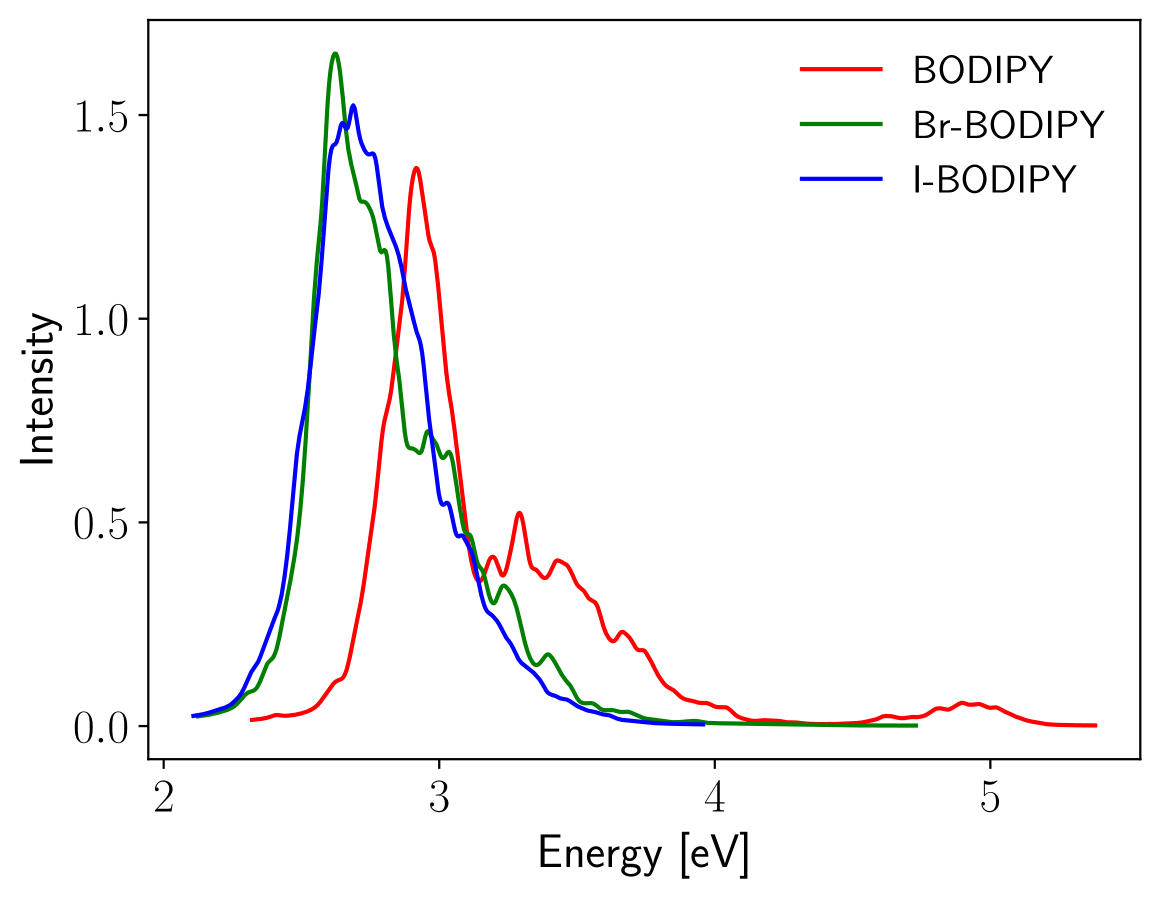}
\caption{Simulated absorption spectra of three BODIPY derivatives in the gas phase calculated from 1000 nuclear configurations sampled from the Wigner distribution at
room temperature.}
\label{fig:absorption_spectra} 
\end{figure}

Simulated absorption spectra of BODIPY and Br-BODIPY from our previous study~\cite{wasif2021theoretical}, and of I-BODIPY studied here, all calculated at the TD-DFT/B3LYP/dhf-TZVP level of theory, are displayed in Fig.~\ref{fig:absorption_spectra}.
The simulations have been carried out using the nuclear ensemble approach (NEA), in which the photoabsorption cross sections
are \emph{averaged} over a set of configuration-space points either sampled from an appropriate phase-space distribution or collected in the course of a ground-state MD simulation.
The cross section for a given geometry is a continuous function (of photon energy) proportional to the sum over the included excited states (i.e., transitions) of the products of the oscillator strength of each transition and a sharp (usually Gaussian
) line-shape function centered at the vertical excitation energy of the transition.~\cite{crespo2012spectrum}
Absorption spectra obtained in this way clearly do not show vibrational structure of the bands.
Each calculation was based on 1000 nuclear configurations sampled from the Wigner distribution of a canonical ensemble of the noninteracting quantum harmonic oscillators~\cite{barbatti2010uv} in the optimized ground-state geometry at 298 K while the three lowest-lying excited singlets of I-BODIPY were included.

It is evident that the spectrum of I-BODIPY shows a remarkable red shift with respect to the spectrum of the unsubstituted BODIPY, quite similar to that of Br-BODIPY, and the spectra of I-BODIPY and Br-BODIPY thus closely overlap. 
This can be explained by the fact that the (essentially $\pi$ bonding) AO 5p of iodine and AO 4p of bromine, both perpendicular to the plane of the rings, destabilize HOMO and LUMO of the pyrrole radical (unequally, increasing the energy of HOMO slightly more than the energy of LUMO) by their positive mesomeric effect, as we already discussed previously~\cite{wasif2021theoretical}.
Despite the fact that the mesomeric effect is somewhat stronger for the bromine than for the iodine substituent (mainly due to the smaller covalent radius of the bromine atom), the calculated HOMO/LUMO gap and red shift are virtually the same.
The overall stabilization of the frontier MOs, roughly equal for HOMO and LUMO, by the negative inductive effect of the halogen substituents is nevertheless greater for bromine than for iodine due to the higher electronegativity of the former.
The shapes and energies of HOMO and LUMO of I-BODIPY are given in Figs.~\ref{fig:homo_ibodipy} and \ref{fig:lumo_ibodipy}.

We have also plotted the electron density in the ground state and the differential electron density (with respect to S${}_0$) in the lowest excited singlet of Br-BODIPY and I-BODIPY at the sf-X2C-S-TD-DFT/B3LYP/x2c-TZVPPall level of theory, as shown in Figs.~S39 through S42.
As can be seen, Br-BODIPY and I-BODIPY exhibit very similar ground-state electron densities as well as differential electron densities for the excitation from S${}_0$ to S${}_1$.
Thanks to this (and the similar geometries of the molecules), we can assume that the behavior of I-BODIPY when embedded in a phospholipid membrane will be quite similar to that of Br-BODIPY, both in the ground state and after photoexcitation to S$_1$.
We thus did not repeat the purely classical MD study of Br-BODIPY in the membrane~\cite{pederzoli2019photophysics} with I-BODIPY.

The calculated vertical excitation energies and oscillator strengths suggest that I-BODIPY could be an efficient photosensitizer. 
It has two spectroscopically active excited states, the first and second excited singlets
, the optical transitions to which exhibit considerable oscillator strengths regardless of whether they are calculated by ADC(2) or TD-DFT (cf. Tables~\ref{table1}, S2a and S3a).
Moreover, the vertical $\Delta E$(T${}_1$--S${}_0$) energy gap is by a decent margin larger than 0.9 eV, the energy necessary to activate O${}_2$ from its triplet ground state ${}^3\Sigma_g$ to its excited singlet state ${}^1 \Delta_g$.~\cite{ogilby2010singlet} 

\subsection{Ultrafast excited-state dynamics}

To model the slow spin-forbidden relaxation processes in I-BODIPY, we employed the accelerated NAMD technique \cite{nijamudheen2017excited} as described above.
Three sets of FSSH MD trajectories were calculated, each set using a different scaling factor of either 2, 3.5 or 5. 
Each set consisted of 20 trajectories with the initial conditions (nuclear coordinates and
momenta%
sampled from the Wigner distribution of a canonical ensemble of the noninteracting quantum harmonic oscillators~\cite{barbatti2010uv} in the optimized ground-state geometry at 298 K.
All trajectories were started in the first excited singlet $\text{S}_1$ (or, more precisely, in a quasirelativistic state which had the greatest overlap with the scalar relativistic singlet S${}_1$).
The quantum chemical method used was TD-DFT(TDA)/BHLYP-D3/dhf-TZVP.
The D3 empirical correction for dispersion was applied as dispersion effects may become significant outside the Franck--Condon region~\cite{grimme2010consistent}.
We included in the simulations the singlets S${}_0$ through S${}_6$ and triplets T${}_1$ through T${}_{10}$ to cover all the low-lying electronic states that could possibly affect the radiationless deactivation of I-BODIPY in its first excited singlet, see Tables S7a--c.
The unusually large number of the embraced higher excited states dominated mainly by $(n,\pi^*)$, $(\pi,\sigma^*)$, and $(n,\sigma^*)$ excitations is due to their extraordinarily strong SOC with the lowest-lying $(\pi,\pi^*)$ states and/or among themselves, in some cases up to 3 orders of magnitude stronger than the average SOC between the lowest-lying $(\pi,\pi^*)$ singlets and triplets, see Table S19.

Strong SOC between $(n,\pi^*)$ and $(\pi,\pi^*)$ states in unsaturated organic molecules containing nitrogen, oxygen, or halogen atoms is a notorious phenomenon described by the well-known El-Sayed rules.~\cite{elsayed1962radiationless,elsayed1963spin,elsayed1968triplet}
For organic molecules in which iodine atoms are directly attached to carbon atoms building cyclic $\pi$-electron rings it has already been reported that electronic states with the largest SOC are those which involve transition of an electron into the C--I $\sigma$ antibonding MO either from a $\pi$ bonding MO of the $\pi$-electron ring or from the nonbonding orbital $n$, essentially the AO 5p of the iodine atom, residing on the iodine atom itself~\cite{ajitha2002photodissociation, baig2024relativistic}. 
Embracing the first six excited singlets and ten triplets in the dynamics will thus get into the game not only the lowest-lying $(\pi,\pi^*)$ and $(n,\pi^*)$ states, but also all the other important, mainly $(\pi,\sigma^*)$ and $(n,\sigma^*)$, states that can participate in successful ISCs.

The (classical) Newton's equations of motion for nuclei were integrated with a time step of 0.5~fs by the velocity-Verlet algorithm while the (quantum) time-dependent Schrödinger equation for electrons was integrated with a (20 times shorter) time step of 0.025~fs by a unitary propagator algorithm using linearly interpolated energies and couplings in the substeps.
The simulations were caried out in the diagonal representation, i.e., in the \emph{spin-adiabatic} basis of the eigenfunctions of the two-component pseudorelativistic Hamiltonian computed by quasidegenerate perturbation theory (QDPT) with the spin-orbit Hamiltonian taken for the perturbation operator, using the 3-step integrator approach.~\cite{mai2015general, pederzoli2017new}
The employed first-order QDPT is nothing but a severely limited spin--orbit CI, namely the diagonalization of the matrix of the two-component pseudorelativistic Hamiltonian in the truncated \emph{spin-diabatic} basis of the eigenfunctions of the (scalar relativistic) spinless Hamiltonian, as a result of which the zero-order QDPT wave functions are just linear combinations of the singlets and (individual components of) triplets included in the basis.

Let, at each 0.5 fs time step,
$${\bf HU}={\bf UE},$$
where {\bf H} is the complex Hermitian matrix of the two-component pseudorelativistic Hamiltonian $\hat H$ in the truncated (and possibly orthonormalized) basis of spin-diabatic states (singlets and individual components of triplets), {\bf E} is the real diagonal matrix of its eigenvalues (quasirelativistic state energies), and {\bf U} is the complex unitary matrix of its eigenvectors.
Then, obviously,
$${\bf U}^{-1}{\bf HU}=
{\bf U}^+\langle{\bf\Theta}|\hat H|{\bf\Theta}\rangle{\bf U}=
\langle{\bf\Phi}|\hat H|{\bf\Phi}\rangle={\bf E}$$
while
$${\bf\Phi}={\bf\Theta U},$$
where $\bf\Theta$ and $\bf\Phi$ are the row vectors of the spin-diabatic and spin-adiabatic bases, respectively.

The FSSH MD simulation
provides a sequence of normalized column vectors ${\bf c}(t)$ of the complex expansion coefficients of the sought time-dependent wave function $\Psi(t)$ in the spin-adiabatic basis ${\bf\Phi}(t)$,
$$\Psi(t)={\bf\Phi}(t){\bf c}(t)={\bf\Theta}(t){\bf U}(t){\bf c}(t)={\bf\Theta}(t){\bf b}(t),$$
where
$${\bf b}(t)={\bf U}(t){\bf c}(t)$$
must be the normalized column vector of the complex expansion coefficients of
$\Psi(t)$ in the spin-diabatic basis ${\bf\Theta}(t)$.
The squares $|b_i(t)|^2
\leq 1$ of the absolute values of the elements of ${\bf b}(t)$ are then the singlet and triplet state populations along a certain trajectory, and averaged over the ensemble of trajectories (while summed over the three components of each triplet) they give the final populations shown below.

From the exposition it is clear that not only $\Psi(t)$ and ${\bf\Phi}(t)$, but also ${\bf\Theta}(t)$ evolves in time, and spin-diabatic states of the same spin multiplicity but different characters may thus swap ordinal indices (or just mix) during the simulation, cf. Tables \hbox{S7a--c}.
This must always be borne in mind when interpreting the results.

\begin{figure}
\centering
\includegraphics[width=7cm]{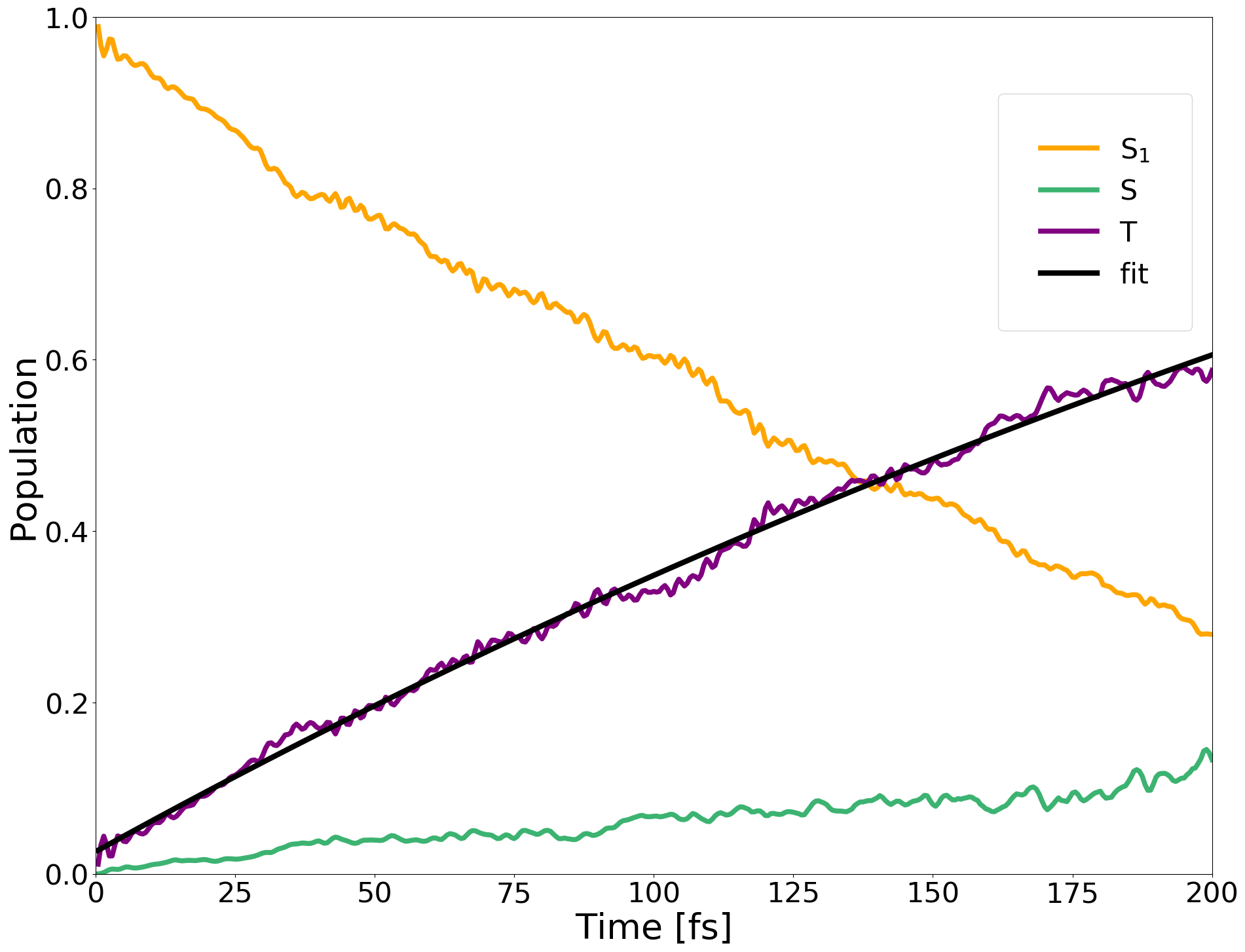}
\caption{Average populations of singlets and triplets (back-transformed from the spin-adiabatic basis) in the 
dynamics of I-BODIPY started in S${}_1$ for the scaling factor $\alpha=2$.}
\label{fig:populations_scaling_2} 
\end{figure}
\begin{figure}
\centering
\includegraphics[width=7cm]{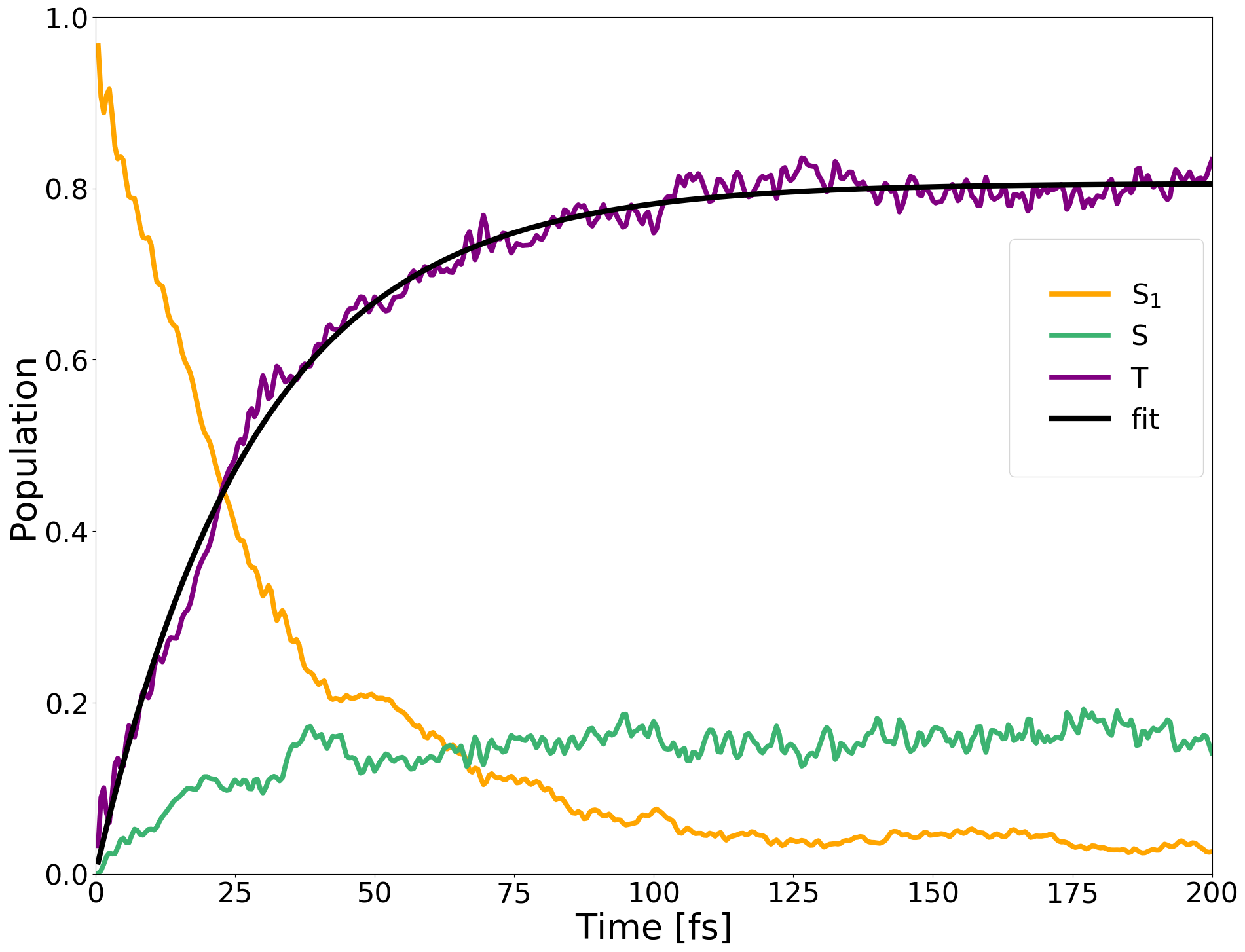}
\caption{Average populations of singlets and triplets (back-transformed from the spin-adiabatic basis) in the 
dynamics of I-BODIPY started in S${}_1$ for the scaling factor $\alpha=3.5$.}
\label{fig:populations_scaling_3.5} 
\end{figure}
\begin{figure}
\centering
\includegraphics[width=7cm]{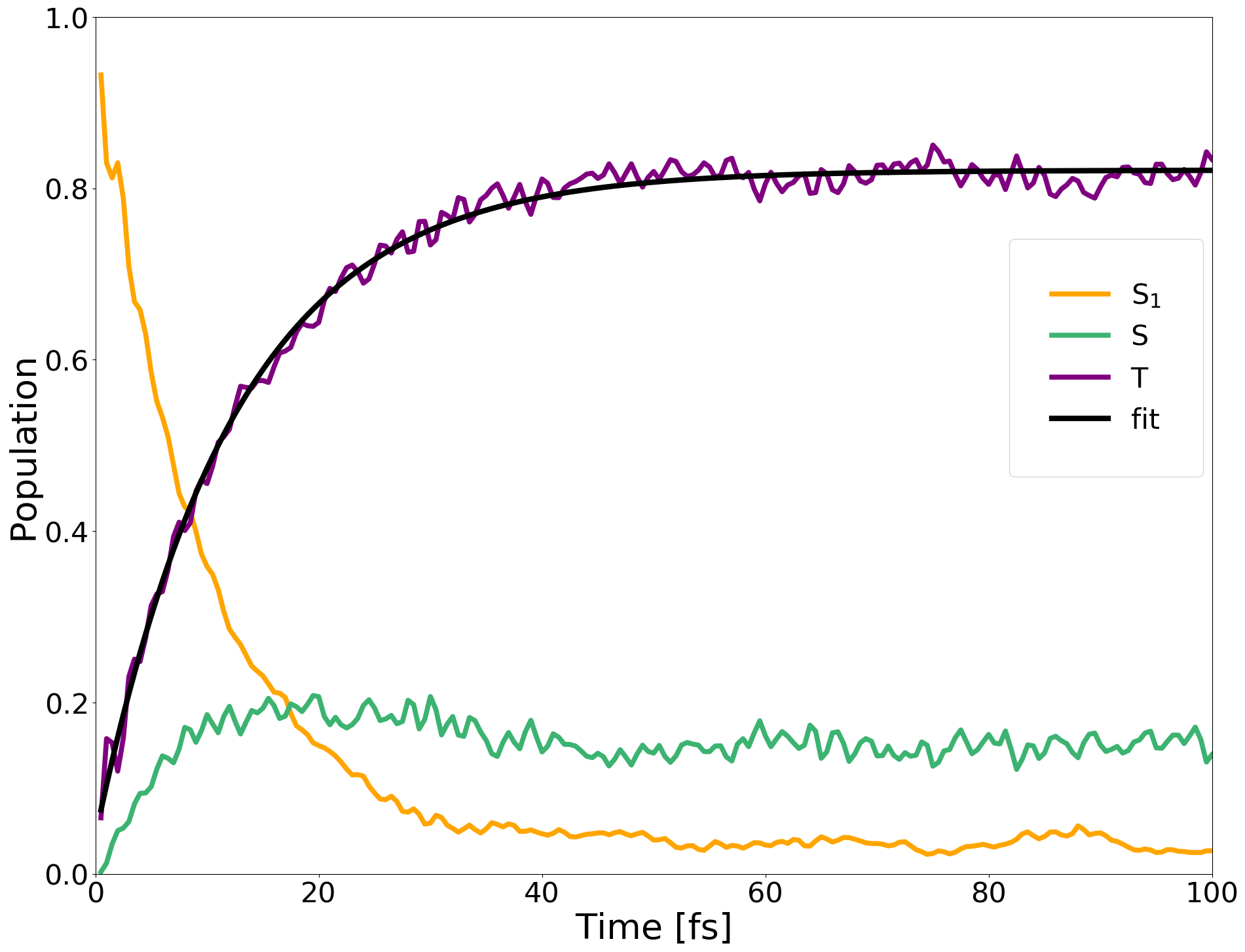}
\caption{Average populations of singlets and triplets (back-transformed from the spin-adiabatic basis) in the 
dynamics of I-BODIPY started in S${}_1$ for the scaling factor $\alpha=5$.}
\label{fig:populations_scaling_5} 
\end{figure}

The maximum simulation time was 200~fs for the trajectories calculated using the scaling factors of 2 and 3.5, and 100~fs for the trajectories calculated using the scaling factor of 5.
The resulting average populations of singlets and triplets (back-transformed from the spin-adiabatic basis) in the excited-state
dynamics of I-BODIPY started in S${}_1$ are depicted in Figs.~\ref{fig:populations_scaling_2}, \ref{fig:populations_scaling_3.5} and \ref{fig:populations_scaling_5} (for the scaling factors of 2, 3.5 and 5).

For the scaling factor $\alpha=2$, about 60\% of the $\text{S}_1$ population decays into triplet states within 200~fs, with a still growing tendency. 
For the scaling factors 3.5 and 5, after a certain time the overall triplet population becomes approximately constant, at the levels of roughly 80 and 82 percents, respectively.
Once this triplet saturation is reached, the populations continue to oscillate between different excited states, maintaining, however, for a limited time, a constant net triplet population.

\begin{figure}
\centering
\includegraphics[width=7cm]{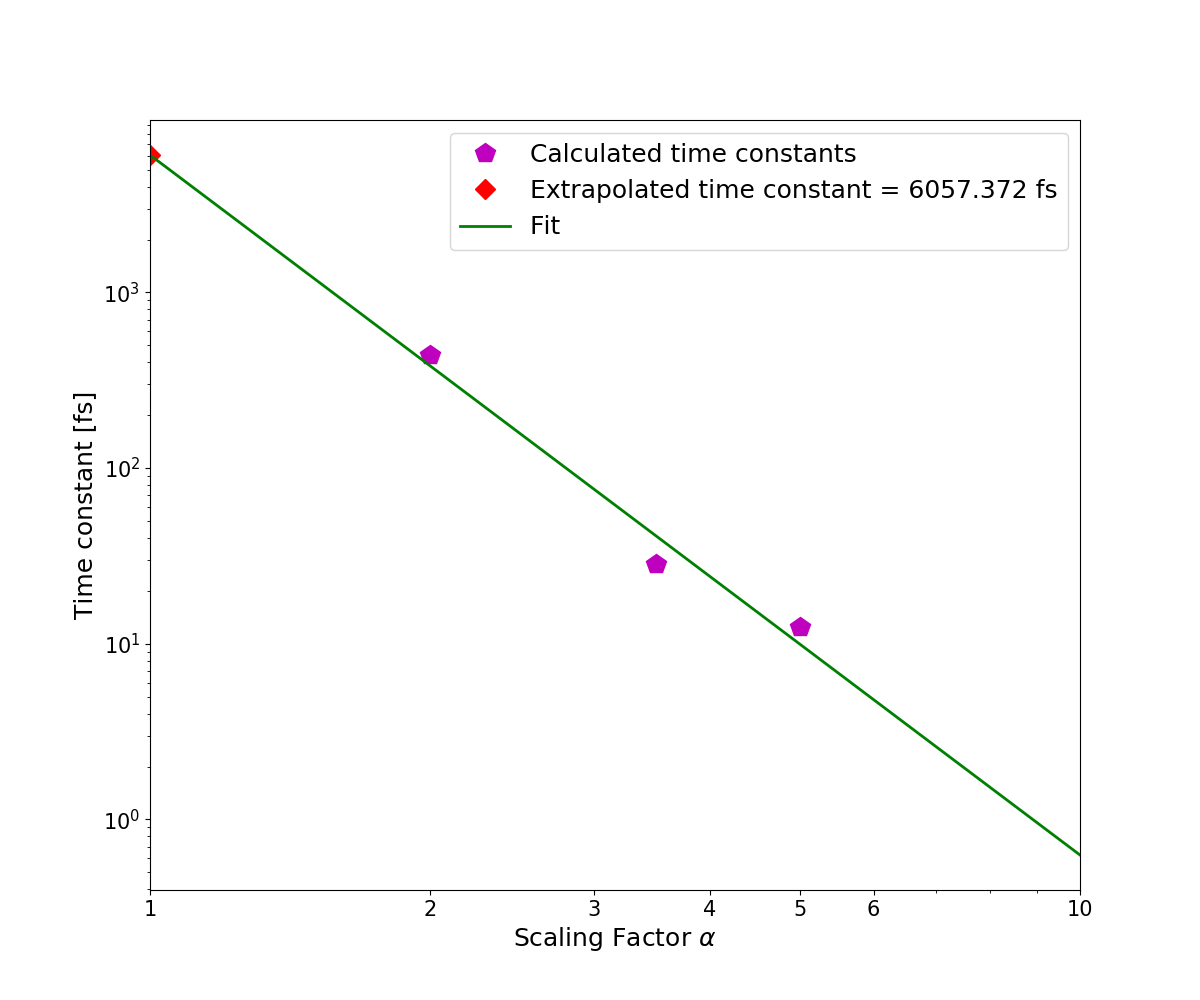}
\caption{Extrapolation of the time constant of the overall triplet population to unit scaling factor.}
\label{fig:lifetime_extrapolation} 
\end{figure}

The time constant of the overall triplet population has been estimated by fitting scaled and shifted complimentary exponential decay curves to the purple data points in Figs.~\ref{fig:populations_scaling_2}, \ref{fig:populations_scaling_3.5} and \ref{fig:populations_scaling_5}, yielding $\tau_2^{\text{T}} = 441.66$ fs, $\tau_{3.5}^{\text{T}} = 28.44$ fs, and $\tau_5^{\text{T}} = 12.44$ fs.
Extrapolation of these time constants to $\alpha=1$ gives $\tau_1^{\text{T}} = 6.06$ ps (as shown in Fig.~\ref{fig:lifetime_extrapolation}). 
The inverse of $\tau_1^{\text{T}}$ is the ISC rate constant $k_{\text{ISC}}=1.73 \times 10^{11}$ s${}^{-1}$.

The lifetime of $\text{S}_1$ has been estimated by fitting scaled and shifted exponential decay curves to the yellow data points in Figs.~\ref{fig:populations_scaling_2}, \ref{fig:populations_scaling_3.5} and \ref{fig:populations_scaling_5} (see also Figs.~S25, S26 and S27), yielding $\tau_2^{\text{S}_1} = 179.1$ fs, $\tau_{3.5}^{\text{S}_1} = 26.3$ fs, and $\tau_5^{\text{S}_1} = 8.8$ fs. 
Extrapolation of these lifetimes to $\alpha=1$ gives $\tau_1^{\text{S}_1} = 1.73$ ps (as shown in Fig.~S28).

Similarly, the lifetime of the overall excited singlet population has been estimated by fitting scaled and shifted exponential decay curves to the yellow data points in Figs.~S29, S30 and S31, yielding $\tau_2^{\text{S}} = 232.8$ fs, $\tau_{3.5}^{\text{S}} = 27.9$ fs, and $\tau_5^{\text{S}} = 13.2$ fs (the tiny amounts of the S${}_0$ populations embraced in the yellow data points were neglected, cf. Fig. S37). 
Extrapolation of these lifetimes to $\alpha=1$ gives $\tau_1^{\text{S}} = 1.93$ ps (as shown in Fig.~S32).

It is quite interesting that the population increase of all singlets except S$_1$ (green data points in Figs.~\ref{fig:populations_scaling_2}--\ref{fig:populations_scaling_5}) is also accelerated by the scaling factor $\alpha$, in spite of the fact that the scaling is only applied to SOC.
This indicates that at least a part of the observed spin-allowed transitions may actually be double ISCs via an intermediate excited triplet state rather than ``true'' internal conversions (ICs) via a conical intersection.

Further analysis of the trajectories revealed that the areas of the $\text{S}_1/\text{S}_0$ conical intersection, where TD-DFT(TDA) might fail, were mostly not accessed during the time span of our simulations.
The populations of the ground state remained below 0.4, 2.5 and 1.5 percents for the scaling factors of 2, 3.5 and 5, respectively (cf. Fig.~S37).
Transitions to $\text{S}_0$ thus occurred infrequently and, apparently, only via the triplet states.

\begin{figure}
\centering
\includegraphics[width=7cm]{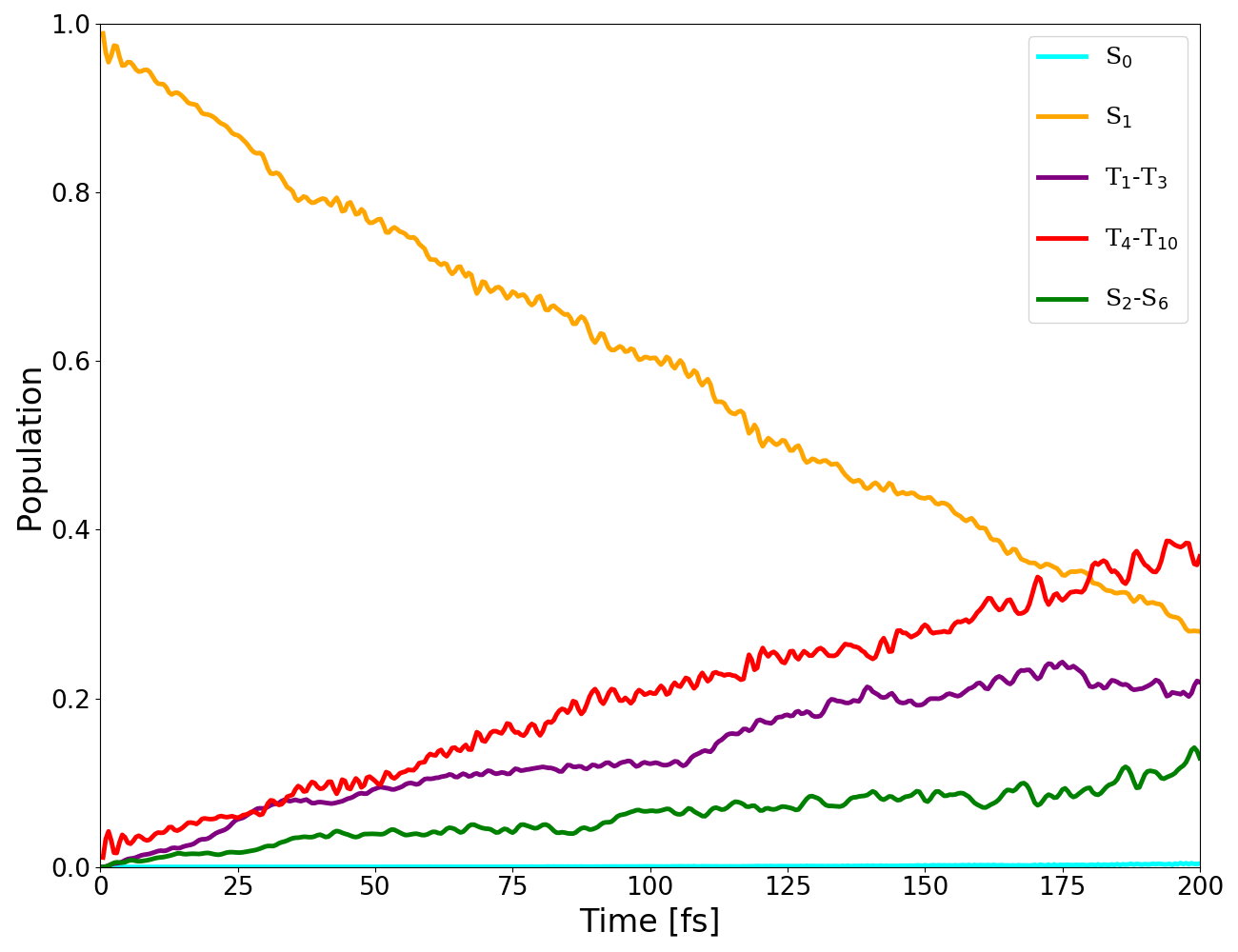}
\caption{Average populations of singlets and triplets (back-transformed from the spin-adiabatic basis) in the excited-state
dynamics of I-BODIPY started in S${}_1$ for the scaling factor $\alpha=2$.}
\label{fig:separatelow_high_T_2}
\end{figure}

\begin{figure}
\centering
\includegraphics[width=7cm]{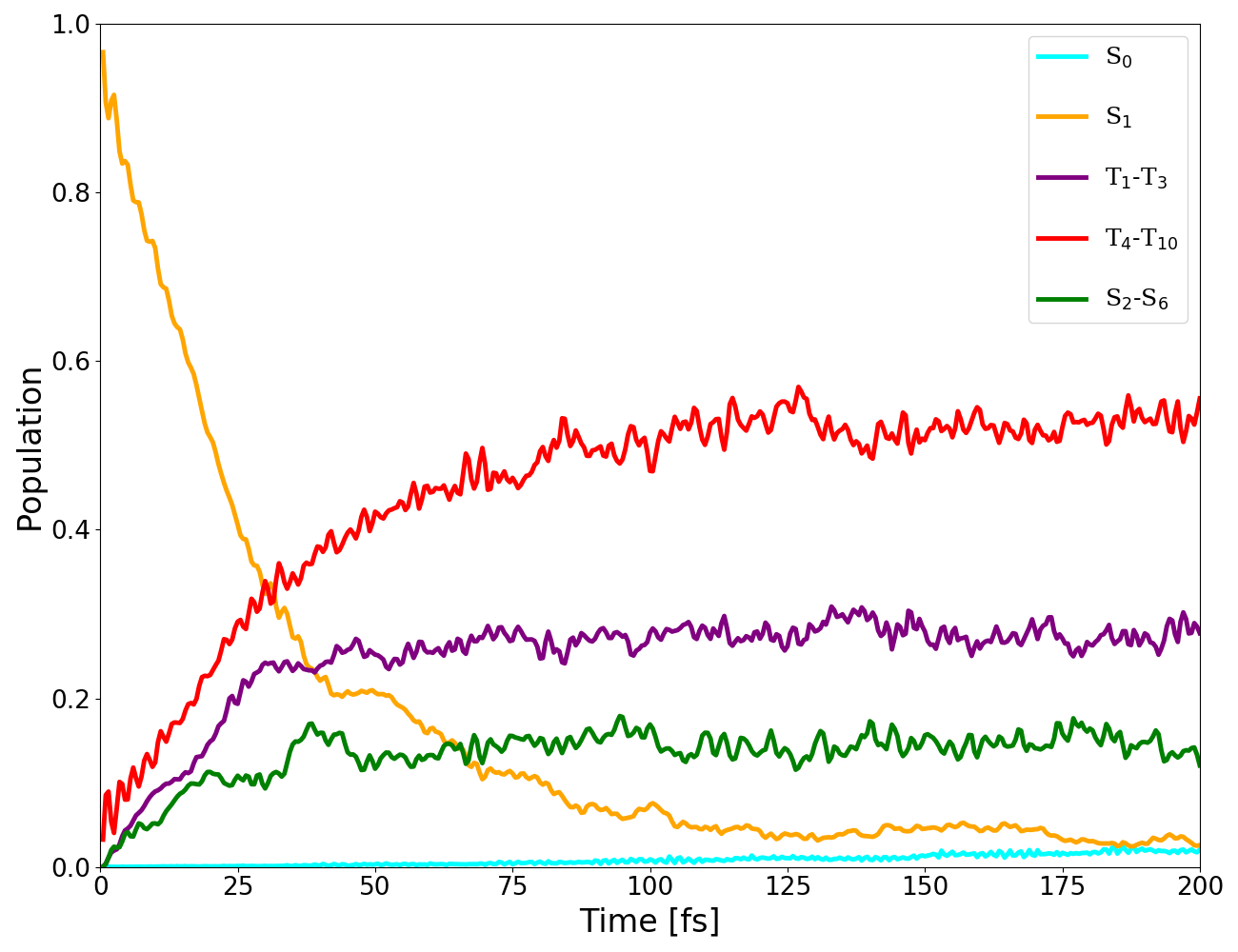}
\caption{Average populations of singlets and triplets (back-transformed from the spin-adiabatic basis) in the excited-state
dynamics of I-BODIPY started in S${}_1$ for the scaling factor $\alpha=3.5$.}
\label{fig:separatelow_high_T_3}
\end{figure}

\begin{figure}
\centering
\includegraphics[width=7cm]{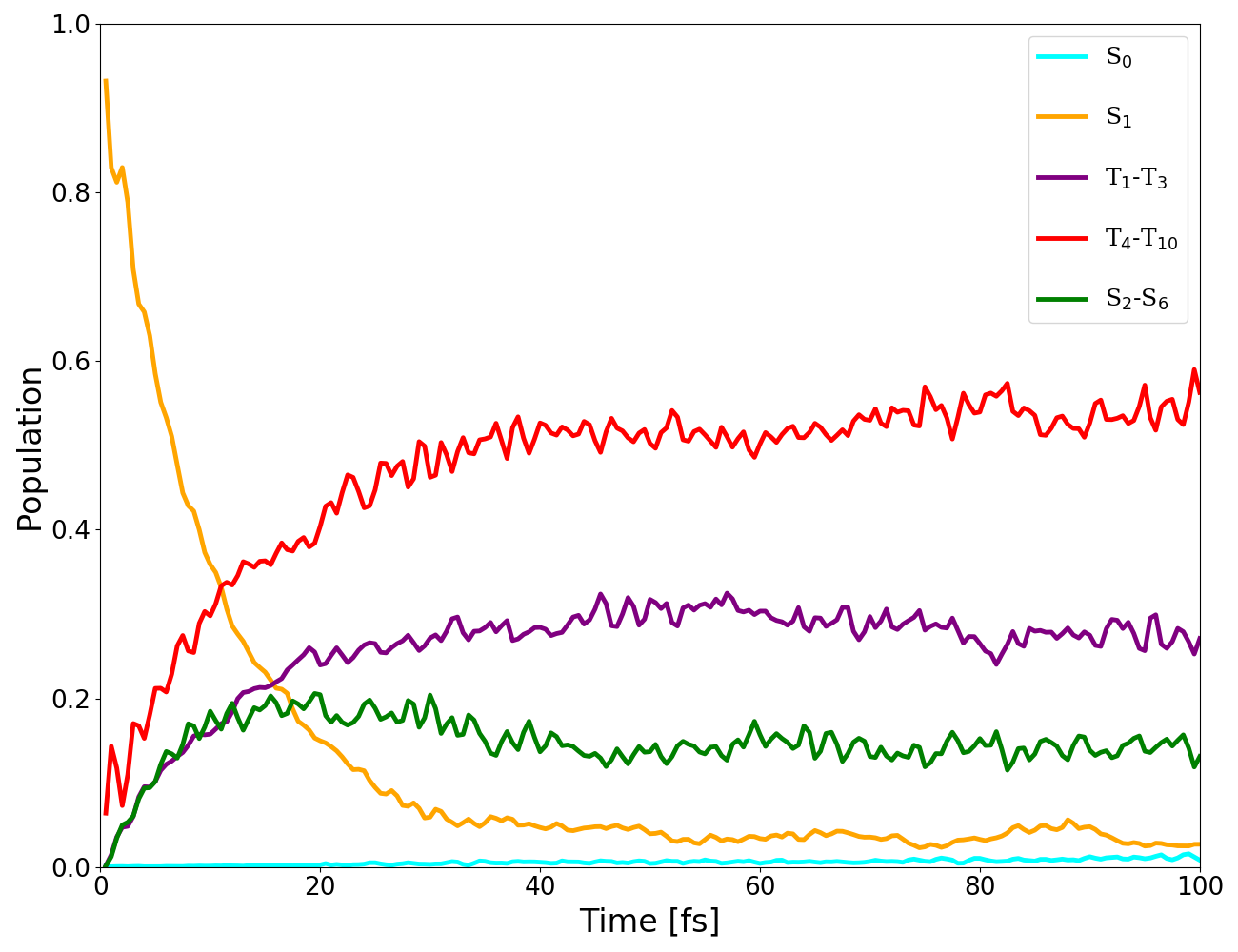}
\caption{Average populations of singlets and triplets (back-transformed from the spin-adiabatic basis) in the excited-state
dynamics of I-BODIPY started in S${}_1$ for the scaling factor $\alpha=5$.}
\label{fig:separatelow_high_T_5}
\end{figure}

In order to explicitly show the importance of the included higher-lying excited states, the average population of the first three triplets
has been separated from that of the higher-lying triplets (T${}_4$ through T${}_{10}$) in Figs.~\ref{fig:separatelow_high_T_2}, \ref{fig:separatelow_high_T_3} and \ref{fig:separatelow_high_T_5}, one for each value of the scaling factor $\alpha$.
From the figures it can be seen that the higher-lying triplets are becoming strongly populated on the same time scale as the lower-lying
ones,
while the higher-lying singlets gain a non-negligible population as well.

The large populations of the higher-lying singlets and triplets may seem surprising and can even raise suspicions that the law of conservation of energy has been violated. Actually, the average kinetic energy of about 3.2 eV distributed among the atomic nuclei of I-BODIPY at the beginning of each trajectory, the energy of an absorbed photon, and also some part of the average potential energy of the molecule in its ground state (in the initial geometry) formally transferred into the first excited singlet is more than enough to power nonadiabatic transitions to all the \emph{pure} singlets and triplets included in our multielectron basis, at least at the employed TD-DFT(TDA)/BHLYP level of theory, cf. Tables S7a--c. Moreover, these and (hypothetically) even higher-lying spin-diabatic (i.e., scalar relativistic) states can be partially populated through their admixtures in some lower-lying spin-adiabatic (i.e., quasirelativistic) states. However, from Tables S2a--c it follows that if we switch from TD-DFT to ADC(2), the energy of S${}_1$ with respect to S${}_0$ decreases as a result of which all the vertical energy gaps between S${}_1$ and the higher-lying states increase. That is why we think that the average populations of the higher-lying spin-diabatic states observed in our TSH MD simulations
are likely to
be
somewhat
overestimated.

Eventually, all the simulations have been carried out on a single (nonrotating) molecule (whose center of mass is at rest) put in an empty space.
In reality, however, the molecule interacts with the environment, which leads, though probably on a longer timescale than is the length of our trajectories, to a dissipation of its vibrational energy into its rotational and translational degrees of freedom as well as into all forms of energy of the surrounding solvent molecules not included in our model.
As a result, the average singlet and triplet populations in the \emph{real} molecule are assumed to continuously flow from the higher-lying to the lower-lying electronic states ``after we stop watching'', supposedly on a longer than picosecond timescale.

\begin{figure}
\centering
\includegraphics[width=7cm]{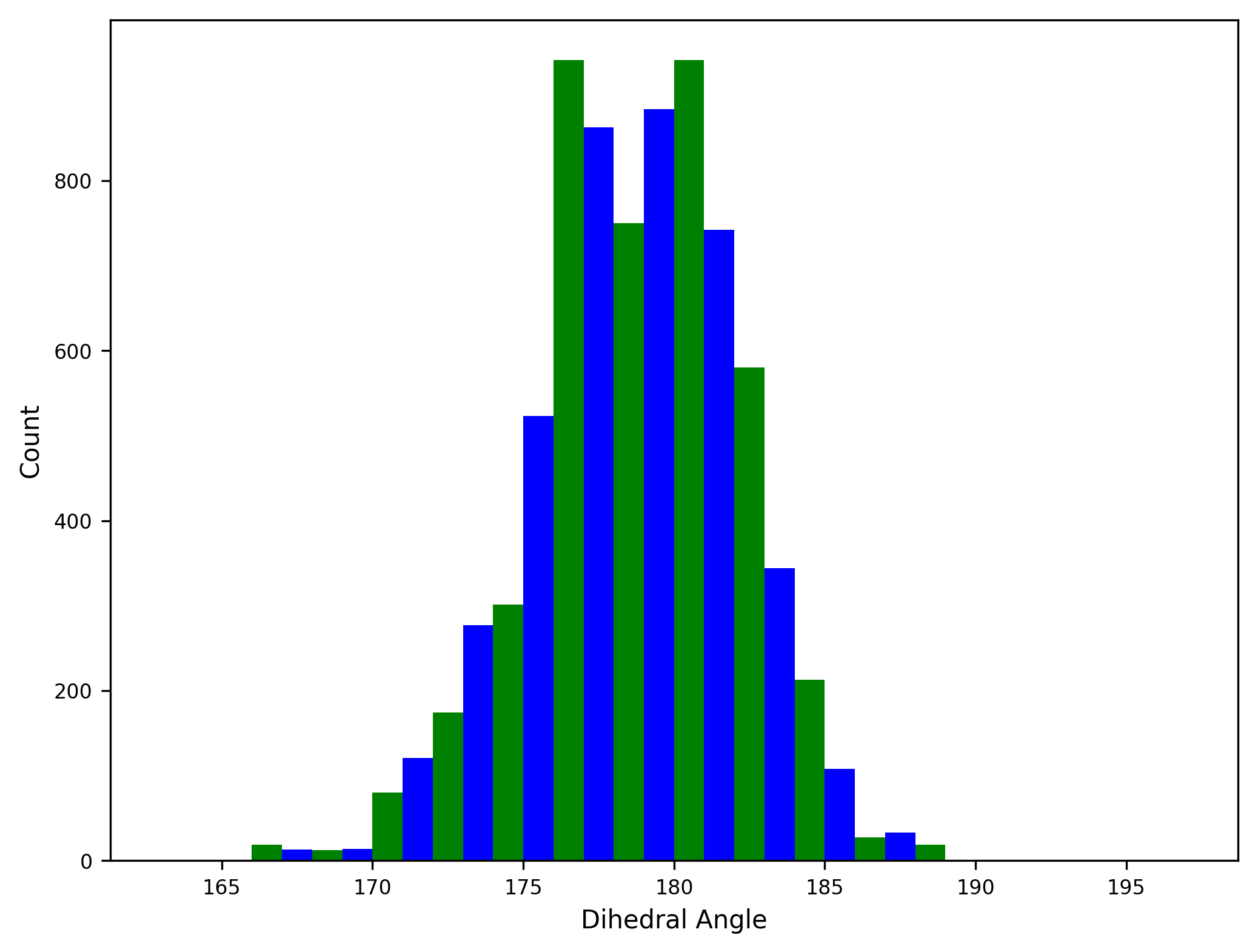}
\caption{Histogram of the frequency of occurrence of the C(2)--B--C(8)--C(6) dihedral angle in I-BODIPY at all time steps of all trajectories.}
\label{fig:dihedral_angles_all} 
\end{figure}

Based on the observation that the magnitudes of SOC between the lowest-lying singlets and triplets of iodinated BODIPY derivatives are very likely to increase with the increasing deviation of the C(2)–B–C(8)–C(6) dihedral angle from 180 degrees (cf. Fig~\ref{fig:soc_variation}), we first, in Fig.~\ref{fig:dihedral_angles_all}, created a histogram of the frequency of occurrence of this dihedral angle in I-BODIPY at all time steps of all trajectories.
From the figure it follows that the dihedral angle naturally varies between some 170 and 186 degrees, while for the vast majority of structures it lies between 175 and 183 degrees. Later it turns out that this range is probably too narrow even for the rapidly growing SOC to significantly affect the excited-state dynamics of the molecule. The small dimple on top of the histogram may indicate there is a saddle point for the almost planar structure (at 178 degrees) and two approximately symmetric local minima nearby (at 176 and 180 degrees) in some of the low-lying excited states.

\begin{figure}
\centering
\includegraphics[width=7cm]{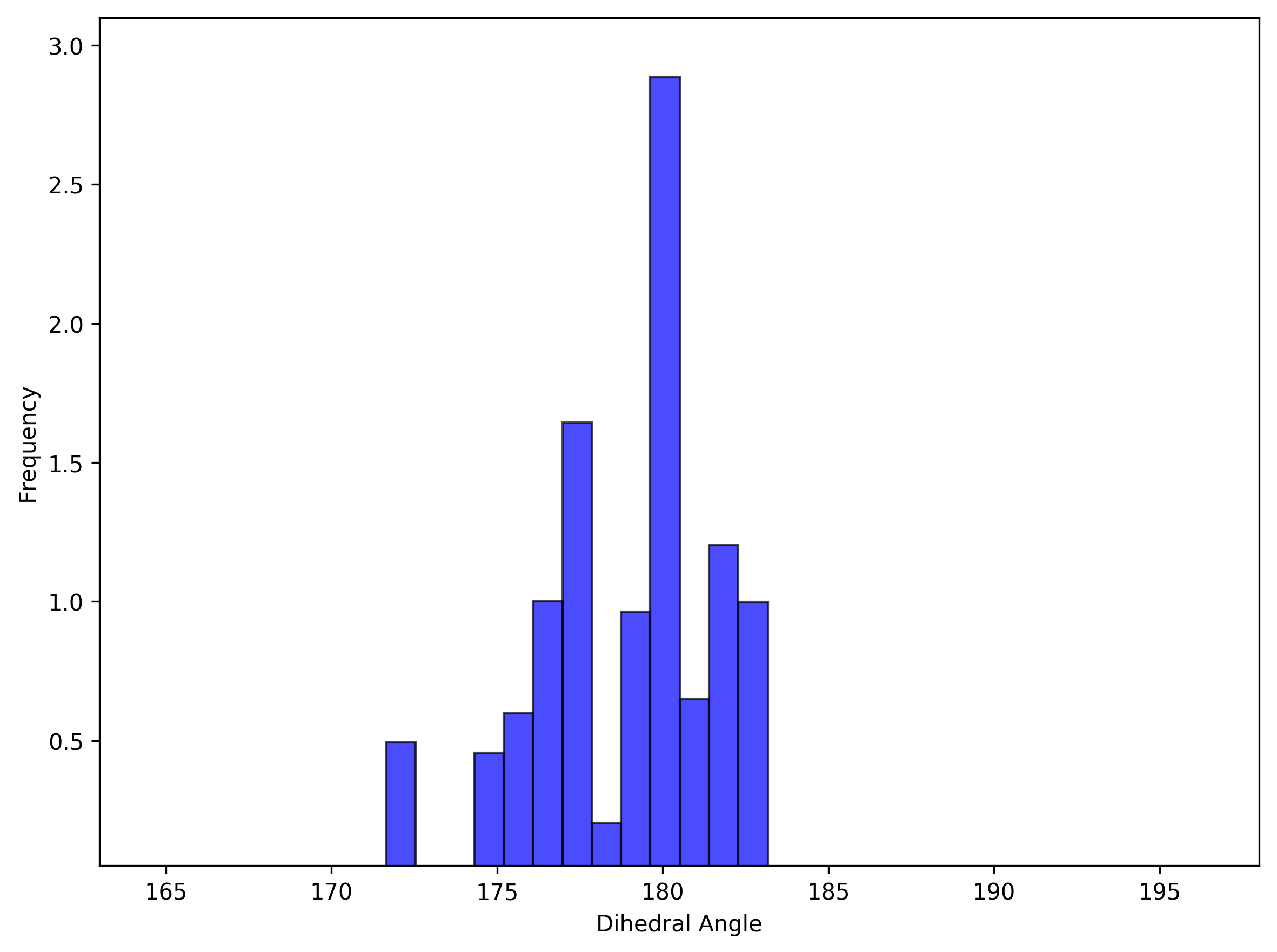}
\caption{Histogram of the frequency of occurrence of the C(2)--B--C(8)--C(6) dihedral angle in I-BODIPY at the time steps at which surface hops with the (possibly partial) $\text{S}_1\to\text{T}_3$ character took place.}
\label{fig:s1_t3_histogram} 
\end{figure}

Next, we searched all the (0.5 fs) time steps of all trajectories for surface hops (between two spin-adiabatic states) and, for every hop found, added (in Fig.~S38) the product of the squares of the absolute values of the
CI expansion coefficients for each pair of the included spin-diabatic states
(in the order S${}_0$--S${}_6$, T${}_1$--T${}_{10}$)%
, the first one from the initial and the other from the final spin-adiabatic state, to the corresponding window and bar of the 17-by-17 matrix of frequency histograms similar to that depicted in Fig.~\ref{fig:dihedral_angles_all}.
From the figure it can be inferred that the greatest number of surface hops has the (possibly partial) $\text{S}_1\to\text{T}_3$ character.
Recently, the transition from $\text{S}_1$ to $\text{T}_3$ have been reported to be the most important among the considered ISCs also for some other halogenated BODIPY derivatives.\cite{pomogaev2020computational}
The histogram for this particular case (taken from the 2nd row and 10th column of the matrix in Fig. S38) is given in Fig.~\ref{fig:s1_t3_histogram}.
No clear preference for bent structures of I-BODIPY can be seen, rather the opposite.
The highest peak is at the C(2)--B--C(8)--C(6) dihedral angle of 180 degrees, despite the fact that SOC between $(\pi,\pi^*)$ singlets and triplets must be stronger in nonplanar geometries.
That is why we decided to focus, in our next study, on iodinated BODIPY derivatives which are planar in their ground state, but strongly bent along the B--C(8) line in their lowest excited singlet state.

However, it should be noted that the data shown in Figs.~S38 and \ref{fig:s1_t3_histogram} are not completely accurate or even correct.
The problem is that by picking out only the time steps at which surface hops between spin-adiabatic states take place we obviously miss the transitions between spin-diabatic states which occur by continuous change of their 
CI expansion coefficients in the (one and only) active spin-adiabatic state.
Actually, sufficiently close to a conical intersection of two spin-adiabatic states (crossing seam of two spin-diabatic states) a hop between spin-adiabatic states may in fact represent staying in a given spin-diabatic state (either singlet or triplet) and \emph{vice versa}, staying in a given spin-adiabatic state may represent a hop between spin-diabatic states (from singlet to triplet or the other way around).
Another undesired consequence of our slightly biased approach are the non-empty histograms on the diagonal of the matrix on Fig.~S38. 

\subsection{Triplet quantum yield}

Triplet quantum yield is a useful concept to quantitatively assess the efficiency of triplet states generation in photophysical processes.~\cite{reindl1997quantum, bachilo2000determination} 
It can be expressed as
\begin{equation}
\nonumber
\Phi_\text{T}=\frac{k_\text{ISC}}{k_\text{ISC}+k_\text{IC}+k_\text{F}},
\end{equation} 
where $k_\text{ISC}$ is the ISC rate constant, $k_\text{IC}$  is the IC rate constant, and  $k_\text{F}$ is the fluorescence rate constant corresponding to the Einstein's coefficient of spontaneous emission
\begin{equation}
\nonumber
k_\text{F}= \frac{2\pi e^2 E_\text{F}^2}{h^2\varepsilon_0 m c^3} f
\end{equation} 
where
$h$ is the Planck constant,
$\varepsilon_0$ is the permittivity of vacuum,
$c$ is the velocity of light in vacuum,
$e$ is the charge and $m$ is the mass of an electron,
$E_\text{F}$ is the fluorescence emission energy and $f$ is the oscillator strength whose values are given in Table S22. 

The rate constants obtained for all the three photophysical processes in I-BODIPY have been summarized in Table S23.
The value of $k_\text{IC}$ has been calculated for all transitions starting in S$_1$ and ending in any other singlet (cf. Figs.~S33--S36).
The resulting triplet quantum yield is 0.85.
This theoretical value is in qualitative agreement with the experimental value of the singlet oxygen generation quantum yield of 0.99$\pm$0.06.~\cite{sanchez2017towards}
However, the comparison of these two quantities is not strictly rigorous as they represent two mutually related but different photophysical processes.
To explicitly describe the latter, we would have to include in the model, in addition to I-BODIPY, also ground-state (triplet) molecular oxygen.
Then, on a longer timescale inaccessible to our simulation, the population of excited singlets should continuously decrease in favor of triplets to maintain the ISC-driven excited singlet-triplet dynamic equilibrium if the triplet population were depleted by reaction with molecular oxygen.
This could possibly lead to a higher quantum yield of singlet oxygen generation than that of triplet states in the absence of oxygen.

\section{Conclusions}

A computational study of I-BODIPY (2-ethyl-4,4-difluoro-6,7-diiodo-1,3-dimethyl-4-bora-3a,4a-diaza-s-indacene) was carried out to investigate its key photophysical properties as a potential triplet photosensitizer capable of generating singlet oxygen.
Multireference CASSCF, CASPT2 and DMRG methods were used to verify and prove the applicability of the single-reference ADC(2) method for studying the electronic structure of the excited states of the molecule, at least in the vicinity of its ground- and (lowest-lying) excited-state equilibrium geometries.
Careful benchmarking of several exchange-correlation functionals in the TD-DFT framework was done with respect to ADC(2), CASPT2 and CASSCF with a particular attention paid to the correct description of the excited states dominated by charge-transfer excitations as well as to the ability of TD-DFT employing a basis set with a two-component pseudopotential on the iodine atoms to well capture SOCs between the low-lying singlets and triplets.
The magnitudes of SOC between excited electronic states of all types found were thoroughly discussed using the Slater--Condon rules applied to an arbitrary one-electron one-center effective spin--orbit Hamiltonian.
The geometry dependence of SOCs between the lowest-lying states was also addressed.
Based on these investigations, the TD-DFT/B3LYP and TD-DFT(TDA)/BHLYP approaches were selected as the methods of choice for the subsequent NEA absorption spectra simulations and mixed quantum-classical TSH MD simulations, respectively.

Both ADC(2) and TD-DFT predict two bright states in the electronic spectrum of I-BODIPY. Introducing the iodine substituents induces a pronounced red shift of the main peak in the visible spectrum of the molecule with respect to the unsubsituted BODIPY.
Excited-state MD simulations including both nonadiabatic effects and SOCs of the relaxation processes in I-BODIPY after its photoexcitation to the S$_1$ state were performed using the accelerated NAMD approach to make the simulation of the slower spin-forbidden transitions computationally feasible.
The TSH MD simulations revealed that ISCs occur on a time scale comparable to ICs in I-BODIPY. This leads to a considerably higher population of triplet states than excited singlet states, while the relaxation to the ground state is almost negligible.
After an initial phase of triplet population growth a ``saturation'' is reached where the ratio of the net triplet to singlet populations is about 4:1, which results in a high triplet quantum yield whose calculated value is in qualitative agreement with the experimentally observed high singlet oxygen generation quantum yield.

\begin{acknowledgement}
We dedicate this paper to Professor Hans Lischka on the occasion of his 80th birthday.
This work has been supported by the Czech Science Foundation (Project No. 23-06364S).
A part of the computational time for this work was supported by the Ministry of Education, Youth and Sports of the Czech Republic through the e-INFRA CZ (ID:90254).
\end{acknowledgement}

\begin{suppinfo}

Supporting Information contains (\emph{i}) tables of characters (types), vertical excitation energies, oscillator strengths, and orbital overlaps of 10 lowest excited singlets and 10 triplets of two molecules, monoiodinated BODIPY in the position 2 (in its optimized S${}_0$ geometry) and I-BODIPY (in its optimized S${}_0$, S${}_1$ and T${}_2$ geometries), calculated by CAS(18,15)PT2 (only for the 1st molecule), ADC(2), and TD-DFT/TDA using several functionals, (\emph{ii}) graphs of vertical excitation energies along the S${}_1$ to T${}_2$ PEC for the 2nd molecule calculated by ADC(2), TD-DFT/B3LYP, and TD-DFT(TDA)/BHLYP, (\emph{iii}) tables of non-parallelities between ADC(2) and TD-DFT(TDA)/BHLYP along the PEC for the 2nd molecule, (\emph{iv}) figures with shapes of frontier MOs of the 2nd molecule, (\emph{v}) tables of SOCs between 11 lowest singlets and 10 lowest triplets of both molecules calculated by CAS(18,15)SCF (only for the 1st molecule, using 4 different spin--orbit Hamiltonians), ADC(2), and TD-DFT/TDA using several functionals, (\emph{vi}) average populations of singlets and triplets in the dynamics of the 2nd molecule for three different values of the scaling factor $\alpha$ and extrapolations of lifetimes and time-constants of singlet populations to $\alpha=1$, (\emph{vii}) histograms of the frequency of occurrence of the C(2)--B--C(8)--C(6) dihedral angle in the 2nd molecule at the time steps at which surface hops with a (possibly partial) $\text{S}_i\to\text{S}_j$, $\text{S}_i\to\text{T}_k$, $\text{T}_k\to\text{S}_i$ or $\text{T}_k\to\text{T}_l$ character took place for all pairs of spin-diabatic states, (\emph{viii}) figures with shapes of the ground-state electron density and differential electron density between S${}_1$ and S${}_0$ of the 2nd molecule, (\emph{ix}) tables with optimized S${}_0$, S${}_1$ and T${}_2$ geometries of the 2nd molecule, (\emph{x}) derivation of the formulae for the evaluation of matrix elements of the one-electron one-center effective spin--orbit Hamiltonian implicitly included in a two-component pseudopotential between nonredundant Cartesian Gaussian functions.

\end{suppinfo}

\bibliography{md}

\providecommand{\latin}[1]{#1}
\makeatletter
\providecommand{\doi}
  {\begingroup\let\do\@makeother\dospecials
  \catcode`\{=1 \catcode`\}=2 \doi@aux}
\providecommand{\doi@aux}[1]{\endgroup\texttt{#1}}
\makeatother
\providecommand*\mcitethebibliography{\thebibliography}
\csname @ifundefined\endcsname{endmcitethebibliography}
  {\let\endmcitethebibliography\endthebibliography}{}
\begin{mcitethebibliography}{119}
\providecommand*\natexlab[1]{#1}
\providecommand*\mciteSetBstSublistMode[1]{}
\providecommand*\mciteSetBstMaxWidthForm[2]{}
\providecommand*\mciteBstWouldAddEndPuncttrue
  {\def\EndOfBibitem{\unskip.}}
\providecommand*\mciteBstWouldAddEndPunctfalse
  {\let\EndOfBibitem\relax}
\providecommand*\mciteSetBstMidEndSepPunct[3]{}
\providecommand*\mciteSetBstSublistLabelBeginEnd[3]{}
\providecommand*\EndOfBibitem{}
\mciteSetBstSublistMode{f}
\mciteSetBstMaxWidthForm{subitem}{(\alph{mcitesubitemcount})}
\mciteSetBstSublistLabelBeginEnd
  {\mcitemaxwidthsubitemform\space}
  {\relax}
  {\relax}

\bibitem[Awuah and You(2012)Awuah, and You]{awuah2012boron}
Awuah,~S.~G.; You,~Y. Boron dipyrromethene (BODIPY)-based photosensitizers for
  photodynamic therapy. \emph{RSC Adv.} \textbf{2012}, \emph{2},
  11169--11183\relax
\mciteBstWouldAddEndPuncttrue
\mciteSetBstMidEndSepPunct{\mcitedefaultmidpunct}
{\mcitedefaultendpunct}{\mcitedefaultseppunct}\relax
\EndOfBibitem
\bibitem[Wenger(2020)]{wenger2020bright}
Wenger,~O.~S. A bright future for photosensitizers. \emph{Nat. Chem.}
  \textbf{2020}, \emph{12}, 323--324\relax
\mciteBstWouldAddEndPuncttrue
\mciteSetBstMidEndSepPunct{\mcitedefaultmidpunct}
{\mcitedefaultendpunct}{\mcitedefaultseppunct}\relax
\EndOfBibitem
\bibitem[Mazzone \latin{et~al.}(2016)Mazzone, Alberto, De~Simone, Marino, and
  Russo]{mazzone2016can}
Mazzone,~G.; Alberto,~M.~E.; De~Simone,~B.~C.; Marino,~T.; Russo,~N. Can
  expanded bacteriochlorins act as photosensitizers in photodynamic therapy?
  Good news from density functional theory computations. \emph{Molecules}
  \textbf{2016}, \emph{21}, 288\relax
\mciteBstWouldAddEndPuncttrue
\mciteSetBstMidEndSepPunct{\mcitedefaultmidpunct}
{\mcitedefaultendpunct}{\mcitedefaultseppunct}\relax
\EndOfBibitem
\bibitem[Li \latin{et~al.}(2021)Li, Wang, Pan, Sun, Wang, and
  Zhang]{li2021unexpected}
Li,~J.; Wang,~X.; Pan,~Y.; Sun,~Y.; Wang,~G.; Zhang,~K. Unexpected long
  room-temperature phosphorescence lifetimes of up to 1.0 s observed in
  iodinated molecular systems. \emph{ChemComm} \textbf{2021}, \emph{57},
  8794--8797\relax
\mciteBstWouldAddEndPuncttrue
\mciteSetBstMidEndSepPunct{\mcitedefaultmidpunct}
{\mcitedefaultendpunct}{\mcitedefaultseppunct}\relax
\EndOfBibitem
\bibitem[Churakov \latin{et~al.}(2021)Churakov, Medved’ko, Prikhodchenko,
  Krut’ko, and Vatsadze]{churakov2021first}
Churakov,~A.~V.; Medved’ko,~A.~V.; Prikhodchenko,~P.~V.; Krut’ko,~D.~P.;
  Vatsadze,~S.~Z. First example of peroxosolvate of iodine-containing organic
  molecule. \emph{Mendeleev Commun.} \textbf{2021}, \emph{31}, 352--355\relax
\mciteBstWouldAddEndPuncttrue
\mciteSetBstMidEndSepPunct{\mcitedefaultmidpunct}
{\mcitedefaultendpunct}{\mcitedefaultseppunct}\relax
\EndOfBibitem
\bibitem[Jereb \latin{et~al.}(2004)Jereb, Zupan, and
  Stavber]{jereb2004effective}
Jereb,~M.; Zupan,~M.; Stavber,~S. Effective and selective iodofunctionalisation
  of organic molecules in water using the iodine--hydrogen peroxide tandem.
  \emph{ChemComm} \textbf{2004}, 2614--2615\relax
\mciteBstWouldAddEndPuncttrue
\mciteSetBstMidEndSepPunct{\mcitedefaultmidpunct}
{\mcitedefaultendpunct}{\mcitedefaultseppunct}\relax
\EndOfBibitem
\bibitem[Zou \latin{et~al.}(2017)Zou, Yin, Ding, Tang, Li, Si, Shao, Zhang,
  Huang, and Dong]{zou2017bodipy}
Zou,~J.; Yin,~Z.; Ding,~K.; Tang,~Q.; Li,~J.; Si,~W.; Shao,~J.; Zhang,~Q.;
  Huang,~W.; Dong,~X. BODIPY derivatives for photodynamic therapy: influence of
  configuration versus heavy atom effect. \emph{ACS Appl. Mater. Interfaces}
  \textbf{2017}, \emph{9}, 32475--32481\relax
\mciteBstWouldAddEndPuncttrue
\mciteSetBstMidEndSepPunct{\mcitedefaultmidpunct}
{\mcitedefaultendpunct}{\mcitedefaultseppunct}\relax
\EndOfBibitem
\bibitem[Alberto \latin{et~al.}(2014)Alberto, De~Simone, Mazzone, Quartarolo,
  and Russo]{alberto2014theoretical}
Alberto,~M.~E.; De~Simone,~B.~C.; Mazzone,~G.; Quartarolo,~A.~D.; Russo,~N.
  Theoretical determination of electronic spectra and intersystem spin--orbit
  coupling: the case of isoindole-BODIPY dyes. \emph{J. Chem. Theory Comput.}
  \textbf{2014}, \emph{10}, 4006--4013\relax
\mciteBstWouldAddEndPuncttrue
\mciteSetBstMidEndSepPunct{\mcitedefaultmidpunct}
{\mcitedefaultendpunct}{\mcitedefaultseppunct}\relax
\EndOfBibitem
\bibitem[S{\'a}nchez-Arroyo \latin{et~al.}(2017)S{\'a}nchez-Arroyo, Palao,
  Agarrabeitia, Ortiz, and Garc{\'\i}a-Fresnadillo]{sanchez2017towards}
S{\'a}nchez-Arroyo,~A.~J.; Palao,~E.; Agarrabeitia,~A.~R.; Ortiz,~M.~J.;
  Garc{\'\i}a-Fresnadillo,~D. Towards improved halogenated BODIPY
  photosensitizers: clues on structural designs and heavy atom substitution
  patterns. \emph{Phys. Chem. Chem. Phys.} \textbf{2017}, \emph{19},
  69--72\relax
\mciteBstWouldAddEndPuncttrue
\mciteSetBstMidEndSepPunct{\mcitedefaultmidpunct}
{\mcitedefaultendpunct}{\mcitedefaultseppunct}\relax
\EndOfBibitem
\bibitem[Kamkaew \latin{et~al.}(2013)Kamkaew, Lim, Lee, Kiew, Chung, and
  Burgess]{kamkaew2013bodipy}
Kamkaew,~A.; Lim,~S.~H.; Lee,~H.~B.; Kiew,~L.~V.; Chung,~L.~Y.; Burgess,~K.
  BODIPY dyes in photodynamic therapy. \emph{Chem. Soc. Rev.} \textbf{2013},
  \emph{42}, 77--88\relax
\mciteBstWouldAddEndPuncttrue
\mciteSetBstMidEndSepPunct{\mcitedefaultmidpunct}
{\mcitedefaultendpunct}{\mcitedefaultseppunct}\relax
\EndOfBibitem
\bibitem[Singh and Gayathri(2014)Singh, and Gayathri]{singh2014evolution}
Singh,~S.~P.; Gayathri,~T. Evolution of BODIPY dyes as potential sensitizers
  for dye-sensitized solar cells. \emph{Eur. J. Org. Chem.} \textbf{2014},
  \emph{2014}, 4689--4707\relax
\mciteBstWouldAddEndPuncttrue
\mciteSetBstMidEndSepPunct{\mcitedefaultmidpunct}
{\mcitedefaultendpunct}{\mcitedefaultseppunct}\relax
\EndOfBibitem
\bibitem[Menges(2015)]{menges2015computational}
Menges,~N. Computational study on aromaticity and resonance structures of
  substituted BODIPY derivatives. \emph{Comput. Theor. Chem.} \textbf{2015},
  \emph{1068}, 117--122\relax
\mciteBstWouldAddEndPuncttrue
\mciteSetBstMidEndSepPunct{\mcitedefaultmidpunct}
{\mcitedefaultendpunct}{\mcitedefaultseppunct}\relax
\EndOfBibitem
\bibitem[Ponte \latin{et~al.}(2018)Ponte, Mazzone, Russo, and
  Sicilia]{ponte2018bodipy}
Ponte,~F.; Mazzone,~G.; Russo,~N.; Sicilia,~E. BODIPY for photodynamic therapy
  applications: computational study of the effect of bromine substitution on
  1O2 photosensitization. \emph{J. Mol. Model.} \textbf{2018}, \emph{24},
  1--6\relax
\mciteBstWouldAddEndPuncttrue
\mciteSetBstMidEndSepPunct{\mcitedefaultmidpunct}
{\mcitedefaultendpunct}{\mcitedefaultseppunct}\relax
\EndOfBibitem
\bibitem[Alkhatib \latin{et~al.}(2022)Alkhatib, Helal, and
  Marashdeh]{alkhatib2022accurate}
Alkhatib,~Q.; Helal,~W.; Marashdeh,~A. Accurate predictions of the electronic
  excited states of BODIPY based dye sensitizers using spin-component-scaled
  double-hybrid functionals: a TD-DFT benchmark study. \emph{RSC Adv.}
  \textbf{2022}, \emph{12}, 1704--1717\relax
\mciteBstWouldAddEndPuncttrue
\mciteSetBstMidEndSepPunct{\mcitedefaultmidpunct}
{\mcitedefaultendpunct}{\mcitedefaultseppunct}\relax
\EndOfBibitem
\bibitem[Spiegel \latin{et~al.}(2015)Spiegel, Kleinschmidt, Larbig, Tatchen,
  and Marian]{spiegel2015quantum}
Spiegel,~J.~D.; Kleinschmidt,~M.; Larbig,~A.; Tatchen,~J.; Marian,~C.~M.
  Quantum-chemical studies on excitation energy transfer processes in
  BODIPY-based donor--acceptor systems. \emph{J. Chem. Theory Comput.}
  \textbf{2015}, \emph{11}, 4316--4327\relax
\mciteBstWouldAddEndPuncttrue
\mciteSetBstMidEndSepPunct{\mcitedefaultmidpunct}
{\mcitedefaultendpunct}{\mcitedefaultseppunct}\relax
\EndOfBibitem
\bibitem[Manton \latin{et~al.}(2014)Manton, Long, Vos, and
  Pryce]{manton2014photo}
Manton,~J.~C.; Long,~C.; Vos,~J.~G.; Pryce,~M.~T. A photo-and electrochemical
  investigation of BODIPY--cobaloxime complexes for hydrogen production,
  coupled with quantum chemical calculations. \emph{Phys. Chem. Chem. Phys.}
  \textbf{2014}, \emph{16}, 5229--5236\relax
\mciteBstWouldAddEndPuncttrue
\mciteSetBstMidEndSepPunct{\mcitedefaultmidpunct}
{\mcitedefaultendpunct}{\mcitedefaultseppunct}\relax
\EndOfBibitem
\bibitem[Pogonin \latin{et~al.}(2020)Pogonin, Shagurin, Savenkova, Telegin,
  Marfin, and Vashurin]{pogonin2020quantum}
Pogonin,~A.~E.; Shagurin,~A.~Y.; Savenkova,~M.~A.; Telegin,~F.~Y.;
  Marfin,~Y.~S.; Vashurin,~A.~S. Quantum chemical study aimed at modeling
  efficient aza-BODIPY NIR dyes: molecular and electronic structure,
  absorption, and emission spectra. \emph{Molecules} \textbf{2020}, \emph{25},
  5361\relax
\mciteBstWouldAddEndPuncttrue
\mciteSetBstMidEndSepPunct{\mcitedefaultmidpunct}
{\mcitedefaultendpunct}{\mcitedefaultseppunct}\relax
\EndOfBibitem
\bibitem[Lin \latin{et~al.}(2020)Lin, Kohn, and Van~Voorhis]{lin2020toward}
Lin,~Z.; Kohn,~A.~W.; Van~Voorhis,~T. Toward prediction of nonradiative decay
  pathways in organic compounds II: two internal conversion channels in
  BODIPYs. \emph{J. Phys. Chem. C} \textbf{2020}, \emph{124}, 3925--3938\relax
\mciteBstWouldAddEndPuncttrue
\mciteSetBstMidEndSepPunct{\mcitedefaultmidpunct}
{\mcitedefaultendpunct}{\mcitedefaultseppunct}\relax
\EndOfBibitem
\bibitem[Nyk{\"{a}}nen \latin{et~al.}(2024)Nyk{\"{a}}nen, Thiessen, Borrelli,
  Krishna, Knecht, and Pavo{\v{s}}evi{\'{c}}]{nykanen2024toward}
Nyk{\"{a}}nen,~A.; Thiessen,~L.; Borrelli,~E.-M.; Krishna,~V.; Knecht,~S.;
  Pavo{\v{s}}evi{\'{c}},~F. Toward accurate calculation of excitation energies
  on quantum computers with $\Delta$ADAPT-VQE: A case study of BODIPY
  derivatives. \emph{J. Phys. Chem. Lett.} \textbf{2024}, \emph{15},
  7111--7117\relax
\mciteBstWouldAddEndPuncttrue
\mciteSetBstMidEndSepPunct{\mcitedefaultmidpunct}
{\mcitedefaultendpunct}{\mcitedefaultseppunct}\relax
\EndOfBibitem
\bibitem[Chen \latin{et~al.}(2018)Chen, Liu, Song, Jiang, Liu, Liu, Fu, Xue,
  Liu, Huang, \latin{et~al.} others]{chen2018insight}
Chen,~Y.; Liu,~J.; Song,~M.; Jiang,~L.; Liu,~L.; Liu,~Y.; Fu,~G.; Xue,~J.;
  Liu,~J.-y.; Huang,~M., \latin{et~al.}  Insights into the binding mechanism of
  BODIPY-based photosensitizers to human serum albumin: A combined experimental
  and computational study. \emph{Spectrochim. Acta A Mol. Biomol. Spectrosc.
  SPECTROCHIM ACTA A} \textbf{2018}, \emph{203}, 158--165\relax
\mciteBstWouldAddEndPuncttrue
\mciteSetBstMidEndSepPunct{\mcitedefaultmidpunct}
{\mcitedefaultendpunct}{\mcitedefaultseppunct}\relax
\EndOfBibitem
\bibitem[Song \latin{et~al.}(2011)Song, Livanec, Klauda, Kuczera, Dunn, and
  Im]{song2011orientation}
Song,~K.~C.; Livanec,~P.~W.; Klauda,~J.~B.; Kuczera,~K.; Dunn,~R.~C.; Im,~W.
  Orientation of fluorescent lipid analogue BODIPY-PC to probe lipid membrane
  properties: Insights from molecular dynamics simulations. \emph{J. Phys.
  Chem. B} \textbf{2011}, \emph{115}, 6157--6165\relax
\mciteBstWouldAddEndPuncttrue
\mciteSetBstMidEndSepPunct{\mcitedefaultmidpunct}
{\mcitedefaultendpunct}{\mcitedefaultseppunct}\relax
\EndOfBibitem
\bibitem[Li \latin{et~al.}(2013)Li, Li, Xiao, Qi, Huang, Xie, Jing, and
  Zhang]{li2013iodo}
Li,~W.; Li,~L.; Xiao,~H.; Qi,~R.; Huang,~Y.; Xie,~Z.; Jing,~X.; Zhang,~H.
  Iodo-BODIPY: a visible-light-driven, highly efficient and photostable
  metal-free organic photocatalyst. \emph{RSC Adv.} \textbf{2013}, \emph{3},
  13417--13421\relax
\mciteBstWouldAddEndPuncttrue
\mciteSetBstMidEndSepPunct{\mcitedefaultmidpunct}
{\mcitedefaultendpunct}{\mcitedefaultseppunct}\relax
\EndOfBibitem
\bibitem[Guo \latin{et~al.}(2013)Guo, Zhang, Huang, Guo, Xiong, and
  Zhao]{guo2013porous}
Guo,~S.; Zhang,~H.; Huang,~L.; Guo,~Z.; Xiong,~G.; Zhao,~J. Porous
  material-immobilized iodo-Bodipy as an efficient photocatalyst for photoredox
  catalytic organic reaction to prepare pyrrolo [2, 1-a] isoquinoline.
  \emph{Chem. Commun.} \textbf{2013}, \emph{49}, 8689--8691\relax
\mciteBstWouldAddEndPuncttrue
\mciteSetBstMidEndSepPunct{\mcitedefaultmidpunct}
{\mcitedefaultendpunct}{\mcitedefaultseppunct}\relax
\EndOfBibitem
\bibitem[Huang and Zhao(2013)Huang, and Zhao]{huang2013iodo}
Huang,~L.; Zhao,~J. Iodo-Bodipys as visible-light-absorbing dual-functional
  photoredox catalysts for preparation of highly functionalized organic
  compounds by formation of C--C bonds via reductive and oxidative quenching
  catalytic mechanisms. \emph{RSC Adv.} \textbf{2013}, \emph{3},
  23377--23388\relax
\mciteBstWouldAddEndPuncttrue
\mciteSetBstMidEndSepPunct{\mcitedefaultmidpunct}
{\mcitedefaultendpunct}{\mcitedefaultseppunct}\relax
\EndOfBibitem
\bibitem[Piskorz \latin{et~al.}(2021)Piskorz, Porolnik, Kucinska, Dlugaszewska,
  Murias, and Mielcarek]{piskorz2021bodipy}
Piskorz,~J.; Porolnik,~W.; Kucinska,~M.; Dlugaszewska,~J.; Murias,~M.;
  Mielcarek,~J. BODIPY-based photosensitizers as potential anticancer and
  antibacterial agents: role of the positive charge and the heavy atom effect.
  \emph{ChemMedChem} \textbf{2021}, \emph{16}, 399--411\relax
\mciteBstWouldAddEndPuncttrue
\mciteSetBstMidEndSepPunct{\mcitedefaultmidpunct}
{\mcitedefaultendpunct}{\mcitedefaultseppunct}\relax
\EndOfBibitem
\bibitem[Zhao \latin{et~al.}(2013)Zhao, Wu, Sun, and Guo]{zhao2013triplet}
Zhao,~J.; Wu,~W.; Sun,~J.; Guo,~S. Triplet photosensitizers: from molecular
  design to applications. \emph{Chem. Soc. Rev.} \textbf{2013}, \emph{42},
  5323--5351\relax
\mciteBstWouldAddEndPuncttrue
\mciteSetBstMidEndSepPunct{\mcitedefaultmidpunct}
{\mcitedefaultendpunct}{\mcitedefaultseppunct}\relax
\EndOfBibitem
\bibitem[Pomogaev \latin{et~al.}(2020)Pomogaev, Chiodo, Ruud, Kuznetsova, and
  Avramov]{pomogaev2020computational}
Pomogaev,~V.; Chiodo,~S.; Ruud,~K.; Kuznetsova,~R.; Avramov,~P. Computational
  investigation on the photophysical properties of halogenated tetraphenyl
  BODIPY. \emph{J. Phys. Chem. C} \textbf{2020}, \emph{124}, 11100--11109\relax
\mciteBstWouldAddEndPuncttrue
\mciteSetBstMidEndSepPunct{\mcitedefaultmidpunct}
{\mcitedefaultendpunct}{\mcitedefaultseppunct}\relax
\EndOfBibitem
\bibitem[Lee \latin{et~al.}(2020)Lee, Malamakal, Chenoweth, and
  Anna]{lee2020halogen}
Lee,~Y.; Malamakal,~R.~M.; Chenoweth,~D.~M.; Anna,~J.~M. Halogen bonding
  facilitates intersystem crossing in iodo-BODIPY chromophores. \emph{J. Phys.
  Chem. Lett.} \textbf{2020}, \emph{11}, 877--884\relax
\mciteBstWouldAddEndPuncttrue
\mciteSetBstMidEndSepPunct{\mcitedefaultmidpunct}
{\mcitedefaultendpunct}{\mcitedefaultseppunct}\relax
\EndOfBibitem
\bibitem[Ly \latin{et~al.}(2021)Ly, Presley, Cooper, Baldwin, Dalton, and
  Grusenmeyer]{ly2021impact}
Ly,~J.~T.; Presley,~K.~F.; Cooper,~T.~M.; Baldwin,~L.~A.; Dalton,~M.~J.;
  Grusenmeyer,~T.~A. Impact of iodine loading and substitution position on
  intersystem crossing efficiency in a series of ten
  methylated-meso-phenyl-BODIPY dyes. \emph{Phys. Chem. Chem. Phys.}
  \textbf{2021}, \emph{23}, 12033--12044\relax
\mciteBstWouldAddEndPuncttrue
\mciteSetBstMidEndSepPunct{\mcitedefaultmidpunct}
{\mcitedefaultendpunct}{\mcitedefaultseppunct}\relax
\EndOfBibitem
\bibitem[Bassan \latin{et~al.}(2022)Bassan, Dai, Fazzi, Gualandi, Cozzi, Negri,
  and Ceroni]{bassan2022effect}
Bassan,~E.; Dai,~Y.; Fazzi,~D.; Gualandi,~A.; Cozzi,~P.~G.; Negri,~F.;
  Ceroni,~P. Effect of the iodine atom position on the phosphorescence of
  BODIPY derivatives: a combined computational and experimental study.
  \emph{Photochem. Photobiol. Sci.} \textbf{2022}, 1--10\relax
\mciteBstWouldAddEndPuncttrue
\mciteSetBstMidEndSepPunct{\mcitedefaultmidpunct}
{\mcitedefaultendpunct}{\mcitedefaultseppunct}\relax
\EndOfBibitem
\bibitem[Pordel \latin{et~al.}(2021)Pordel, Pickens, and
  White]{pordel2021release}
Pordel,~S.; Pickens,~R.~N.; White,~J.~K. Release of CO and production of 1O2
  from a Mn-BODIPY photoactivated CO releasing molecule with visible light.
  \emph{Organometallics} \textbf{2021}, \emph{40}, 2983--2994\relax
\mciteBstWouldAddEndPuncttrue
\mciteSetBstMidEndSepPunct{\mcitedefaultmidpunct}
{\mcitedefaultendpunct}{\mcitedefaultseppunct}\relax
\EndOfBibitem
\bibitem[Ziems \latin{et~al.}(2018)Ziems, Gr{\"a}fe, and
  Kupfer]{ziems2018photo}
Ziems,~K.~M.; Gr{\"a}fe,~S.; Kupfer,~S. Photo-induced charge separation vs.
  degradation of a BODIPY-based photosensitizer assessed by TDDFT and RASPT2.
  \emph{Catalysts} \textbf{2018}, \emph{8}, 520\relax
\mciteBstWouldAddEndPuncttrue
\mciteSetBstMidEndSepPunct{\mcitedefaultmidpunct}
{\mcitedefaultendpunct}{\mcitedefaultseppunct}\relax
\EndOfBibitem
\bibitem[Wang \latin{et~al.}(2021)Wang, Toffoletti, Hou, Zhao, Barbon, and
  Dick]{wang2021insight}
Wang,~Z.; Toffoletti,~A.; Hou,~Y.; Zhao,~J.; Barbon,~A.; Dick,~B. Insight into
  the drastically different triplet lifetimes of BODIPY obtained by
  optical/magnetic spectroscopy and theoretical computations. \emph{Chem. Sci.}
  \textbf{2021}, \emph{12}, 2829--2840\relax
\mciteBstWouldAddEndPuncttrue
\mciteSetBstMidEndSepPunct{\mcitedefaultmidpunct}
{\mcitedefaultendpunct}{\mcitedefaultseppunct}\relax
\EndOfBibitem
\bibitem[He \latin{et~al.}(2012)He, Si, Zhong, and Dubey]{he2012iodized}
He,~H.; Si,~L.; Zhong,~Y.; Dubey,~M. Iodized BODIPY as a long wavelength light
  sensitizer for the near-infrared emission of ytterbium (III) ion. \emph{Chem.
  Commun.} \textbf{2012}, \emph{48}, 1886--1888\relax
\mciteBstWouldAddEndPuncttrue
\mciteSetBstMidEndSepPunct{\mcitedefaultmidpunct}
{\mcitedefaultendpunct}{\mcitedefaultseppunct}\relax
\EndOfBibitem
\bibitem[{\"O}zcan \latin{et~al.}(2021){\"O}zcan, Dedeoglu, Chumakov,
  G{\"u}rek, Zorlu, {\c{C}}o{\c{s}}ut, and Menaf~Ayhan]{ozcan2021halogen}
{\"O}zcan,~E.; Dedeoglu,~B.; Chumakov,~Y.; G{\"u}rek,~A.~G.; Zorlu,~Y.;
  {\c{C}}o{\c{s}}ut,~B.; Menaf~Ayhan,~M. Halogen-bonded BODIPY frameworks with
  tunable optical features. \emph{Eur. J. Chem.} \textbf{2021}, \emph{27},
  1603--1608\relax
\mciteBstWouldAddEndPuncttrue
\mciteSetBstMidEndSepPunct{\mcitedefaultmidpunct}
{\mcitedefaultendpunct}{\mcitedefaultseppunct}\relax
\EndOfBibitem
\bibitem[Patalag \latin{et~al.}(2022)Patalag, Hoche, Mitric, Werz, and
  Feringa]{patalag2022transforming}
Patalag,~L.~J.; Hoche,~J.; Mitric,~R.; Werz,~D.~B.; Feringa,~B.~L. Transforming
  dyes into fluorophores: exciton-induced emission with chain-like oligo-BODIPY
  superstructures. \emph{Angewandte Chemie} \textbf{2022}, \emph{134},
  e202116834\relax
\mciteBstWouldAddEndPuncttrue
\mciteSetBstMidEndSepPunct{\mcitedefaultmidpunct}
{\mcitedefaultendpunct}{\mcitedefaultseppunct}\relax
\EndOfBibitem
\bibitem[Dole{\v{z}}el \latin{et~al.}(2024)Dole{\v{z}}el, Poryvai, Slanina,
  Filgas, and Slav{\'{\i}}{\v{c}}ek]{dolezel2023spin}
Dole{\v{z}}el,~J.; Poryvai,~A.; Slanina,~T.; Filgas,~J.;
  Slav{\'{\i}}{\v{c}}ek,~P. Spin--vibronic coupling controls the intersystem
  crossing of iodine-substituted BODIPY triplet chromophores. \emph{Eur. J.
  Chem.} \textbf{2024}, \emph{30}, e202303154--10\relax
\mciteBstWouldAddEndPuncttrue
\mciteSetBstMidEndSepPunct{\mcitedefaultmidpunct}
{\mcitedefaultendpunct}{\mcitedefaultseppunct}\relax
\EndOfBibitem
\bibitem[Wasif~Baig \latin{et~al.}(2021)Wasif~Baig, Pederzoli, K{\'{y}}vala,
  Cwiklik, and Pittner]{wasif2021theoretical}
Wasif~Baig,~M.; Pederzoli,~M.; K{\'{y}}vala,~M.; Cwiklik,~L.; Pittner,~J.
  Theoretical investigation of the effect of alkylation and bromination on
  intersystem crossing in BODIPY-based photosensitizers. \emph{J. Phys. Chem.
  B} \textbf{2021}, \emph{125}, 11617--11627\relax
\mciteBstWouldAddEndPuncttrue
\mciteSetBstMidEndSepPunct{\mcitedefaultmidpunct}
{\mcitedefaultendpunct}{\mcitedefaultseppunct}\relax
\EndOfBibitem
\bibitem[Pederzoli \latin{et~al.}(2019)Pederzoli, Wasif~Baig, K{\'{y}}vala,
  Pittner, and Cwiklik]{pederzoli2019photophysics}
Pederzoli,~M.; Wasif~Baig,~M.; K{\'{y}}vala,~M.; Pittner,~J.; Cwiklik,~L.
  Photophysics of BODIPY-based photosensitizer for photodynamic therapy:
  surface hopping and classical molecular dynamics. \emph{J. Chem. Theory
  Comput.} \textbf{2019}, \emph{15}, 5046--5057\relax
\mciteBstWouldAddEndPuncttrue
\mciteSetBstMidEndSepPunct{\mcitedefaultmidpunct}
{\mcitedefaultendpunct}{\mcitedefaultseppunct}\relax
\EndOfBibitem
\bibitem[Ahlrichs \latin{et~al.}(1989)Ahlrichs, B{\"a}r, H{\"a}ser, Horn, and
  K{\"o}lmel]{ahlrichs1989electronic}
Ahlrichs,~R.; B{\"a}r,~M.; H{\"a}ser,~M.; Horn,~H.; K{\"o}lmel,~C. Electronic
  structure calculations on workstation computers: The program system
  turbomole. \emph{Chem. Phys. Lett.} \textbf{1989}, \emph{162}, 165--169\relax
\mciteBstWouldAddEndPuncttrue
\mciteSetBstMidEndSepPunct{\mcitedefaultmidpunct}
{\mcitedefaultendpunct}{\mcitedefaultseppunct}\relax
\EndOfBibitem
\bibitem[Becke(1993)]{becke1993density}
Becke,~A.~D. Density-functional thermochemistry. III. The role of exact
  exchange. \emph{J. Chem. Phys.} \textbf{1993}, \emph{98}, 5648--5652\relax
\mciteBstWouldAddEndPuncttrue
\mciteSetBstMidEndSepPunct{\mcitedefaultmidpunct}
{\mcitedefaultendpunct}{\mcitedefaultseppunct}\relax
\EndOfBibitem
\bibitem[Stephens \latin{et~al.}(1994)Stephens, Devlin, Chabalowski, and
  Frisch]{stephens1994ab}
Stephens,~P.~J.; Devlin,~F.~J.; Chabalowski,~C.~F.; Frisch,~M.~J. {\emph{Ab}}
  {\emph{initio}} calculation of vibrational absorption and circular dichroism
  spectra using density functional force fields. \emph{J. Phys. Chem.}
  \textbf{1994}, \emph{98}, 11623--11627\relax
\mciteBstWouldAddEndPuncttrue
\mciteSetBstMidEndSepPunct{\mcitedefaultmidpunct}
{\mcitedefaultendpunct}{\mcitedefaultseppunct}\relax
\EndOfBibitem
\bibitem[Becke(1993)]{becke1993new}
Becke,~A.~D. A new mixing of Hartree--Fock and local density-functional
  theories. \emph{J. Chem. Phys.} \textbf{1993}, \emph{98}, 1372--1377\relax
\mciteBstWouldAddEndPuncttrue
\mciteSetBstMidEndSepPunct{\mcitedefaultmidpunct}
{\mcitedefaultendpunct}{\mcitedefaultseppunct}\relax
\EndOfBibitem
\bibitem[Zhao and Truhlar(2008)Zhao, and Truhlar]{zhao2008m06}
Zhao,~Y.; Truhlar,~D.~G. The M06 suite of density functionals for main group
  thermochemistry, thermochemical kinetics, noncovalent interactions, excited
  states, and transition elements: two new functionals and systematic testing
  of four M06-class functionals and 12 other functionals. \emph{Theor. Chem.
  Acc.} \textbf{2008}, \emph{120}, 215--241\relax
\mciteBstWouldAddEndPuncttrue
\mciteSetBstMidEndSepPunct{\mcitedefaultmidpunct}
{\mcitedefaultendpunct}{\mcitedefaultseppunct}\relax
\EndOfBibitem
\bibitem[Weigend \latin{et~al.}(1998)Weigend, H{\"{a}}ser, Patzelt, and
  Ahlrichs]{weigend1998ri}
Weigend,~F.; H{\"{a}}ser,~M.; Patzelt,~H.; Ahlrichs,~R. RI-MP2: optimized
  auxiliary basis sets and demonstration of efficiency. \emph{Chem. Phys.
  Lett.} \textbf{1998}, \emph{294}, 143--152\relax
\mciteBstWouldAddEndPuncttrue
\mciteSetBstMidEndSepPunct{\mcitedefaultmidpunct}
{\mcitedefaultendpunct}{\mcitedefaultseppunct}\relax
\EndOfBibitem
\bibitem[Weigend and Baldes(2010)Weigend, and Baldes]{weigend2010segmented}
Weigend,~F.; Baldes,~A. Segmented contracted basis sets for one- and
  two-component Dirac--Fock effective core potentials. \emph{J. Chem. Phys.}
  \textbf{2010}, \emph{133}, 174102\relax
\mciteBstWouldAddEndPuncttrue
\mciteSetBstMidEndSepPunct{\mcitedefaultmidpunct}
{\mcitedefaultendpunct}{\mcitedefaultseppunct}\relax
\EndOfBibitem
\bibitem[Peterson \latin{et~al.}(2006)Peterson, Shepler, Figgen, and
  Stoll]{peterson2006spectroscopic}
Peterson,~K.~A.; Shepler,~B.~C.; Figgen,~D.; Stoll,~H. On the spectroscopic and
  thermochemical properties of ClO, BrO, IO, and their anions. \emph{J. Phys.
  Chem. A} \textbf{2006}, \emph{110}, 13877--13883\relax
\mciteBstWouldAddEndPuncttrue
\mciteSetBstMidEndSepPunct{\mcitedefaultmidpunct}
{\mcitedefaultendpunct}{\mcitedefaultseppunct}\relax
\EndOfBibitem
\bibitem[Frisch \latin{et~al.}(2009)Frisch, Trucks, Schlegel, Scuseria, Robb,
  Cheeseman, Scalmani, Barone, Mennucci, Petersson, Nakatsuji, Caricato, Li,
  Hratchian, Izmaylov, Bloino, Zheng, Sonnenberg, Hada, Ehara, Toyota, Fukuda,
  Hasegawa, Ishida, Nakajima, Honda, Kitao, Nakai, Vreven, Montgomery, Jr.,
  Ogliaro, Bearpark, Heyd, Brothers, Kudin, Staroverov, Kobayashi, Normand,
  Raghavachari, Rendell, Burant, Iyengar, Tomasi, Cossi, Rega, Millam, Klene,
  Knox, Cross, Bakken, Adamo, Jaramillo, Gomperts, Stratmann, Yazyev, Austin,
  Cammi, Pomelli, Ochterski, Martin, Morokuma, Zakrzewski, Voth, Salvador,
  Dannenberg, Dapprich, Daniels, Farkas, Foresman, Ortiz, Cioslowski, and
  Fox]{frisch2009gaussian}
Frisch,~M.~J. \latin{et~al.}  Gaussian 09 Revision D01. 2009; Gaussian, Inc.,
  Wallingford CT\relax
\mciteBstWouldAddEndPuncttrue
\mciteSetBstMidEndSepPunct{\mcitedefaultmidpunct}
{\mcitedefaultendpunct}{\mcitedefaultseppunct}\relax
\EndOfBibitem
\bibitem[Dunning~Jr(1989)]{dunning1989gaussian}
Dunning~Jr,~T.~H. Gaussian basis sets for use in correlated molecular
  calculations. I. The atoms boron through neon and hydrogen. \emph{J. Chem.
  Phys.} \textbf{1989}, \emph{90}, 1007--1023\relax
\mciteBstWouldAddEndPuncttrue
\mciteSetBstMidEndSepPunct{\mcitedefaultmidpunct}
{\mcitedefaultendpunct}{\mcitedefaultseppunct}\relax
\EndOfBibitem
\bibitem[Kendall \latin{et~al.}(1992)Kendall, Dunning~Jr, and
  Harrison]{kendall1992electron}
Kendall,~R.~A.; Dunning~Jr,~T.~H.; Harrison,~R.~J. Electron affinities of the
  first-row atoms revisited. Systematic basis sets and wave functions. \emph{J.
  Chem. Phys.} \textbf{1992}, \emph{96}, 6796--6806\relax
\mciteBstWouldAddEndPuncttrue
\mciteSetBstMidEndSepPunct{\mcitedefaultmidpunct}
{\mcitedefaultendpunct}{\mcitedefaultseppunct}\relax
\EndOfBibitem
\bibitem[Peterson \latin{et~al.}(2003)Peterson, Figgen, Goll, Stoll, and
  Dolg]{peterson2003systematically}
Peterson,~K.~A.; Figgen,~D.; Goll,~E.; Stoll,~H.; Dolg,~M. Systematically
  convergent basis sets with relativistic pseudopotentials. II. Small-core
  pseudopotentials and correlation consistent basis sets for the post-d group
  16--18 elements. \emph{J. Chem. Phys.} \textbf{2003}, \emph{119},
  11113--11123\relax
\mciteBstWouldAddEndPuncttrue
\mciteSetBstMidEndSepPunct{\mcitedefaultmidpunct}
{\mcitedefaultendpunct}{\mcitedefaultseppunct}\relax
\EndOfBibitem
\bibitem[Grimme \latin{et~al.}(2016)Grimme, Hansen, Brandenburg, and
  Bannwarth]{grimme2016dispersion}
Grimme,~S.; Hansen,~A.; Brandenburg,~J.~G.; Bannwarth,~C. Dispersion-corrected
  mean-field electronic structure methods. \emph{Chem. Rev.} \textbf{2016},
  \emph{116}, 5105--5154\relax
\mciteBstWouldAddEndPuncttrue
\mciteSetBstMidEndSepPunct{\mcitedefaultmidpunct}
{\mcitedefaultendpunct}{\mcitedefaultseppunct}\relax
\EndOfBibitem
\bibitem[Aquilante \latin{et~al.}(2010)Aquilante, de~Vico, Ferr{\'{e}}, Ghigo,
  Malmqvist, Neogr{\'{a}}dy, Pedersen, Pito{\v{n}}{\'{a}}k, Reiher, Roos,
  Serrano-Andr{\'{e}}s, Urban, Veryazov, and Lindh]{aquilante2010molcas}
Aquilante,~F.; de~Vico,~L.; Ferr{\'{e}},~N.; Ghigo,~G.; Malmqvist,~P.-{\AA}.;
  Neogr{\'{a}}dy,~P.; Pedersen,~T.~B.; Pito{\v{n}}{\'{a}}k,~M.; Reiher,~M.;
  Roos,~B.~O.; Serrano-Andr{\'{e}}s,~L.; Urban,~M.; Veryazov,~V.; Lindh,~R.
  MOLCAS 7: The next generation. \emph{J. Comput. Chem.} \textbf{2010},
  \emph{31}, 224--247\relax
\mciteBstWouldAddEndPuncttrue
\mciteSetBstMidEndSepPunct{\mcitedefaultmidpunct}
{\mcitedefaultendpunct}{\mcitedefaultseppunct}\relax
\EndOfBibitem
\bibitem[Roos \latin{et~al.}(2004)Roos, Lindh, Malmqvist, Veryazov, and
  Widmark]{roos2004main}
Roos,~B.~O.; Lindh,~R.; Malmqvist,~P.-{\AA}.; Veryazov,~V.; Widmark,~P.-O. Main
  group atoms and dimers studied with a new relativistic ANO basis set.
  \emph{J. Phys. Chem. A} \textbf{2004}, \emph{108}, 2851--2858\relax
\mciteBstWouldAddEndPuncttrue
\mciteSetBstMidEndSepPunct{\mcitedefaultmidpunct}
{\mcitedefaultendpunct}{\mcitedefaultseppunct}\relax
\EndOfBibitem
\bibitem[Brabec \latin{et~al.}(2021)Brabec, Brandejs, Kowalski, Xantheas,
  Legeza, and Veis]{brabec2021massively}
Brabec,~J.; Brandejs,~J.; Kowalski,~K.; Xantheas,~S.; Legeza,~O.; Veis,~L.
  Massively parallel quantum chemical density matrix renormalization group
  method. \emph{J. Comp. Chem.} \textbf{2021}, \emph{42}, 534--544\relax
\mciteBstWouldAddEndPuncttrue
\mciteSetBstMidEndSepPunct{\mcitedefaultmidpunct}
{\mcitedefaultendpunct}{\mcitedefaultseppunct}\relax
\EndOfBibitem
\bibitem[Liu \latin{et~al.}(1997)Liu, Hong, Dai, Li, and Dolg]{liu1997beijing}
Liu,~W.; Hong,~G.; Dai,~D.; Li,~L.; Dolg,~M. The Beijing four-component density
  functional program package (BDF) and its application to EuO, EuS, YbO and
  YbS. \emph{Theor. Chem. Acc.} \textbf{1997}, \emph{96}, 75--83\relax
\mciteBstWouldAddEndPuncttrue
\mciteSetBstMidEndSepPunct{\mcitedefaultmidpunct}
{\mcitedefaultendpunct}{\mcitedefaultseppunct}\relax
\EndOfBibitem
\bibitem[Zhang \latin{et~al.}(2020)Zhang, Suo, Wang, Zhang, Li, Lei, Zou, Gao,
  Peng, Pu, \latin{et~al.} others]{zhang2020bdf}
Zhang,~Y.; Suo,~B.; Wang,~Z.; Zhang,~N.; Li,~Z.; Lei,~Y.; Zou,~W.; Gao,~J.;
  Peng,~D.; Pu,~Z., \latin{et~al.}  BDF: A relativistic electronic structure
  program package. \emph{J. Chem. Phys.} \textbf{2020}, \emph{152},
  064113\relax
\mciteBstWouldAddEndPuncttrue
\mciteSetBstMidEndSepPunct{\mcitedefaultmidpunct}
{\mcitedefaultendpunct}{\mcitedefaultseppunct}\relax
\EndOfBibitem
\bibitem[Liu and Xiao(2018)Liu, and Xiao]{liu2018relativistic}
Liu,~W.; Xiao,~Y. Relativistic time-dependent density functional theories.
  \emph{Chem. Soc. Rev.} \textbf{2018}, \emph{47}, 4481--4509\relax
\mciteBstWouldAddEndPuncttrue
\mciteSetBstMidEndSepPunct{\mcitedefaultmidpunct}
{\mcitedefaultendpunct}{\mcitedefaultseppunct}\relax
\EndOfBibitem
\bibitem[Pollak and Weigend(2017)Pollak, and Weigend]{pollak2017segmented}
Pollak,~P.; Weigend,~F. Segmented contracted error-consistent basis sets of
  double-and triple-$\zeta$ valence quality for one-and two-component
  relativistic all-electron calculations. \emph{J. Chem. Theory Comput.}
  \textbf{2017}, \emph{13}, 3696--3705\relax
\mciteBstWouldAddEndPuncttrue
\mciteSetBstMidEndSepPunct{\mcitedefaultmidpunct}
{\mcitedefaultendpunct}{\mcitedefaultseppunct}\relax
\EndOfBibitem
\bibitem[Crespo-Otero and Barbatti(2012)Crespo-Otero, and
  Barbatti]{crespo2012spectrum}
Crespo-Otero,~R.; Barbatti,~M. Spectrum simulation and decomposition with
  nuclear ensemble: formal derivation and application to benzene, furan and
  2-phenylfuran. \emph{Theor. Chem. Acc.} \textbf{2012}, \emph{131},
  1237--14\relax
\mciteBstWouldAddEndPuncttrue
\mciteSetBstMidEndSepPunct{\mcitedefaultmidpunct}
{\mcitedefaultendpunct}{\mcitedefaultseppunct}\relax
\EndOfBibitem
\bibitem[Tully(1990)]{tully1990molecular}
Tully,~J.~C. Molecular dynamics with electronic transitions. \emph{J. Chem.
  Phys.} \textbf{1990}, \emph{93}, 1061--1071\relax
\mciteBstWouldAddEndPuncttrue
\mciteSetBstMidEndSepPunct{\mcitedefaultmidpunct}
{\mcitedefaultendpunct}{\mcitedefaultseppunct}\relax
\EndOfBibitem
\bibitem[Wang \latin{et~al.}(2016)Wang, Akimov, and Prezhdo]{wang2016recent}
Wang,~L.; Akimov,~A.; Prezhdo,~O.~V. Recent progress in surface hopping:
  2011--2015. \emph{J. Phys. Chem. Lett.} \textbf{2016}, \emph{7},
  2100--2112\relax
\mciteBstWouldAddEndPuncttrue
\mciteSetBstMidEndSepPunct{\mcitedefaultmidpunct}
{\mcitedefaultendpunct}{\mcitedefaultseppunct}\relax
\EndOfBibitem
\bibitem[Crespo-Otero and Barbatti(2018)Crespo-Otero, and
  Barbatti]{crespo2018recent}
Crespo-Otero,~R.; Barbatti,~M. Recent advances and perspectives on nonadiabatic
  mixed quantum-classical dynamics. \emph{Chem. Rev.} \textbf{2018},
  \emph{118}, 7026--7068\relax
\mciteBstWouldAddEndPuncttrue
\mciteSetBstMidEndSepPunct{\mcitedefaultmidpunct}
{\mcitedefaultendpunct}{\mcitedefaultseppunct}\relax
\EndOfBibitem
\bibitem[Nelson \latin{et~al.}(2020)Nelson, White, Bjorgaard, Sifain, Zhang,
  Nebgen, Fernandez-Alberti, Mozyrsky, Roitberg, and
  Tretiak]{nelson2020nonadiabatic}
Nelson,~T.~R.; White,~A.~J.; Bjorgaard,~J.~A.; Sifain,~A.~E.; Zhang,~Y.;
  Nebgen,~B.; Fernandez-Alberti,~S.; Mozyrsky,~D.; Roitberg,~A.~E.; Tretiak,~S.
  Non-adiabatic excited-state molecular dynamics: Theory and applications for
  modeling photophysics in extended molecular materials. \emph{Chem. Rev.}
  \textbf{2020}, \emph{120}, 2215--2287\relax
\mciteBstWouldAddEndPuncttrue
\mciteSetBstMidEndSepPunct{\mcitedefaultmidpunct}
{\mcitedefaultendpunct}{\mcitedefaultseppunct}\relax
\EndOfBibitem
\bibitem[Richter \latin{et~al.}(2011)Richter, Marquetand, González-Vázquez,
  Sola, and González]{richter2011sharc}
Richter,~M.; Marquetand,~P.; González-Vázquez,~J.; Sola,~I.; González,~L.
  SHARC: \emph{ab} \emph{initio} molecular dynamics with surface hopping in the
  adiabatic representation including arbitrary couplings. \emph{J. Chem. Theory
  Comput.} \textbf{2011}, \emph{7}, 1253--1258\relax
\mciteBstWouldAddEndPuncttrue
\mciteSetBstMidEndSepPunct{\mcitedefaultmidpunct}
{\mcitedefaultendpunct}{\mcitedefaultseppunct}\relax
\EndOfBibitem
\bibitem[Cui and Thiel(2014)Cui, and Thiel]{cui2014generalized}
Cui,~G.; Thiel,~W. Generalized trajectory surface-hopping method for internal
  conversion and intersystem crossing. \emph{J. Chem. Phys.} \textbf{2014},
  \emph{141}, 124101\relax
\mciteBstWouldAddEndPuncttrue
\mciteSetBstMidEndSepPunct{\mcitedefaultmidpunct}
{\mcitedefaultendpunct}{\mcitedefaultseppunct}\relax
\EndOfBibitem
\bibitem[Barbatti \latin{et~al.}(2014)Barbatti, Ruckenbauer, Plasser, Pittner,
  Granucci, Persico, and Lischka]{barbatti2014newton}
Barbatti,~M.; Ruckenbauer,~M.; Plasser,~F.; Pittner,~J.; Granucci,~G.;
  Persico,~M.; Lischka,~H. Newton-X: a surface-hopping program for nonadiabatic
  molecular dynamics. \emph{Wiley Interdiscip. Rev. Comput. Mol. Sci.}
  \textbf{2014}, \emph{4}, 26--33\relax
\mciteBstWouldAddEndPuncttrue
\mciteSetBstMidEndSepPunct{\mcitedefaultmidpunct}
{\mcitedefaultendpunct}{\mcitedefaultseppunct}\relax
\EndOfBibitem
\bibitem[Barbatti \latin{et~al.}(2022)Barbatti, Bondanza, Crespo-Otero,
  Demoulin, Dral, Granucci, Kossoski, Lischka, Mennucci, Mukherjee, Pederzoli,
  Persico, Jr., Pittner, Plasser, Gil, and Stojanovic]{barbatti2022newton}
Barbatti,~M. \latin{et~al.}  The Newton-X platform: new software developments
  for surface hopping and nuclear ensembles. \emph{J. Chem. Theor. Comp.}
  \textbf{2022}, \emph{18}, 6851--6865\relax
\mciteBstWouldAddEndPuncttrue
\mciteSetBstMidEndSepPunct{\mcitedefaultmidpunct}
{\mcitedefaultendpunct}{\mcitedefaultseppunct}\relax
\EndOfBibitem
\bibitem[Pittner \latin{et~al.}(2009)Pittner, Lischka, and
  Barbatti]{pittner2009optimization}
Pittner,~J.; Lischka,~H.; Barbatti,~M. Optimization of mixed quantum-classical
  dynamics: time-derivative coupling terms and selected couplings. \emph{Chem.
  Phys.} \textbf{2009}, \emph{356}, 147--152\relax
\mciteBstWouldAddEndPuncttrue
\mciteSetBstMidEndSepPunct{\mcitedefaultmidpunct}
{\mcitedefaultendpunct}{\mcitedefaultseppunct}\relax
\EndOfBibitem
\bibitem[Pederzoli and Pittner(2017)Pederzoli, and Pittner]{pederzoli2017new}
Pederzoli,~M.; Pittner,~J. A new approach to molecular dynamics with
  non-adiabatic and spin-orbit effects with applications to QM/MM simulations
  of thiophene and selenophene. \emph{J. Chem. Phys.} \textbf{2017},
  \emph{146}, 114101\relax
\mciteBstWouldAddEndPuncttrue
\mciteSetBstMidEndSepPunct{\mcitedefaultmidpunct}
{\mcitedefaultendpunct}{\mcitedefaultseppunct}\relax
\EndOfBibitem
\bibitem[Mai \latin{et~al.}(2015)Mai, Marquetand, and
  Gonz{\'a}lez]{mai2015general}
Mai,~S.; Marquetand,~P.; Gonz{\'a}lez,~L. A general method to describe
  intersystem crossing dynamics in trajectory surface hopping. \emph{Int. J.
  Quantum Chem.} \textbf{2015}, \emph{115}, 1215--1231\relax
\mciteBstWouldAddEndPuncttrue
\mciteSetBstMidEndSepPunct{\mcitedefaultmidpunct}
{\mcitedefaultendpunct}{\mcitedefaultseppunct}\relax
\EndOfBibitem
\bibitem[Granucci and Persico(2007)Granucci, and Persico]{granucci2007critical}
Granucci,~G.; Persico,~M. Critical appraisal of the fewest switches algorithm
  for surface hopping. \emph{J. Chem. Phys.} \textbf{2007}, \emph{126},
  134114\relax
\mciteBstWouldAddEndPuncttrue
\mciteSetBstMidEndSepPunct{\mcitedefaultmidpunct}
{\mcitedefaultendpunct}{\mcitedefaultseppunct}\relax
\EndOfBibitem
\bibitem[Granucci \latin{et~al.}(2010)Granucci, Persico, and
  Zoccante]{granucci2010including}
Granucci,~G.; Persico,~M.; Zoccante,~A. Including quantum decoherence in
  surface hopping. \emph{J. Chem. Phys.} \textbf{2010}, \emph{133},
  134111\relax
\mciteBstWouldAddEndPuncttrue
\mciteSetBstMidEndSepPunct{\mcitedefaultmidpunct}
{\mcitedefaultendpunct}{\mcitedefaultseppunct}\relax
\EndOfBibitem
\bibitem[Jain \latin{et~al.}(2016)Jain, Alguire, and
  Subotnik]{jain2016efficient}
Jain,~A.; Alguire,~E.; Subotnik,~J.~E. An efficient, augmented surface hopping
  algorithm that includes decoherence for use in large-scale simulations.
  \emph{J. Chem. Theory Comput.} \textbf{2016}, \emph{12}, 5256\relax
\mciteBstWouldAddEndPuncttrue
\mciteSetBstMidEndSepPunct{\mcitedefaultmidpunct}
{\mcitedefaultendpunct}{\mcitedefaultseppunct}\relax
\EndOfBibitem
\bibitem[Granucci \latin{et~al.}(2012)Granucci, Persico, and
  Spighi]{granucci2012surface}
Granucci,~G.; Persico,~M.; Spighi,~G. Surface hopping trajectory simulations
  with spin-orbit and dynamical couplings. \emph{J. Chem. Phys.} \textbf{2012},
  \emph{137}, 22A501\relax
\mciteBstWouldAddEndPuncttrue
\mciteSetBstMidEndSepPunct{\mcitedefaultmidpunct}
{\mcitedefaultendpunct}{\mcitedefaultseppunct}\relax
\EndOfBibitem
\bibitem[Heller \latin{et~al.}(2021)Heller, Joswig, and
  Seifert]{heller2021exploring}
Heller,~E.~R.; Joswig,~J.-O.; Seifert,~G. Exploring the effects of quantum
  decoherence on the excited-state dynamics of molecular systems. \emph{Theor.
  Chem. Acc.} \textbf{2021}, \emph{140}, 42\relax
\mciteBstWouldAddEndPuncttrue
\mciteSetBstMidEndSepPunct{\mcitedefaultmidpunct}
{\mcitedefaultendpunct}{\mcitedefaultseppunct}\relax
\EndOfBibitem
\bibitem[He{\ss} \latin{et~al.}(1995)He{\ss}, Marian, and
  Peyerimhoff]{hess1995abinitio}
He{\ss},~B.~A.; Marian,~C.~M.; Peyerimhoff,~S.~D. In \emph{Modern electronic
  structure theory, part I}; Yarkony,~D.~R., Ed.; Advanced Series in Physical
  Chemistry; World Scientific: Singapore, 1995; Vol.~2; Chapter 4, pp
  152--278\relax
\mciteBstWouldAddEndPuncttrue
\mciteSetBstMidEndSepPunct{\mcitedefaultmidpunct}
{\mcitedefaultendpunct}{\mcitedefaultseppunct}\relax
\EndOfBibitem
\bibitem[He{\ss} \latin{et~al.}(1996)He{\ss}, Marian, Wahlgren, and
  Gropen]{hess1996mean}
He{\ss},~B.~A.; Marian,~C.~M.; Wahlgren,~U.; Gropen,~O. A mean-field
  spin--orbit method applicable to correlated wavefunctions. \emph{Chem. Phys.
  Lett.} \textbf{1996}, \emph{251}, 365--371\relax
\mciteBstWouldAddEndPuncttrue
\mciteSetBstMidEndSepPunct{\mcitedefaultmidpunct}
{\mcitedefaultendpunct}{\mcitedefaultseppunct}\relax
\EndOfBibitem
\bibitem[Chalupsk{\'{y}} and Yanai(2013)Chalupsk{\'{y}}, and
  Yanai]{chalupsky2013flexible}
Chalupsk{\'{y}},~J.; Yanai,~T. Flexible nuclear screening approximation to the
  two-electron spin--orbit coupling based on ab initio parameterization.
  \emph{J. Chem. Phys.} \textbf{2013}, \emph{139}, 204106\relax
\mciteBstWouldAddEndPuncttrue
\mciteSetBstMidEndSepPunct{\mcitedefaultmidpunct}
{\mcitedefaultendpunct}{\mcitedefaultseppunct}\relax
\EndOfBibitem
\bibitem[Lee \latin{et~al.}(1977)Lee, Ermler, and Pitzer]{lee1977ab}
Lee,~Y.~S.; Ermler,~W.~C.; Pitzer,~K.~S. {\emph{Ab}} {\emph{Initio}} effective
  core potentials including relativistic effects. I. Formalism and applications
  to the Xe and Au atoms. \emph{J. Chem. Phys.} \textbf{1977}, \emph{67},
  5861--5876\relax
\mciteBstWouldAddEndPuncttrue
\mciteSetBstMidEndSepPunct{\mcitedefaultmidpunct}
{\mcitedefaultendpunct}{\mcitedefaultseppunct}\relax
\EndOfBibitem
\bibitem[Pitzer and Winter(1988)Pitzer, and Winter]{pitzer1988electronic}
Pitzer,~R.~M.; Winter,~N.~W. Electronic-structure methods for heavy-atom
  molecules. \emph{J. Phys. Chem.} \textbf{1988}, \emph{92}, 3061--3063\relax
\mciteBstWouldAddEndPuncttrue
\mciteSetBstMidEndSepPunct{\mcitedefaultmidpunct}
{\mcitedefaultendpunct}{\mcitedefaultseppunct}\relax
\EndOfBibitem
\bibitem[Schwerdtfeger(2011)]{schwerdtfeger2011pseudopotential}
Schwerdtfeger,~P. The pseudopotential approximation in electronic structure
  theory. \emph{ChemPhysChem} \textbf{2011}, \emph{12}, 3143--3155\relax
\mciteBstWouldAddEndPuncttrue
\mciteSetBstMidEndSepPunct{\mcitedefaultmidpunct}
{\mcitedefaultendpunct}{\mcitedefaultseppunct}\relax
\EndOfBibitem
\bibitem[Dolg and Cao(2012)Dolg, and Cao]{dolg2012relativistic}
Dolg,~M.; Cao,~X. Relativistic pseudopotentials: their development and scope of
  applications. \emph{Chem. Rev.} \textbf{2012}, \emph{112}, 403--480\relax
\mciteBstWouldAddEndPuncttrue
\mciteSetBstMidEndSepPunct{\mcitedefaultmidpunct}
{\mcitedefaultendpunct}{\mcitedefaultseppunct}\relax
\EndOfBibitem
\bibitem[Pitzer and Winter(1991)Pitzer, and Winter]{pitzer1991spin}
Pitzer,~R.~M.; Winter,~N.~W. Spin-orbit (core) and core potential integrals.
  \emph{Int. J. Quantum Chem.} \textbf{1991}, \emph{40}, 773--780\relax
\mciteBstWouldAddEndPuncttrue
\mciteSetBstMidEndSepPunct{\mcitedefaultmidpunct}
{\mcitedefaultendpunct}{\mcitedefaultseppunct}\relax
\EndOfBibitem
\bibitem[Lingerfelt \latin{et~al.}(2016)Lingerfelt, Williams-Young, Petrone,
  and Li]{lingerfelt2016direct}
Lingerfelt,~D.~B.; Williams-Young,~D.~B.; Petrone,~A.; Li,~X. Direct ab initio
  (meta-)surface-hopping dynamics. \emph{J. Chem. Theory Comput.}
  \textbf{2016}, \emph{12}, 935--945\relax
\mciteBstWouldAddEndPuncttrue
\mciteSetBstMidEndSepPunct{\mcitedefaultmidpunct}
{\mcitedefaultendpunct}{\mcitedefaultseppunct}\relax
\EndOfBibitem
\bibitem[Nijamudheen and Akimov(2017)Nijamudheen, and
  Akimov]{nijamudheen2017excited}
Nijamudheen,~A.; Akimov,~A.~V. Excited-state dynamics in two-dimensional
  heterostructures: SiR/TiO2 and GeR/TiO2 (R= H, Me) as promising
  photocatalysts. \emph{The Journal of Physical Chemistry C} \textbf{2017},
  \emph{121}, 6520--6532\relax
\mciteBstWouldAddEndPuncttrue
\mciteSetBstMidEndSepPunct{\mcitedefaultmidpunct}
{\mcitedefaultendpunct}{\mcitedefaultseppunct}\relax
\EndOfBibitem
\bibitem[Momeni and Brown(2015)Momeni, and Brown]{momeni2015why}
Momeni,~M.~R.; Brown,~A. Why do TD-DFT excitation energies of BODIPY/aza-BODIPY
  families largely deviate from experiment? Answers from electron correlated
  and multireference methods. \emph{J. Chem. Theory Comput.} \textbf{2015},
  \emph{11}, 2619--2632\relax
\mciteBstWouldAddEndPuncttrue
\mciteSetBstMidEndSepPunct{\mcitedefaultmidpunct}
{\mcitedefaultendpunct}{\mcitedefaultseppunct}\relax
\EndOfBibitem
\bibitem[De~Vetta \latin{et~al.}(2019)De~Vetta, Gonz{\'{a}}lez, and
  Corral]{devetta2019role}
De~Vetta,~M.; Gonz{\'{a}}lez,~L.; Corral,~I. The role of electronic triplet
  states and high-lying singlet states in the deactivation mechanism of the
  parent BODIPY: An ADC (2) and CASPT2 study. \emph{ChemPhotoChem}
  \textbf{2019}, \emph{3}, 727--738\relax
\mciteBstWouldAddEndPuncttrue
\mciteSetBstMidEndSepPunct{\mcitedefaultmidpunct}
{\mcitedefaultendpunct}{\mcitedefaultseppunct}\relax
\EndOfBibitem
\bibitem[Vetta and Corral(2019)Vetta, and Corral]{devetta2019ctc}
Vetta,~M.~D.; Corral,~I. Insight into the optical properties of
  meso-pentaﬂuorophenyl(PFP)-BODIPY: An attractive platform for
  functionalization of BODIPY dyes. \emph{Comp. Theor. Chem.} \textbf{2019},
  \emph{1150}, 110--120\relax
\mciteBstWouldAddEndPuncttrue
\mciteSetBstMidEndSepPunct{\mcitedefaultmidpunct}
{\mcitedefaultendpunct}{\mcitedefaultseppunct}\relax
\EndOfBibitem
\bibitem[Postils \latin{et~al.}(2021)Postils, Ruip{\'{e}}rez, and
  Casanova]{postils2021mild}
Postils,~V.; Ruip{\'{e}}rez,~F.; Casanova,~D. Mild open-shell character of
  BODIPY and its impact on singlet and triplet excitation energies. \emph{J.
  Chem. Theory Comput.} \textbf{2021}, \emph{17}, 5825--5838\relax
\mciteBstWouldAddEndPuncttrue
\mciteSetBstMidEndSepPunct{\mcitedefaultmidpunct}
{\mcitedefaultendpunct}{\mcitedefaultseppunct}\relax
\EndOfBibitem
\bibitem[Boguslawski \latin{et~al.}(2013)Boguslawski, Tecmer, Barcza, Legeza,
  and Reiher]{boguslawski2013orbital}
Boguslawski,~K.; Tecmer,~P.; Barcza,~G.; Legeza,~{\"{O}}.; Reiher,~M. Orbital
  entanglement in bond-formation processes. \emph{J. Chem. Theory Comput.}
  \textbf{2013}, \emph{9}, 2959--2973\relax
\mciteBstWouldAddEndPuncttrue
\mciteSetBstMidEndSepPunct{\mcitedefaultmidpunct}
{\mcitedefaultendpunct}{\mcitedefaultseppunct}\relax
\EndOfBibitem
\bibitem[Peach \latin{et~al.}(2008)Peach, Benfield, Helgaker, and
  Tozer]{peach2008excitation}
Peach,~M.~J.; Benfield,~P.; Helgaker,~T.; Tozer,~D.~J. Excitation energies in
  density functional theory: an evaluation and a diagnostic test. \emph{J.
  Chem. Phys.} \textbf{2008}, \emph{128}, 044118\relax
\mciteBstWouldAddEndPuncttrue
\mciteSetBstMidEndSepPunct{\mcitedefaultmidpunct}
{\mcitedefaultendpunct}{\mcitedefaultseppunct}\relax
\EndOfBibitem
\bibitem[Dreuw \latin{et~al.}(2003)Dreuw, Weisman, and
  Head-Gordon]{dreuw2003long}
Dreuw,~A.; Weisman,~J.~L.; Head-Gordon,~M. Long-range charge-transfer excited
  states in time-dependent density functional theory require non-local
  exchange. \emph{J. Chem. Phys.} \textbf{2003}, \emph{119}, 2943--2946\relax
\mciteBstWouldAddEndPuncttrue
\mciteSetBstMidEndSepPunct{\mcitedefaultmidpunct}
{\mcitedefaultendpunct}{\mcitedefaultseppunct}\relax
\EndOfBibitem
\bibitem[Dreuw and Head-Gordon(2004)Dreuw, and Head-Gordon]{dreuw2004failure}
Dreuw,~A.; Head-Gordon,~M. Failure of time-dependent density functional theory
  for long-range charge-transfer excited states: the
  zincbacteriochlorin--bacteriochlorin and bacteriochlorophyll--spheroidene
  complexes. \emph{J. Am. Chem. Soc.} \textbf{2004}, \emph{126},
  4007--4016\relax
\mciteBstWouldAddEndPuncttrue
\mciteSetBstMidEndSepPunct{\mcitedefaultmidpunct}
{\mcitedefaultendpunct}{\mcitedefaultseppunct}\relax
\EndOfBibitem
\bibitem[Jacquemin \latin{et~al.}(2010)Jacquemin, Perp{\`{e}}te, Ciofini,
  Adamo, Valero, Zhao, and Truhlar]{jacquemin2010on}
Jacquemin,~D.; Perp{\`{e}}te,~E.~A.; Ciofini,~I.; Adamo,~C.; Valero,~R.;
  Zhao,~Y.; Truhlar,~D.~G. On the performance of the M06 family of density
  functionals for electronic excitation energies. \emph{J. Chem. Theory
  Comput.} \textbf{2010}, \emph{6}, 2071--2085\relax
\mciteBstWouldAddEndPuncttrue
\mciteSetBstMidEndSepPunct{\mcitedefaultmidpunct}
{\mcitedefaultendpunct}{\mcitedefaultseppunct}\relax
\EndOfBibitem
\bibitem[Dev \latin{et~al.}(2012)Dev, Agrawal, and English]{dev2012determining}
Dev,~P.; Agrawal,~S.; English,~N.~J. Determining the appropriate
  exchange-correlation functional for time-dependent density functional theory
  studies of charge-transfer excitations in organic dyes. \emph{J. Chem. Phys.}
  \textbf{2012}, \emph{136}, 224301\relax
\mciteBstWouldAddEndPuncttrue
\mciteSetBstMidEndSepPunct{\mcitedefaultmidpunct}
{\mcitedefaultendpunct}{\mcitedefaultseppunct}\relax
\EndOfBibitem
\bibitem[Peverati and Truhlar(2012)Peverati, and
  Truhlar]{peverati2012performance}
Peverati,~R.; Truhlar,~D.~G. Performance of the M11 and M11-L density
  functionals for calculations of electronic excitation energies by adiabatic
  time-dependent density functional theory. \emph{Phys. Chem. Chem. Phys.}
  \textbf{2012}, \emph{14}, 11363--11370\relax
\mciteBstWouldAddEndPuncttrue
\mciteSetBstMidEndSepPunct{\mcitedefaultmidpunct}
{\mcitedefaultendpunct}{\mcitedefaultseppunct}\relax
\EndOfBibitem
\bibitem[Li \latin{et~al.}(2014)Li, Nieman, Aquino, Lischka, and
  Tretiak]{li2014comparison}
Li,~H.; Nieman,~R.; Aquino,~A.~J.; Lischka,~H.; Tretiak,~S. Comparison of
  LC-TDDFT and ADC(2) methods in computations of bright and charge transfer
  states in stacked oligothiophenes. \emph{J. Chem. Theory Comput.}
  \textbf{2014}, \emph{10}, 3280--3289\relax
\mciteBstWouldAddEndPuncttrue
\mciteSetBstMidEndSepPunct{\mcitedefaultmidpunct}
{\mcitedefaultendpunct}{\mcitedefaultseppunct}\relax
\EndOfBibitem
\bibitem[Shao \latin{et~al.}(2020)Shao, Mei, Sundholm, and
  Kaila]{shao2020benchmarking}
Shao,~Y.; Mei,~Y.; Sundholm,~D.; Kaila,~V. R.~I. Benchmarking the performance
  of time-dependent density functional theory methods on biochromophores.
  \emph{J. Chem. Theory Comput.} \textbf{2020}, \emph{16}, 587--600\relax
\mciteBstWouldAddEndPuncttrue
\mciteSetBstMidEndSepPunct{\mcitedefaultmidpunct}
{\mcitedefaultendpunct}{\mcitedefaultseppunct}\relax
\EndOfBibitem
\bibitem[Hirata and Head-Gordon(1999)Hirata, and Head-Gordon]{hirata1999time}
Hirata,~S.; Head-Gordon,~M. Time-dependent density functional theory within the
  Tamm–Dancoff approximation. \emph{Chem. Phys. Lett.} \textbf{1999},
  \emph{314}, 291--299\relax
\mciteBstWouldAddEndPuncttrue
\mciteSetBstMidEndSepPunct{\mcitedefaultmidpunct}
{\mcitedefaultendpunct}{\mcitedefaultseppunct}\relax
\EndOfBibitem
\bibitem[El-Sayed(1962)]{elsayed1962radiationless}
El-Sayed,~M.~A. The radiationless processes involving change of multiplicity in
  the diazenes. \emph{J. Chem. Phys.} \textbf{1962}, \emph{36}, 573--574\relax
\mciteBstWouldAddEndPuncttrue
\mciteSetBstMidEndSepPunct{\mcitedefaultmidpunct}
{\mcitedefaultendpunct}{\mcitedefaultseppunct}\relax
\EndOfBibitem
\bibitem[El-Sayed(1963)]{elsayed1963spin}
El-Sayed,~M.~A. Spin--orbit coupling and the radiationless processes in
  nitrogen heterocyclics. \emph{J. Chem. Phys.} \textbf{1963}, \emph{38},
  2834--2838\relax
\mciteBstWouldAddEndPuncttrue
\mciteSetBstMidEndSepPunct{\mcitedefaultmidpunct}
{\mcitedefaultendpunct}{\mcitedefaultseppunct}\relax
\EndOfBibitem
\bibitem[El-Sayed(1968)]{elsayed1968triplet}
El-Sayed,~M.~A. The triplet state: Its radiative and nonradiative properties.
  \emph{Acc. Chem. Res.} \textbf{1968}, \emph{1}, 8--16\relax
\mciteBstWouldAddEndPuncttrue
\mciteSetBstMidEndSepPunct{\mcitedefaultmidpunct}
{\mcitedefaultendpunct}{\mcitedefaultseppunct}\relax
\EndOfBibitem
\bibitem[Perun \latin{et~al.}(2008)Perun, Tatchen, and
  Marian]{perun2008singlet}
Perun,~S.; Tatchen,~J.; Marian,~C.~M. Singlet and triplet excited states and
  intersystem crossing in free-base porphyrin: TDDFT and DFT/MRCI study.
  \emph{Chem. Phys. Chem.} \textbf{2008}, \emph{9}, 282--292\relax
\mciteBstWouldAddEndPuncttrue
\mciteSetBstMidEndSepPunct{\mcitedefaultmidpunct}
{\mcitedefaultendpunct}{\mcitedefaultseppunct}\relax
\EndOfBibitem
\bibitem[Penfold and Worth(2010)Penfold, and Worth]{penfold2010effect}
Penfold,~T.; Worth,~G. The effect of molecular distortions on spin--orbit
  coupling in simple hydrocarbons. \emph{Chem. Phys.} \textbf{2010},
  \emph{375}, 58--66\relax
\mciteBstWouldAddEndPuncttrue
\mciteSetBstMidEndSepPunct{\mcitedefaultmidpunct}
{\mcitedefaultendpunct}{\mcitedefaultseppunct}\relax
\EndOfBibitem
\bibitem[Marian(2012)]{marian2012spin}
Marian,~C.~M. Spin--orbit coupling and intersystem crossing in molecules.
  \emph{Wiley Interdiscip. Rev. Comput. Mol. Sci.} \textbf{2012}, \emph{2},
  187--203\relax
\mciteBstWouldAddEndPuncttrue
\mciteSetBstMidEndSepPunct{\mcitedefaultmidpunct}
{\mcitedefaultendpunct}{\mcitedefaultseppunct}\relax
\EndOfBibitem
\bibitem[Penfold \latin{et~al.}(2018)Penfold, Gindensperger, Daniel, and
  Marian]{penfold2018spin}
Penfold,~T.~J.; Gindensperger,~E.; Daniel,~C.; Marian,~C.~M. Spin-vibronic
  mechanism for intersystem crossing. \emph{Chem. Rev.} \textbf{2018},
  \emph{118}, 6975--7025\relax
\mciteBstWouldAddEndPuncttrue
\mciteSetBstMidEndSepPunct{\mcitedefaultmidpunct}
{\mcitedefaultendpunct}{\mcitedefaultseppunct}\relax
\EndOfBibitem
\bibitem[Yang \latin{et~al.}(2016)Yang, Shen, Zhang, and Yang]{yang2016conical}
Yang,~Y.; Shen,~L.; Zhang,~D.; Yang,~W. Conical intersections from
  particle--particle random phase and Tamm--Dancoff approximations. \emph{J.
  Phys. Chem. Lett.} \textbf{2016}, \emph{7}, 2407--2411\relax
\mciteBstWouldAddEndPuncttrue
\mciteSetBstMidEndSepPunct{\mcitedefaultmidpunct}
{\mcitedefaultendpunct}{\mcitedefaultseppunct}\relax
\EndOfBibitem
\bibitem[Matsika(2021)]{matsika2021electronic}
Matsika,~S. Electronic structure methods for the description of nonadiabatic
  effects and conical intersections. \emph{Chem. Rev.} \textbf{2021},
  \emph{121}, 9407--9449\relax
\mciteBstWouldAddEndPuncttrue
\mciteSetBstMidEndSepPunct{\mcitedefaultmidpunct}
{\mcitedefaultendpunct}{\mcitedefaultseppunct}\relax
\EndOfBibitem
\bibitem[Peach \latin{et~al.}(2011)Peach, Williamson, and
  Tozer]{peach2011influence}
Peach,~M. J.~G.; Williamson,~M.~J.; Tozer,~D.~J. Influence of triplet
  instabilities in TDDFT. \emph{J. Chem. Theory Comput.} \textbf{2011},
  \emph{7}, 3578--3585\relax
\mciteBstWouldAddEndPuncttrue
\mciteSetBstMidEndSepPunct{\mcitedefaultmidpunct}
{\mcitedefaultendpunct}{\mcitedefaultseppunct}\relax
\EndOfBibitem
\bibitem[Dutta and Sherrill(2003)Dutta, and Sherrill]{dutta2003full}
Dutta,~A.; Sherrill,~C.~D. Full configuration interaction potential energy
  curves for breaking bonds to hydrogen: An assessment of single-reference
  correlation methods. \emph{J. Chem. Phys.} \textbf{2003}, \emph{118},
  1610--1619\relax
\mciteBstWouldAddEndPuncttrue
\mciteSetBstMidEndSepPunct{\mcitedefaultmidpunct}
{\mcitedefaultendpunct}{\mcitedefaultseppunct}\relax
\EndOfBibitem
\bibitem[Barbatti \latin{et~al.}(2010)Barbatti, Aquino, and
  Lischka]{barbatti2010uv}
Barbatti,~M.; Aquino,~A.~J.; Lischka,~H. The UV absorption of nucleobases:
  semi-classical ab initio spectra simulations. \emph{Phys. Chem. Chem. Phys.}
  \textbf{2010}, \emph{12}, 4959--4967\relax
\mciteBstWouldAddEndPuncttrue
\mciteSetBstMidEndSepPunct{\mcitedefaultmidpunct}
{\mcitedefaultendpunct}{\mcitedefaultseppunct}\relax
\EndOfBibitem
\bibitem[Ogilby(2010)]{ogilby2010singlet}
Ogilby,~P.~R. Singlet oxygen: there is indeed something new under the sun.
  \emph{Chem. Soc. Rev.} \textbf{2010}, \emph{39}, 3181--3209\relax
\mciteBstWouldAddEndPuncttrue
\mciteSetBstMidEndSepPunct{\mcitedefaultmidpunct}
{\mcitedefaultendpunct}{\mcitedefaultseppunct}\relax
\EndOfBibitem
\bibitem[Grimme \latin{et~al.}(2010)Grimme, Antony, Ehrlich, and
  Krieg]{grimme2010consistent}
Grimme,~S.; Antony,~J.; Ehrlich,~S.; Krieg,~H. A consistent and accurate ab
  initio parametrization of density functional dispersion correction (DFT-D)
  for the 94 elements H-Pu. \emph{J. Chem. Phys.} \textbf{2010}, \emph{132},
  154104\relax
\mciteBstWouldAddEndPuncttrue
\mciteSetBstMidEndSepPunct{\mcitedefaultmidpunct}
{\mcitedefaultendpunct}{\mcitedefaultseppunct}\relax
\EndOfBibitem
\bibitem[Ajitha \latin{et~al.}(2002)Ajitha, Fedorov, Finley, and
  Hirao]{ajitha2002photodissociation}
Ajitha,~D.; Fedorov,~D.; Finley,~J.; Hirao,~K. Photodissociation of alkyl and
  aryl iodides and effect of fluorination: Analysis of proposed mechanisms and
  vertical excitations by spin--orbit ab initio study. \emph{J. Chem. Phys.}
  \textbf{2002}, \emph{117}, 7068--7076\relax
\mciteBstWouldAddEndPuncttrue
\mciteSetBstMidEndSepPunct{\mcitedefaultmidpunct}
{\mcitedefaultendpunct}{\mcitedefaultseppunct}\relax
\EndOfBibitem
\bibitem[Baig \latin{et~al.}(2024)Baig, Mehmood, and
  Akhtar]{baig2024relativistic}
Baig,~M.~W.; Mehmood,~H.; Akhtar,~T. Relativistic two-component density
  functional study of ethyl 2-(2-iodobenzylidenehydrazinyl)
  thiazole-4-carboxylate. \emph{Comput. Theor. Chem.} \textbf{2024},
  \emph{1237}, 114670\relax
\mciteBstWouldAddEndPuncttrue
\mciteSetBstMidEndSepPunct{\mcitedefaultmidpunct}
{\mcitedefaultendpunct}{\mcitedefaultseppunct}\relax
\EndOfBibitem
\bibitem[Reindl \latin{et~al.}(1997)Reindl, Penzkofer, Gong, Landthaler,
  Szeimies, Abels, and B{\"a}umler]{reindl1997quantum}
Reindl,~S.; Penzkofer,~A.; Gong,~S.-H.; Landthaler,~M.; Szeimies,~R.;
  Abels,~C.; B{\"a}umler,~W. Quantum yield of triplet formation for indocyanine
  green. \emph{J. Photochem. Photobiol. A} \textbf{1997}, \emph{105},
  65--68\relax
\mciteBstWouldAddEndPuncttrue
\mciteSetBstMidEndSepPunct{\mcitedefaultmidpunct}
{\mcitedefaultendpunct}{\mcitedefaultseppunct}\relax
\EndOfBibitem
\bibitem[Bachilo and Weisman(2000)Bachilo, and
  Weisman]{bachilo2000determination}
Bachilo,~S.~M.; Weisman,~R.~B. Determination of triplet quantum yields from
  triplet- triplet annihilation fluorescence. \emph{J. Phys. Chem. A}
  \textbf{2000}, \emph{104}, 7711--7714\relax
\mciteBstWouldAddEndPuncttrue
\mciteSetBstMidEndSepPunct{\mcitedefaultmidpunct}
{\mcitedefaultendpunct}{\mcitedefaultseppunct}\relax
\EndOfBibitem
\end{mcitethebibliography}

\end{document}